\documentclass[twocolumn]{aastex631}

\newcommand\cgsunits{{\rm erg\,s^{-1}\,cm^{-2}\,arcsec^{-2}}}

\usepackage{color}
\definecolor{mygreen}{HTML}{009900}

\newcommand\comment[1]{\textcolor{red}{({\tiny remove this comment--} #1)}}

\newcommand\joao[1]{\textcolor{mygreen}{#1}}

\newcommand\muse{{MUSE}}
\newcommand\musew{MUSE-Wide}
\usepackage{pifont}
\newcommand{\cmark}{\textcolor{mygreen}{\Large\text{\ding{51}}}}
\newcommand{\xmark}{\textcolor{red}{\Large\text{\ding{55}}}}
\newcommand\qmark{\textcolor{blue}{\large\fontfamily{cyklop}\selectfont \textit{?}}}
\received{\today} %March 1, 2021}
\submitjournal{ApJ}

\shorttitle{Double peak H$\alpha$ emission in galaxy halos}
\shortauthors{S\'anchez Almeida et al.}
\graphicspath{{./}{figures/}}
\begin{document}

\title{
%Discovery of faint H$\alpha$ emission clumps in the circumgalactic medium of nearby galaxies\\
%Discovery of ultra-faint double-peak H$\alpha$ emission clumps in the circumgalactic medium of nearby galaxies}
%Discovery of faint double-peak H$\alpha$ emission in the circumgalactic medium of low redshift galaxies\\
Discovery of faint double-peak H$\alpha$ emission in the halo of low redshift galaxies
}

\correspondingauthor{JSA}
\email{jos@iac.es}

\author[0000-0003-1123-6003]{J. S\'anchez Almeida} \affil{Instituto de Astrof\'\i sica de Canarias, La Laguna, Tenerife, E-38200, Spain} \affil{Departamento de Astrof\'\i sica, Universidad de La Laguna, Spain}

\author[0000-0003-1803-6899]{J. Calhau} \affil{Instituto de Astrof\'\i sica de Canarias, La Laguna, Tenerife, E-38200, Spain} \affil{Departamento de Astrof\'\i sica, Universidad de La Laguna, Spain}

\author[0000-0001-8876-4563]{C. Mu\~noz-Tu\~n\'on} \affil{Instituto de Astrof\'\i sica de Canarias, La Laguna, Tenerife, E-38200, Spain} \affil{Departamento de Astrof\'\i sica, Universidad de La Laguna, Spain}

\author[0000-0002-2733-0670]{A. L.  Gonz\'alez-Mor\'an} \affil{Instituto de Astrof\'\i sica de Canarias, La Laguna, Tenerife, E-38200, Spain} \affil{Departamento de Astrof\'\i sica, Universidad de La Laguna, Spain}

\author[0000-0002-0674-1470]{J. M.  Rodr\'\i guez-Espinosa} \affil{Instituto de Astrof\'\i sica de Canarias, La Laguna, Tenerife, E-38200, Spain} \affil{Departamento de Astrof\'\i sica, Universidad de La Laguna, Spain} 
\affil{Instituto de Astrof\'\i sica de Andaluc\'\i a, Granada,  E-18008, Spain}

%\author{August Muench}
%\affiliation{American Astronomical Society \\
%1667 K Street NW, Suite 800 \\

%\collaboration{6}{(AAS Journals Data Editors)}

%\author{Butler Burton}
%\affiliation{Leiden University}
%\affiliation{AAS Journals Associate Editor-in-Chief}

%\altaffiliation{AASTeX v6+ programmer}

%\affiliation{AAS Director of Publishing}
%\affiliation{American Astronomical Society \\
%1667 K Street NW, Suite 800 \\

%\author{Magaret Donnelly}
%\affiliation{IOP Publishing, Washington, DC 20005}



\begin{abstract}
Aiming at the detection of cosmological gas being accreted onto galaxies of the local Universe, we examined the H$\alpha$ emission in the halo of 164 galaxies in the field of view of the Multi-Unit Spectroscopic Explorer Wide survey (\musew ) with observable H$\alpha$  (redshift $< 0.42$). An exhaustive screening of the corresponding H$\alpha$ images led us to select 118 reliable H$\alpha$ emitting gas clouds. The signals are faint, with a surface brightness of ${\rm 10^{-17.3\pm 0.3}\,erg\,s^{-1}\,cm^{-2}\,arcsec^{-2}}$. 
Through statistical tests and other arguments, we ruled out that they are created by instrumental artifacts, telluric line residuals, or high redshift interlopers. 
Around 38\,\%\ of the time, the H$\alpha$ line profile shows a double peak with the drop in intensity at the rest-frame of the central galaxy, and  with a typical peak-to-peak separation of the order of $\pm 200\,{\rm km\,s^{-1}}$. 
%
%Only a few of the H$\alpha$ emitting clumps also show  H$\beta$ or [O{\sc iii}]$\lambda$5007. 
%
%The non-detection of [N{\sc ii}]$\lambda$6583 sets a loose upper limit on the metallicity of the emitting gas,  ${\rm 12+\log(O/H) < 8.5}$.
%
Most line emission clumps are spatially unresolved. %at the resolution limit of the observation. %, with  diameters of ${\rm 0.9\pm 0.4\,arcsec}$  equivalent to ${\rm 4.3\pm 2.0\, kpc}$ at the redshift of the central galaxy.
%
%The inferred H$\alpha$ luminosities are around $\rm 10^{38.5\pm 0.5} \, erg \, s^{-1}$. %\comment{Joao: revise scattet. I got it by eye from Fig. 8} \joao{Bang on. You have better eyes than I do.}
%
The mass of emitting gas is estimated to be between  one and $10^{-3}$ times the stellar mass of the central galaxy. 
The signals are not isotropically distributed; their azimuth tends to be aligned with the major axis of the corresponding galaxy.
The distances to the central galaxies are not random either. The counts drop at a distance $> 50$ galaxy radii, which roughly corresponds to the virial radius of the central galaxy.
%
%
%We explore several physical mechanisms to explain these observations 
%
%\jorge{(either)} (accretion disks around rogue intermediate mass black holes, expanding super nova bubbles, cosmological gas accretion, and various kinds of shocks),  but none of them fully work. %
%
%including the cosmological gas accretion that motivated the study. 
%Of them all, accretion disks around rogue intermediate mass black holes seem to fit the observations best.
We explore several physical scenarios to explain this H$\alpha$ emission, among which accretion disks around rogue intermediate
mass black holes fit the observations best.
\end{abstract}

\keywords{
Extragalactic astronomy (506) ---
Circumgalactic medium (1879) ---
Galaxy accretion (575) ---
Galaxy fountains (596) ---
Galaxy winds (626) ---
Galaxy formation (595) ---
Intergalactic medium (813) ---
Intergalactic gas (812) ---
Astrophysical black holes (98) --
Intermediate-mass black holes (816)
}


%\addtocounter{section}{-1} % so that this section is number zero
%\section{TBD}
%\begin{itemize}
%\item[-] What is the typical velocity separation between the two peaks of the double peak profiles? \joao{Done}
%\item [-]  Is there any relation between the properties of the Ha emitting clums (e.g., mass, flux, size, central distance) and the properties of the central galaxy? The topic came up after a discussion with Ana Luisa and, maybe, this is a work for her to carry out.
%\item[-] [N{\sc ii}]$\lambda$6585 $\longrightarrow$ [N{\sc ii}]$\lambda$6583 \joao{ Done}
%\end{itemize}

%{\em When you have eliminated the impossible, whatever remains, however improbable, must be the truth} Sherlock Holmes in {\em The Sign of the Four}
%\joao{Nice!}
%%%%%%%%%%%%%%
\section{Introduction} \label{sec:intro}

According to theoretical considerations and current numerical simulations, the accretion of gas from the cosmic web is the fundamental driver of star formation in galaxies \citep[e.g.,][]{2009Natur.457..451D,2012RAA....12..917S,2017ASSL..430.....F,2020ARA&A..58..363P}. The process is particularly important in low-mass galaxies (dark matter halo mass $< 10^{12}\,{\rm M}_\odot$; e.g.,
\citeauthor{2013MNRAS.435..999D}\citeyear{2013MNRAS.435..999D}), hence, in the early Universe when most galaxies had low masses, but also in the present Universe around isolated dwarfs. 
The preference for low-mass galaxies has to do with the physics of the cosmological accretion.  When the gas encounters a massive halo ($> 10^{12}\,{\rm M}_\odot$), it becomes shock heated and requires long time to cool down and settle. However, gas streams reach the inner halo of low-mass galaxies directly, in a process called cold-flow accretion \citep[][]{2003MNRAS.345..349B}.  
The cosmological gas is predicted to be tenuous, patchy, partly ionized, multi-phase, and of very low metallicity \citep[e.g.,][]{2012MNRAS.423.2991V}, and it becomes intertwined with gas from the outflows produced by stellar feedback, i.e., by the collective effect of stellar winds and SN (super nova) explosions from short-lived stars. Inflow and outflow rates are often comparable, which implies that the gas ending up in stars corresponds to only a small fraction of the gas involved in large-scale gas flows \citep[e.g.,][]{2012MNRAS.421...98D,2012ApJ...760...50S}. Shocks are to be expected either when gas accreted from the inter-galactic medium (IGM) meets the gas already existing in the galaxy halo, or when SN driven outflows encounter the gas in the halo \citep[see][and references therein]{2014A&ARv..22...71S}.

The above picture is as clear in simulations as it has been difficult to confirm observationally. The search for cosmic gas inflows has only had partial and/or indirect success \citep[see][and references therein]{2008A&ARv..15..189S,2014A&ARv..22...71S,2017ASSL..430...67S}. For example, some of the Ly$\alpha$ and metallic line absorptions on quasar spectra seem to be created  by  intervening gas in the IGM \citep[see, e.g.,][]{2011Sci...334.1245F,2019MNRAS.485.1595P}. Likewise, chemical inhomogeneities in disk galaxies appear to be the result of pristine gas accretion \citep[][]{2010Natur.467..811C,2013ApJ...767...74S,2015ApJ...810L..15S,2018MNRAS.476.4765S,Hwang_pp_2018,2019ApJ...882....9S,2021MNRAS.505.4655S}.  Cosmic web gas may explain some of the HI filaments found in blind HI surveys \citep[][]{2011A&A...527A..90P}, as well as some of the high velocity clouds detected around our galaxy \citep[e.g.,][]{2008A&ARv..15..189S}.  Finally, Ly$\alpha$ emission has been detected in cosmic web filaments \citep[][]{2014ApJ...786..107M,2014Natur.506...63C,2016A&A...587A..98W,2021A&A...647A.107B}, around quasars \citep[][]{2016ApJ...829....3A}, and such emission becomes pervasive when going faint enough, so that any line-of-sight intercepts emitting gas in between redshift 6 and us \citep{2018Natur.562..229W}. All these findings are rather indirect, in the sense that connecting them with the predictions of numerical simulations still requires a significant dose of interpretation.  Therefore, it is important to develop alternative yet complementary approaches to detect and characterize gas flowing around galaxies.

Our work is aimed at exploring a new way to detect and characterize this gas in the local Universe through its emission in H$\alpha$.  As we mentioned above, there is much evidence for Ly$\alpha$ emission around high-redshift galaxies tracing gas in the IGM and CGM (circum-galactic medium). The Ly$\alpha$ emission of nearby galaxies is not accessible from ground-based observations but, fortunately, most physical processes generating Ly$\alpha$ photons produce  H$\alpha$ as well (details in App.~\ref{app:predictions}). The expected H$\alpha$ flux is very low, typically $\lesssim\,10^{-18}\cgsunits$   (App.~\ref{app:predictions}; Table~\ref{tab:summary}), but this flux limit is within reach of the IFU Multi-Unit Spectroscopic Explorer \citep[MUSE;][]{2014Msngr.157...13B} in one hour integration \citep[][]{2015A&A...575A..75B}. We take advantage of the existing \musew\ ancillary data set \citep{2019A&A...624A.141U}, which has the required depth and field-of-view (FOV; see Sect.~\ref{sec:data}) to look for faint H$\alpha$ signals around low redshift galaxies. The FOV has to be significant since the virial radius of one of these galaxies, where the transition between CGM and IGM occurs, is often of the order of one arcmin around its center (further details in Sect.~\ref{sec:area}). In addition, a large FOV also provides enough central galaxies to have a proper statistics.

The paper is organized as follows: 
Sect.~\ref{sec:data} describes the \musew\ data set employed in the work. We present the selected central galaxies (Sect.~\ref{sec:central_galax}) and analyze the effectiveness of the telluric line correction (Sect.~\ref{sec:telluric}).  
The data analysis is detailed in Sect.~\ref{sec:data_analysis}. 
These days when AI-assisted data analysis is all the rage, our signal detection algorithm was guided  by human intuition (Sect.~\ref{candidate_classification}). The reliability of the detection was assesses later on using statistical tests. We did not have enough previous knowledge on the expected signals to be able to devise an automated algorithm. %\joao{I just know some AI/ML enthusiasts are going to see red after reading this paragraph :-D}
Various sub-sections are devoted to explain the searching area (Sect.~\ref{sec:area}), the extraction of images from the original data-cubes (Sects.~\ref{datacubes} and \ref{Ha,R,G images}), and to discard various potential biases (Sect.~\ref{skylines}).
The outcome of the search is given in Sect.~\ref{sec:results}, where we put forward the main physical properties of the detected H$\alpha$ emiting clumps; their number density (Sect.~\ref{sec:statistics}), line shapes (Sect.~\ref{sec:line_shapes}), redshifts, fluxes and sizes (Sect.~\ref{sec:physical_properties}), masses (Sect.~\ref{sec:gas_mass}), and spatial distribution (Sects.~\ref{sec:Azimuths} and \ref{sec:distances}). A cross-match with external catalogs is described in Sect.~\ref{sec:Xrays_radio_match}. Section~\ref{sec:origin} is devoted to explore various physical scenarios to explain the observation. 
Finally, the main results are summarized and discussed in Sect.~\ref{sec:conclusions}. The appendixes are devoted to estimate the magnitude of the expected H$\alpha$ signals (App.~\ref{app:predictions}), the contamination by interlopers (App.~\ref{app:a}), the shape of the residuals left by insufficient telluric line correction (App.~\ref{app:b}), and the search for X-ray counterparts by stacking (App.~\ref{sec:xray_stacking}). 
%\comment{We may want to add another appendix on the stacking of X-ray signals.}

Whenever it was required, we adopted the cosmological parameters $H_0 = 70\, {\rm km\,s^{-1}\,Mpc^{-1}}$, $\Omega_M$ = 0.3, $\Omega_\Lambda$ = 0.7. None of the results presented in the paper depends significantly on this assumption. %\comment{Joao: plug in the values that you use} 

%%%%%%%%%%%%%%%

\section{Description of the MUSE Wide dataset}\label{sec:data}

\begin{figure*}
\centering
\includegraphics[width=0.9\linewidth]{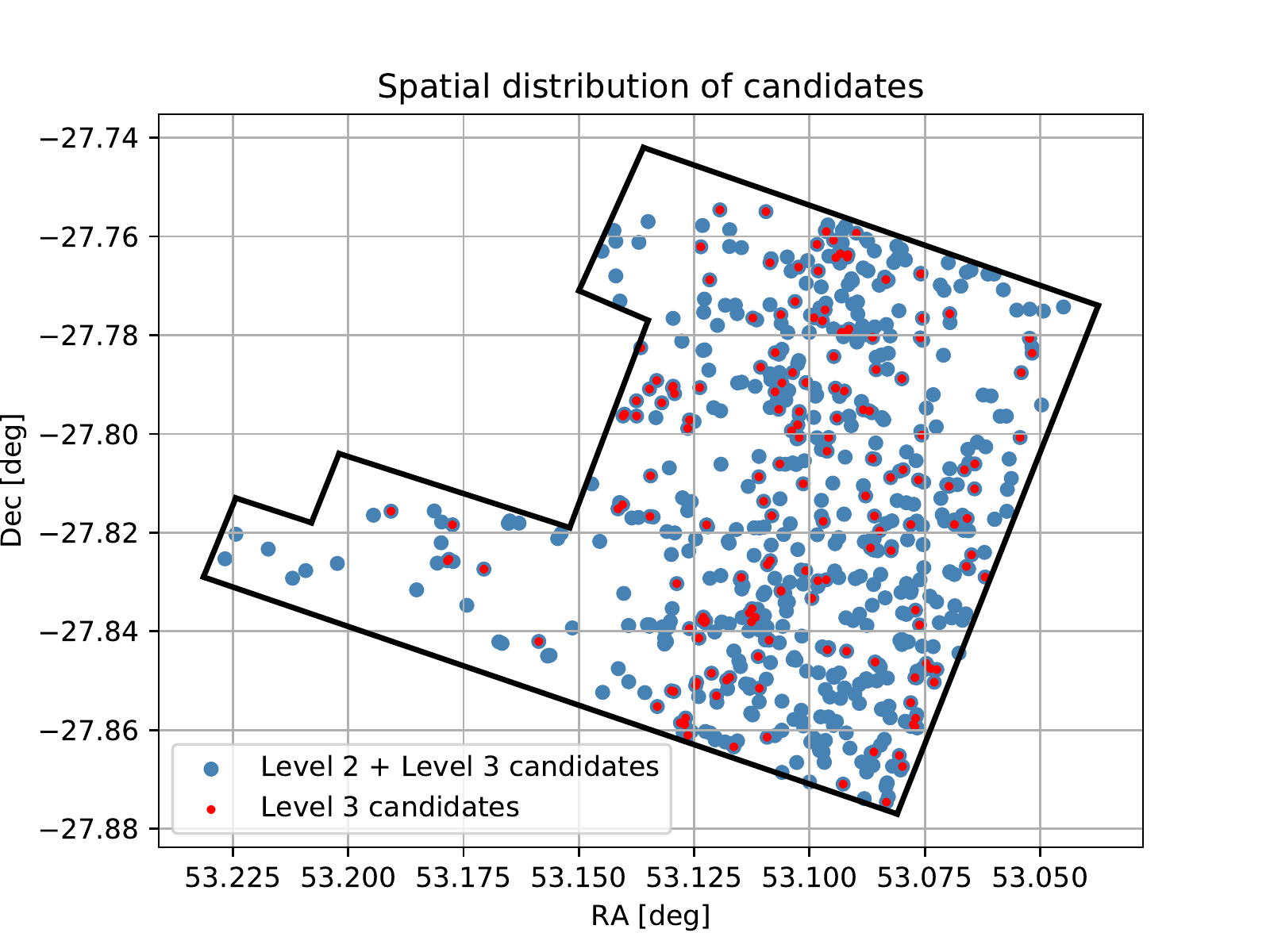}
\caption{Spatial distribution of all H$\alpha$ emitting candidates classified as level 2 and 3. These are the candidates that proceed to the second stage of scrutiny, having their spectra inspected for possible telluric contamination. The outline illustrates the footprint of the \musew\ survey in its current state. The color code is explained in the inset.}
\label{fig:spatial_distro}
\end{figure*}
The Multi-Unit Spectroscopic Explorer \citep[MUSE;][]{2014Msngr.157...13B} is a second generation instrument on the Very Large Telescope (VLT) designed for integral field spectroscopy in the optical band (4750\,\AA\,--\,9350\,\AA). In its wide field mode, it has a $\rm 1\arcmin \times 1\arcmin$ FOV with a spatial sampling of 0.2\arcsec . The spectral  sampling is 1.25\,\AA, resulting in a resolution of around $\rm 2.5\,\AA$. It is capable of taking around 90000 spectra in a single exposure. %The instrument was tested on the Hubble Deep Field South through a single 27 hour integration on a 1$\rm \, arcmin^2$ region and was able to provide integrations for almost 200 objects simultaneously, including some without HST data \citep[][]{2015A&A...575A..75B}. 
We direct the interested reader to \citet{2014Msngr.157...13B,2015A&A...575A..75B} for an in-depth description of the instrument.

The \musew\ survey is a blind spectroscopic survey encompassing the CANDELS/GOODS-S and CANDELS/COSMOS regions. We include the outline of the current state of the \musew\ survey in Fig.~\ref{fig:spatial_distro}. In its complete form, it will cover ${\rm 100\,arcmin^{2}}$ of the sky and will provide spectroscopic information for thousands of galaxies. It complements the already existing MUSE-Deep survey in the Hubble Deep Field South \citep[][]{2017A&A...608A...1B} by enlarging the search area at the cost of increasing noise level. Each pointing has an integration time of 1 hours.
The current release (DR1) comprises 44 ${\rm 1\arcmin \times 1\arcmin}$ contiguous fields (Fig.~\ref{fig:spatial_distro}) and includes catalogues of detected galaxies with information on emission lines, stellar masses, and redshifts. 
%Cross-matching with other catalogues provides additional information such as photometry in the optical and infrared bands. 
We direct the interested reader to the data release paper by \cite{2019A&A...624A.141U}.

%
%%%%%%%%%%%%%%%%%%
\subsection{Selection of target galaxies}\label{sec:central_galax}
Our target galaxies were selected for H$\alpha$ to appear within the \muse\ wavelength range. Thus, all galaxies in the \musew\ DR1 catalog with redshift $<0.42$ were chosen to search for their H$\alpha$ emission, resulting in a total of 164 possible host galaxies with an average redshift of $\rm 0.29 \pm 0.08$. Here and throughout the paper, error bars refer to the standard deviation of the named quantity.

The properties of these galaxies are obtained by cross-matching, within 1\arcsec, the \musew\  Survey DR1 catalog with the multi-wavelength catalogue from \cite{2013ApJS..207...24G}. This allows us to obtain information on the radius of the galaxies
%20\%, 50\% and 80\% light radius of our galaxies, 
as well as on their HST fluxes in the WFC3 filters including UV, optical, and infrared data. We also cross-match the \musew\ catalog with \cite{2012ApJS..203...24V} to obtain parameters like S\'ersic index and orientation.
Neither \cite{2013ApJS..207...24G} nor \cite{2012ApJS..203...24V} completely overlap with the \musew\ catalog. The match includes around 80\,\% of the galaxies. We attempt to recover as many galaxies as possible by cross-matching them with other available catalogues %\comment{Joao: is this true? Correct if required} 
but, when this is not possible, the missing data are excluded from the analysis.  %\comment{Joao: comment what did we do to replace the data when not available.}

\begin{figure}
\centering
\includegraphics[width=0.9\linewidth]{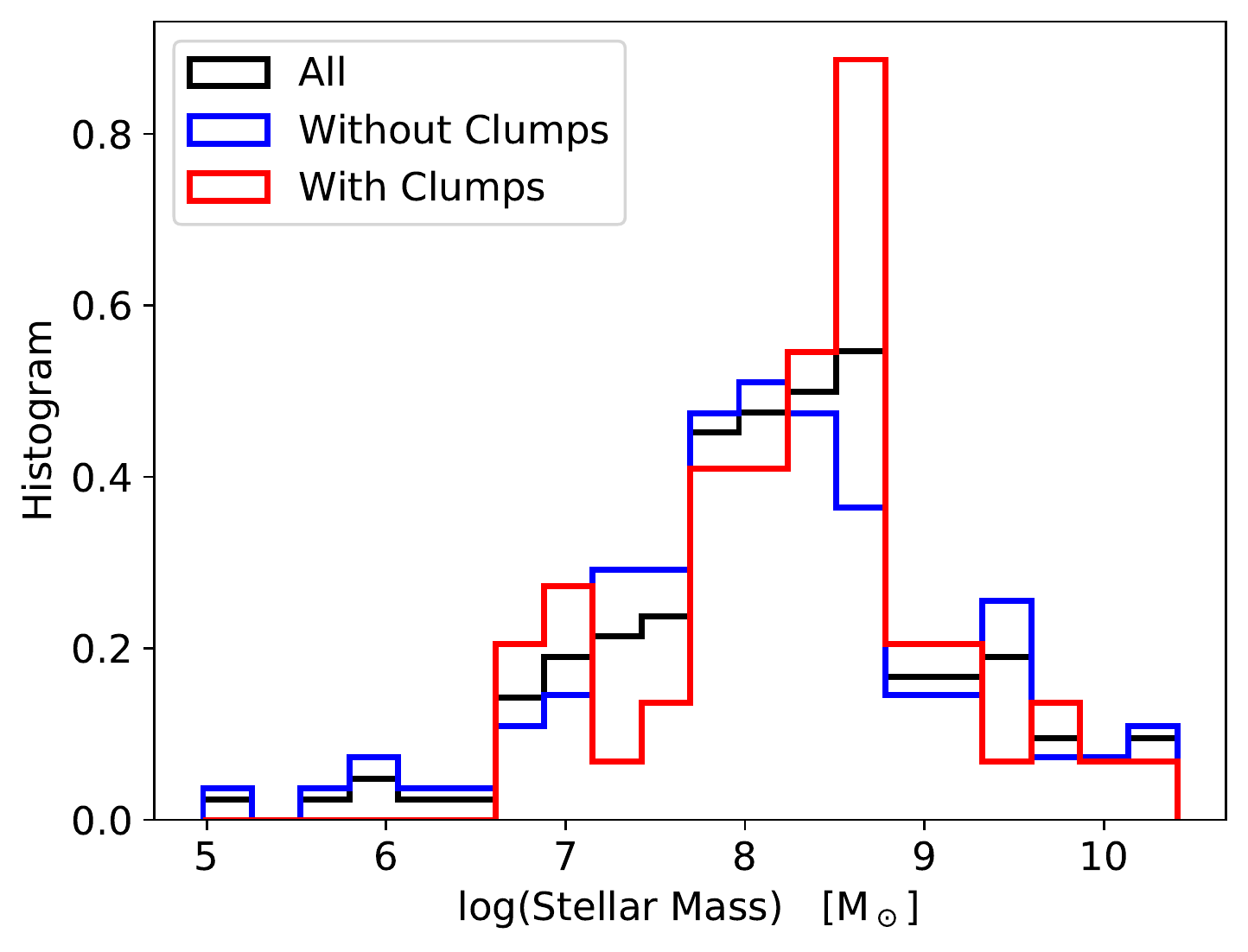}\\ % I did the plots reproducin Joao's format
\includegraphics[width=0.9\linewidth]{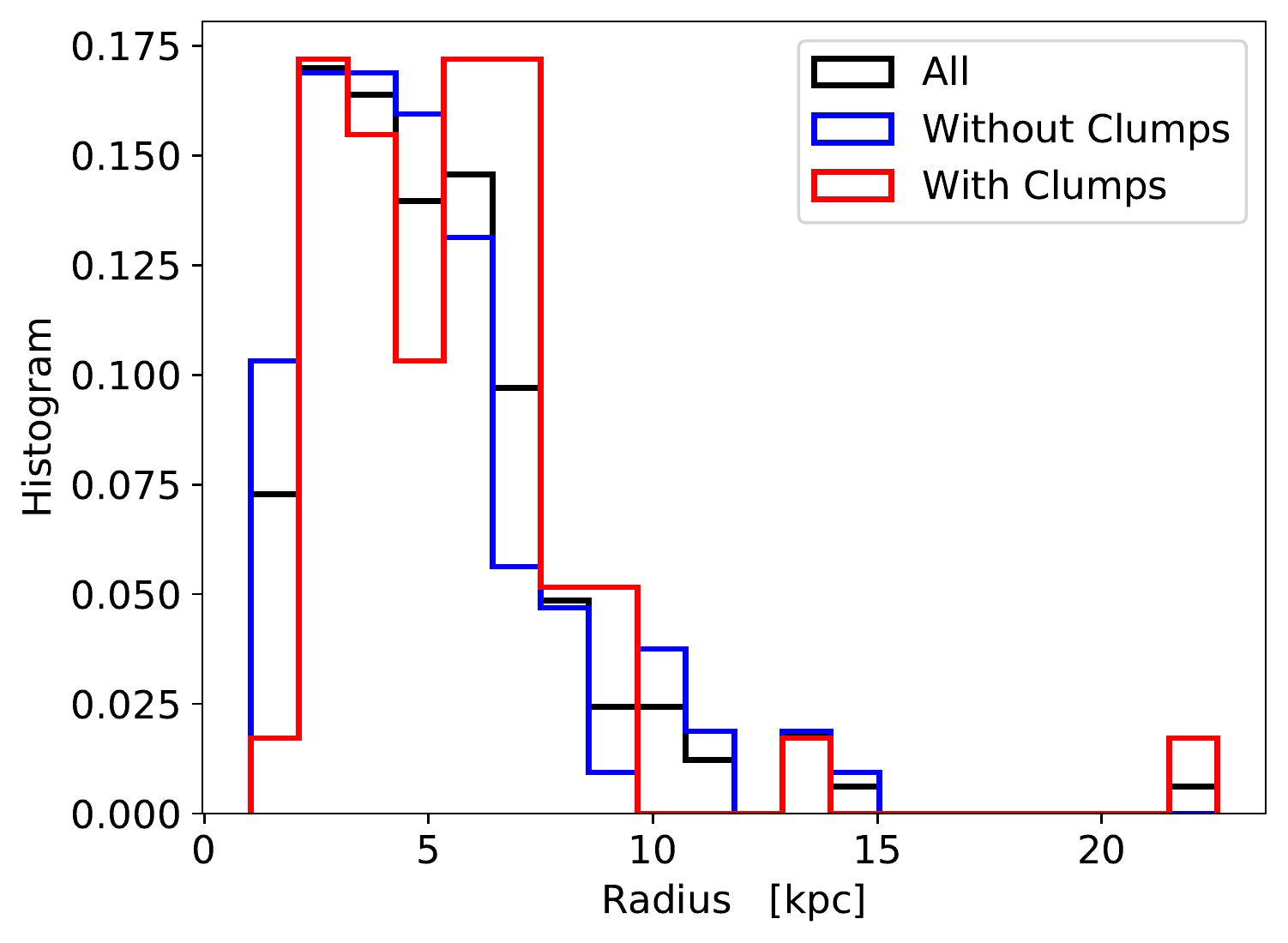}\\
\includegraphics[width=0.9\linewidth]{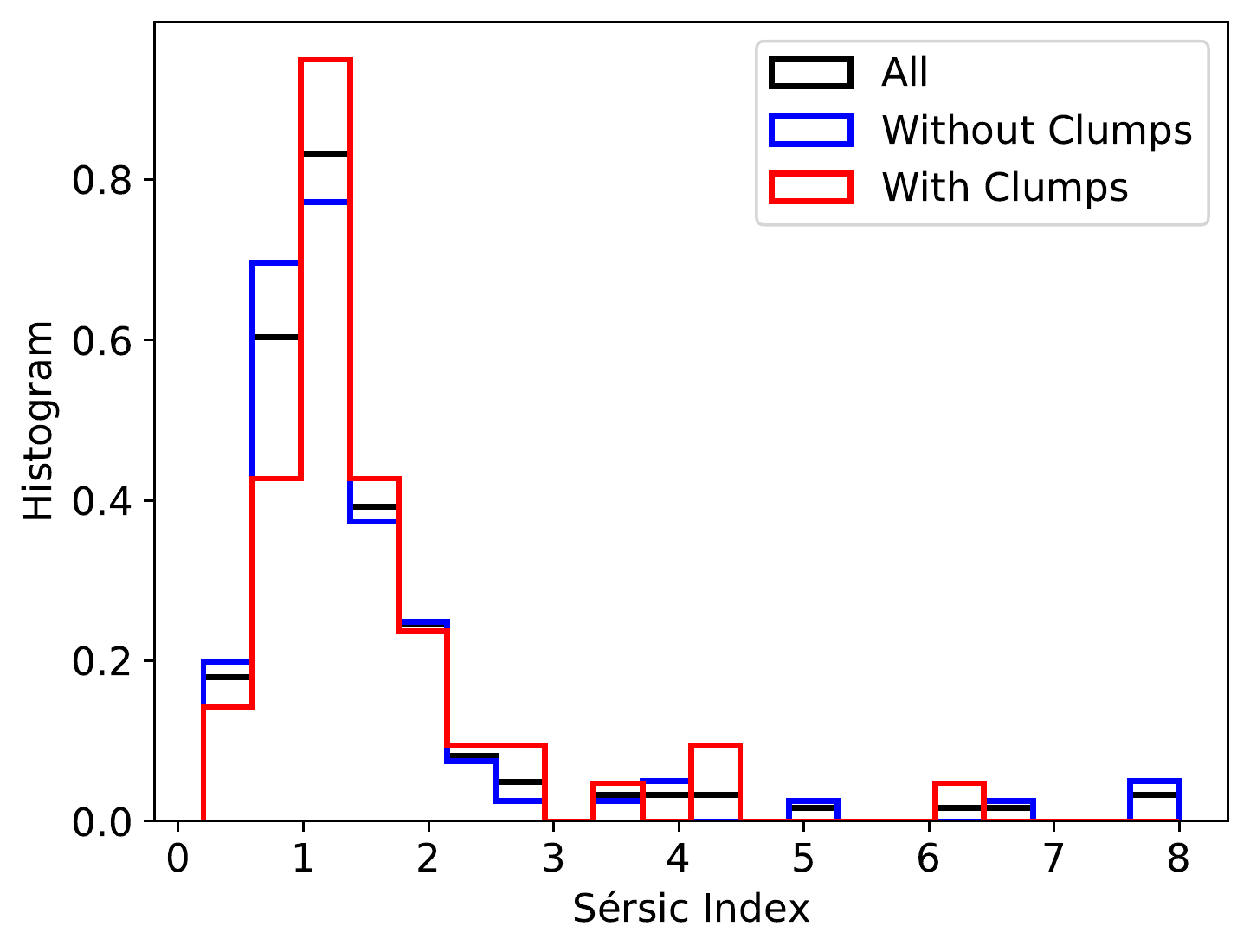}
\caption{
Global physical parameters of the central galaxies considered in our study.
Top panel: distribution of stellar masses of the parent sample (the black line), of those galaxies having H$\alpha$ emitting clumps (the red line), and of those without emitting clumps (the blue line).  All histograms are normalized to area equals one.
Middle panel: same color code as in the upper panel but for galaxy radii. Bottom panel: same color code as in the upper panel but for S\'ersic indexes.
% \joao{Histograms showing the distribution of the S\'{e}rsic indexes of the populations of galaxies considered in this study against the distribution of the galaxies which possess valid candidates for H$\alpha$ emission clumps. The galaxies scrutinized in this work have average stellar mass of $\rm 10^{9} \, M_{\odot}$ and an average S\'{e}rsic index of $\rm \sim 2$. The average radius of these galaxies is $\rm \sim 7 \, kpc$. As can be seen from the histograms, the sub-sample containing only galaxies with valid gas clumps candidates have lower averages for all quantities, suggesting gas clumps have a 'preference' for smaller, low mass spirals (see Sect. \ref{sec:Candidates v Galaxy}). The histograms have been normalized by area.}
%Galaxies with H$\alpha$ clumps tend to have smaller masses, radii, and, perhaps, S\'ersic indexes (Sect.~\ref{sec:Candidates v Galaxy}).
References for the used S\'ersic indexes, masses and radii are given in the main text, Sect.~\ref{sec:central_galax}.
}
\label{fig:host_histograms}
\end{figure}

The galaxies selected for scrutiny
have an average stellar mass ($M_\star$) of $\rm 10^{8.2 \pm 0.9} \, M_{\odot}$ %$\rm 10^{8.5 \pm 1.8} \, M_{\odot}$ 
with the most massive galaxies reaching $\rm \sim 10^{12} \, M_{\odot}$. They present a S\'ersic index of $\rm \sim 1.5 \pm 1.2$, about half-way between the expected values for low-mass spirals ($\sim 1$) and massive ellipticals  ($\gtrsim 4$). Their average radius and dispersion are $\rm 5.2 \pm 3.0 \, kpc$. The actual values of these three parametrers are distributed as shown in Fig.~\ref{fig:host_histograms} (the black lines).
%
%\comment{these numbers have to be double checked before submission} % Done.
%\comment{joao. I keep here the properties of the whole set. The other subsets have not been mentioned yet.} \joao{Understood.}
%I copy most of this paragraph to Sect.~\ref{sec:Candidates v Galaxy}. Revise whether what I interpreted is correct.}
%If we consider only galaxies without candidates, the values are similar to the overall population, although slightly higher (see also Sect. \ref{sec:Candidates v Galaxy} for a comparison with the galaxies that only host valid gas candidates.} %\joao{Add information on the galaxy radius and represent everything in histogram plots. Try to see if it is easy to show the difference between those which have and have not clumps.}

%
%%%%%%%%%%%%%%
\subsection{Sky emission subtraction}\label{sec:telluric}

Molecules in the terrestrial atmosphere,  particularly ${\rm O_2}$  and ${\rm H_2O}$, absorb and emit light on their own, with many lines in the red and near-infrared. Ground based observations have therefore to be decontaminated from these telluric lines. This correction is particularly important in our case since we are looking for weak emission lines in the observed spectra. The sky subtraction is handled by \musew\ using principal component analysis (PCA) through the Zurich Atmospheric Purge algorithm \citep[\textsc{zap} - ][]{2016MNRAS.458.3210S}. \textsc{zap} is particularly effective in our case, when the astronomical sources are weak and small. Regions in the science data-cubes devoid of astronomical signal are used to reconstruct the shape of the telluric line spectrum to be removed. The shape is the same but the amplitude differs from one spaxel to another. There is no need for a list of telluric lines to carry out the correction, and both weak and strong emission lines are taken into account simultaneously and self-consistently. However, the procedure has the inherent risk of the model sky spectrum also including faint, extended, true astronomical signals, which will be erroneously removed from the science spectra.  \textsc{zap} minimizes this confusion problem through spectral filtering and by segmentation of the sky spectrum at particular wavelengths \citep[for details, see][]{2016MNRAS.458.3210S}. This reduces or eliminates the number of astronomical signals included in the sky eigenspectra.

While \textsc{zap} does a very good job correcting for sky lines, there is still the possibility of contamination at low-noise level, which is where we expect the signals to be.
This residual contamination is particularly deceitful since averaging signal-free regions of the cube does not result in the residuals being enhanced, as one would normally expect from a stacking process. Therefore, we cannot depend on stacking to identify residuals. 
Nonetheless, the existence of sky line residuals does not seem to be affecting the conclusions of the work. The detected H$\alpha$ signals do not overlap with significant sky emission lines (Sect.~\ref{skylines}) and the shape of the detected signals does not agree with the shape of the residuals left by {\sc zap} (App.~\ref{app:b}). Additional arguments are collected in Sect.~\ref{skylines}.

%%%%%%%%%%%%%
%\section{\jorge{Procedure used to detect faint H$\alpha$ emission}}\label{sec:data_analysis}
\section{Detection of Faint H$\alpha$ Emission}\label{sec:data_analysis}

This section presents the steps taken to find and select faint H$\alpha$ emission around the galaxies introduced in Sect.~\ref{sec:central_galax}. The searching area around the galaxies is defined in Sect.~\ref{sec:area}. Section~\ref{datacubes} explains how we merged the original data-cubes as required to carry out the actual search. The search is based on the inspection of H$\alpha$ and broad-band images produced out of the \musew\ data-cubes (Sect.~\ref{Ha,R,G images}). Candidates are selected as features with excess H$\alpha$ signal and no broad-band counterpart. The spectrum of each candidate is then inspected individually to identify bona-fide H$\alpha$ emission clumps (Sect.~\ref{candidate_classification}). Finally, we devote Sect.~\ref{skylines} to argue that neither artifacts from the reduction pipeline nor contamination from  telluric lines explain the detected signals, indicating that they must have an astronomical origin.

%
%%%%%%%
%
\subsection{Searching area}\label{sec:area}
In order to identify structures with faint H$\alpha$ emission in the CGM of the selected galaxies, we define the searching area as a disk having 100~times the radius of the host galaxy, taken to be the radius containing 80\,\% of the light measured in the HST F160W band \citep[1.54\,$\mu m$ central wavelength;][]{2013ApJS..207...24G}. The transition between the CGM and the IGM is thought to occur at distances of approximately 70\% to 90\% of the optical effective radius \citep[e.g.,][]{2013ApJ...764L..31K}. Therefore, searching an area equivalent to $100\,\times$ the IR radius allows us to examine both the full CGM and the IGM closest to the central galaxy.

While using such a large searching radius increases the chances of detection, it complicates the exploration since the searching region is often larger than the area covered by a single MUSE data-cube ($1\arcmin\times 1\arcmin$). Even if the 44
MUSE-Wide data-cubes cover a contiguous FOV (Fig.~\ref{fig:spatial_distro}), each one of them has different spatial and wavelength samplings and have to be interpolated onto a common scale to be used together. In principle, one can either construct the H$\alpha$ and continuum  images for the individual data-cubes and merge them later on, or merge the original data-cubes and build the images from the resulting merged {\em super} cube. We take the second approach, as detailed in Sect.~\ref{datacubes}.

%
%%%%%%%%%%%%%
%
\subsection{Preparation of super data-cubes}\label{datacubes}
From the original \musew\ data-cubes, we build new data-cubes large enough to encompass the searching area around each central galaxies (Sect.~\ref{sec:area}). These {\em super} cubes are used not only to search for H$\alpha$ emission, but also for all subsequent analysis. The approach of building super cubes ensures that the artifacts introduced by merging the original data-cubes is the same throughout the analysis, making it easier to control potential biases.

The original \musew\ cubes were observed at different heliocentric velocity, so we are forced not only to match them in RA and DEC but also to interpolate the cubes into a common wavelength grid. We define the new common grid as that of the central cube in each super cube. In order to assemble the super cubes, we use the standard software \textsc{Montage}\,\footnote{http://montage.ipac.caltech.edu/index.html}, which is a toolkit for assembling FITS images and cubes into custom mosaics. We use it in the so-called fast reprojection. 
\textsc{Montage} has the option to correct the background of the cubes. While this makes visualization easier, it also carries the drawback of altering the fluxes of the faintest noise-level emission. We made sure to assemble the data-cubes with this feature disabled, thus keeping the original background. 

While merging all MUSE-Wide cubes into a single super cube would solve our problem, it has the disadvantage of requiring prohibitively large amounts of memory. As a compromise, we build three different super cubes, which together encompass the totality of the MUSE-Wide FOV and guarantee the searching area of each galaxy to be included in at least one of them. Cutouts of these cubes centered in the galaxies are then employed to create the H$\alpha$ and broad-band images employed to search for weak H$\alpha$ signals, as explained in the next subsection.

%
%%%%%
\begin{figure*}
\centering
\includegraphics[width=1.\linewidth]{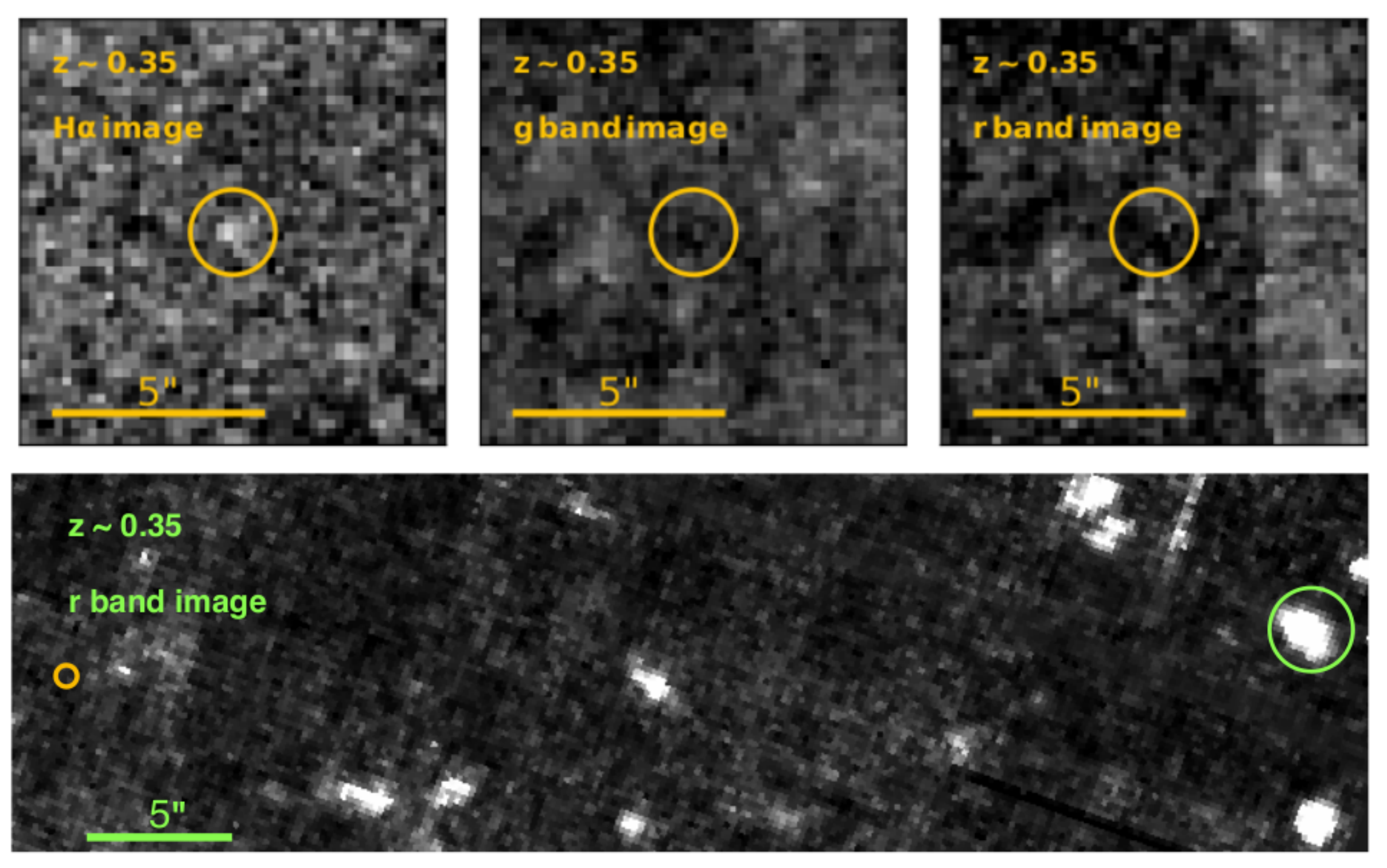}
\caption{Images derived from the \musew\ data-cubes and involved in the process of selecting line emission candidates. The upper panels show the H$\alpha$ image and the broad band images in $g$ and $r$, as indicated in the inset. H$\alpha$ emission candidates must have a distinguishable signal in H$\alpha$ and no clear emission in the broad bands. The yellow circle marks the candidate detected in this particular region, with its radius giving the size assigned to it. 
The lower panel shows the candidate in context, with the yellow circle indicating its position with respect to the corresponding
  central galaxy (the green circle). 5\arcsec\ scales are included in all panels.
}
\label{fig:narrow_broad_bands}
\end{figure*}
%
%%%%%%%%%%%
%
\subsection{Building the H$\alpha$ and broad band images}\label{Ha,R,G images}
We aim at identifying faint extended H$\alpha$ emission features within the searching radius around the target galaxies. For this, we need to find structures showing up in H$\alpha$ images but without broad band counterpart. Using the \musew\ {\em super} data-cubes, we construct the required narrow H$\alpha$ images tuned to the redshift of the central galaxy, as well as pseudo broad band $r$ and $g$ images. As an additional piece of information, we also built color images following the recipe by \cite{2004PASP..116..133L}. 

We compute the H$\alpha$ image by summing the data-cube within a wavelength range of $\rm \pm\, 10\, \AA$ around the H$\alpha$ wavelength of the central galaxy. This range represents a trade off: it is narrow  enough to avoid including too much noise into the image while, at the same time, it allows the H$\alpha$ emission to be shifted from the central galaxy with a generous proper motion of $\pm\, 400\,{\rm km \, s^{-1}}$.  We make no attempt to remove the continuum emission from the H$\alpha$ image.
A similar approach was taken with the broad bands, except that their bandpasses were defined following the central wavelength and full width half-maximum (FWHM) of the SDSS $g$ and $r$ filters \citep[][]{1996AJ....111.1748F}. 
We add up the data-cube in a range of $\rm \pm\,FWHM/2$ around the central wavelength of the broad band filter. Examples of H$\alpha$ and broad band images are given in Fig.~\ref{fig:narrow_broad_bands}.

The broad band images thus constructed allow us to reach a limiting surface brightness of 27.6 and 27.2\,mag\,arcsec$^{-2}$ for the $g$ and $r$ bands, respectively. These values correspond to three times the RMS (root mean square) fluctuation of the total flux per unit area considering regions of the data-cube without sources. Following \citet{2016ApJ...823..123T}, the areas used for the calculation were squares of $10\arcsec\times 10\arcsec$. The same procedure applied to the H$\alpha$ image yields a limiting surface brightness flux of $\rm 2.4 \times 10^{-19} \,erg \, s^{-1} \, cm^{-2}\, arcsec^{-2}$. These limits are in good agreement with estimates from the literature based on \muse\ data. \citet{2016MNRAS.462.1978F} obtain a $\rm 2\sigma$ limiting magnitude of 26.6\,mag\,arcsec$^{-2}$ in the $r$~band, comparing areas of 1 arcsec radius with an integration time of 4.1\,h.  %\comment{integration time? give it explicitly}
%\comment{Maybe I am wrong, but Fumagalli+ refer to integrated magnitudes rather than surface brightness and taken with another instrument LRIS@Keck. Please, check. If so this reference has to be taken out.}\joao{Checked. The magnitudes are indeed obtained with LRIS@Keck, however, in section 2.1 and 2.2, the authors compare the magnitudes with the ones obtained from MUSE in photometric conditions and conclude the two are in agreement within calibration errors. So I think we are good comparing with them.}
Using that same band and area, we find a $\rm 2\sigma$ limiting magnitude of $\rm 25.6\,mag \, arcsec^{-2}$. Assuming the signal-to-noise ratio to scale with the square root of the integration time, our limiting magnitude after 4.1\,h should be around $\rm 26.4\,mag \, arcsec^{-2}$, in fair agreement with  \citet{2016MNRAS.462.1978F}.
%\comment{reworded to make the comparison more clear. Is it correct?}
%\comment{Joao: I still see an inconsistency here. if the integration is 4.1 hours then the magnitude difference should be -2.5*log(sqrt(4.1))= -0.77, while we quote 0.3. We should talk about it.}
%
We also compare our results with \cite{2021A&A...647A.107B}, which analyze 120\,--\,140\,h of integration in the \muse\ Deep-Field. \musew\ has an integration time of 1\,h for each data-cube so, considering the square root scaling of the signal-to-noise ratio with  integration time, this translates into our limiting surface brightness fluxes being expected to be 10 to 12 times worse than those of \cite{2021A&A...647A.107B}.  As the standard to characterize noise, they use five times the RMS fluctuations of the signal in areas of 1\,arcsec${^2}$ averaging the spectra at 7000\,\AA\ over 3.75\,\AA. Their detection limit turns out to be   $\rm 1.3 \times 10^{-19} \,erg \, s^{-1} \, cm^{-2}\, arcsec^{-2}$. We replicate their estimate using \musew\ data to obtain a 5\,$\sigma$ limiting flux of $\rm 1.4 \times 10^{-18}\,erg \, s^{-1} \, cm^{-2}\, arcsec^{-2}$, meaning that our limiting flux is $\sim 11$ times worse than that in \cite{2021A&A...647A.107B}, thus right within the expected range.

%
%%%%%%%
\begin{figure*}
\centering
\includegraphics[width=0.3\linewidth]{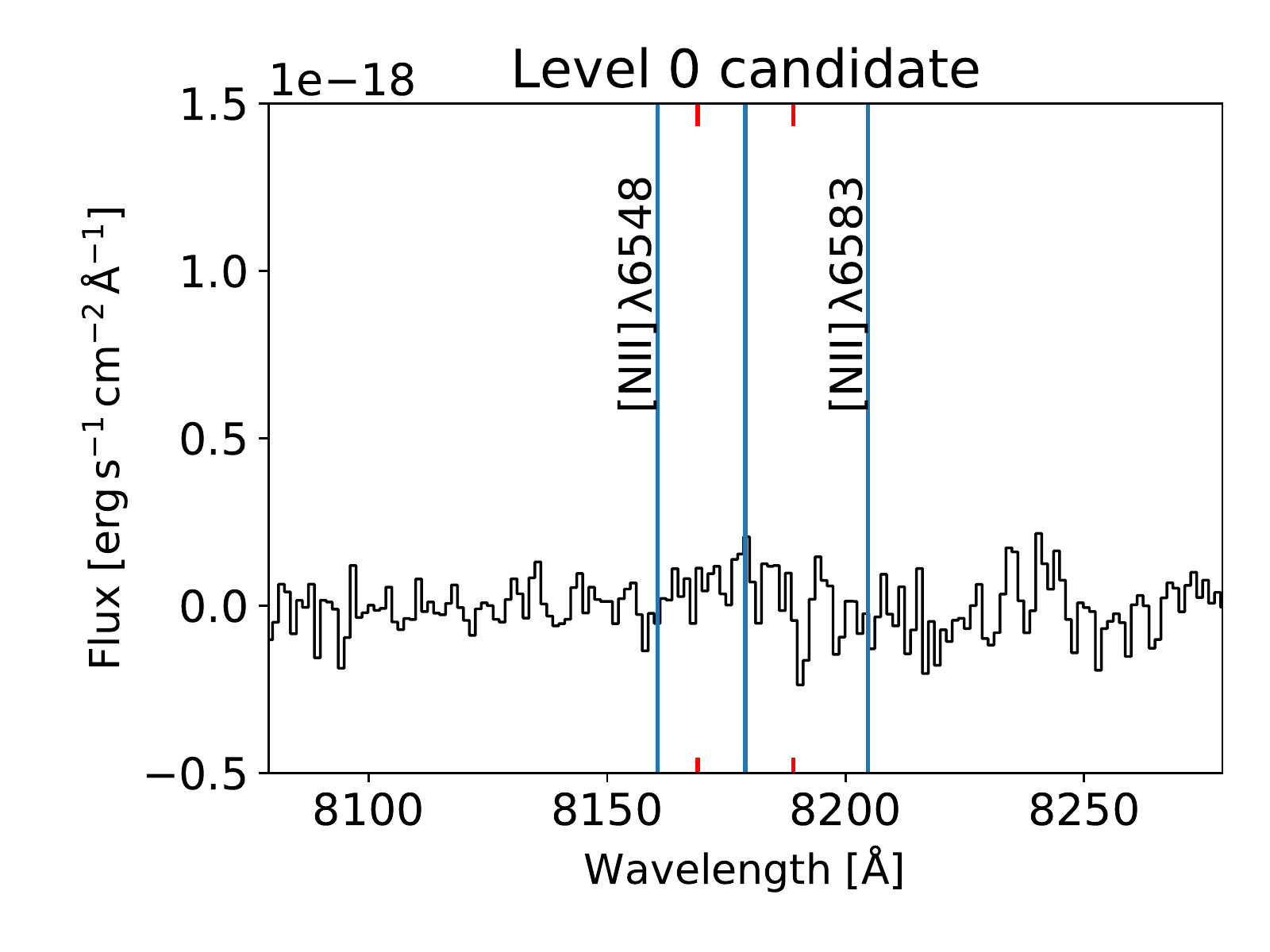}
\includegraphics[width=0.3\linewidth]{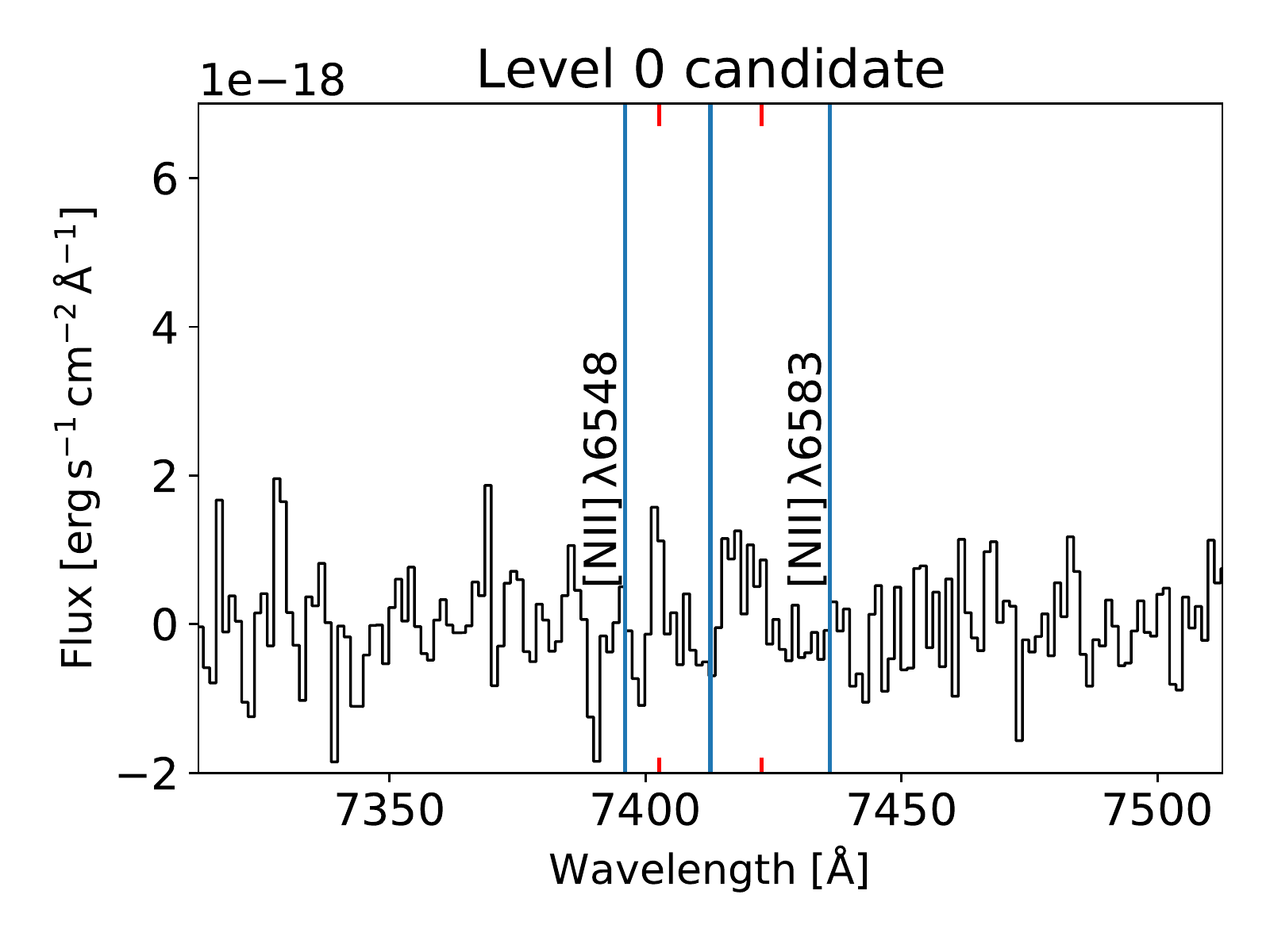}
\includegraphics[width=0.3\linewidth]{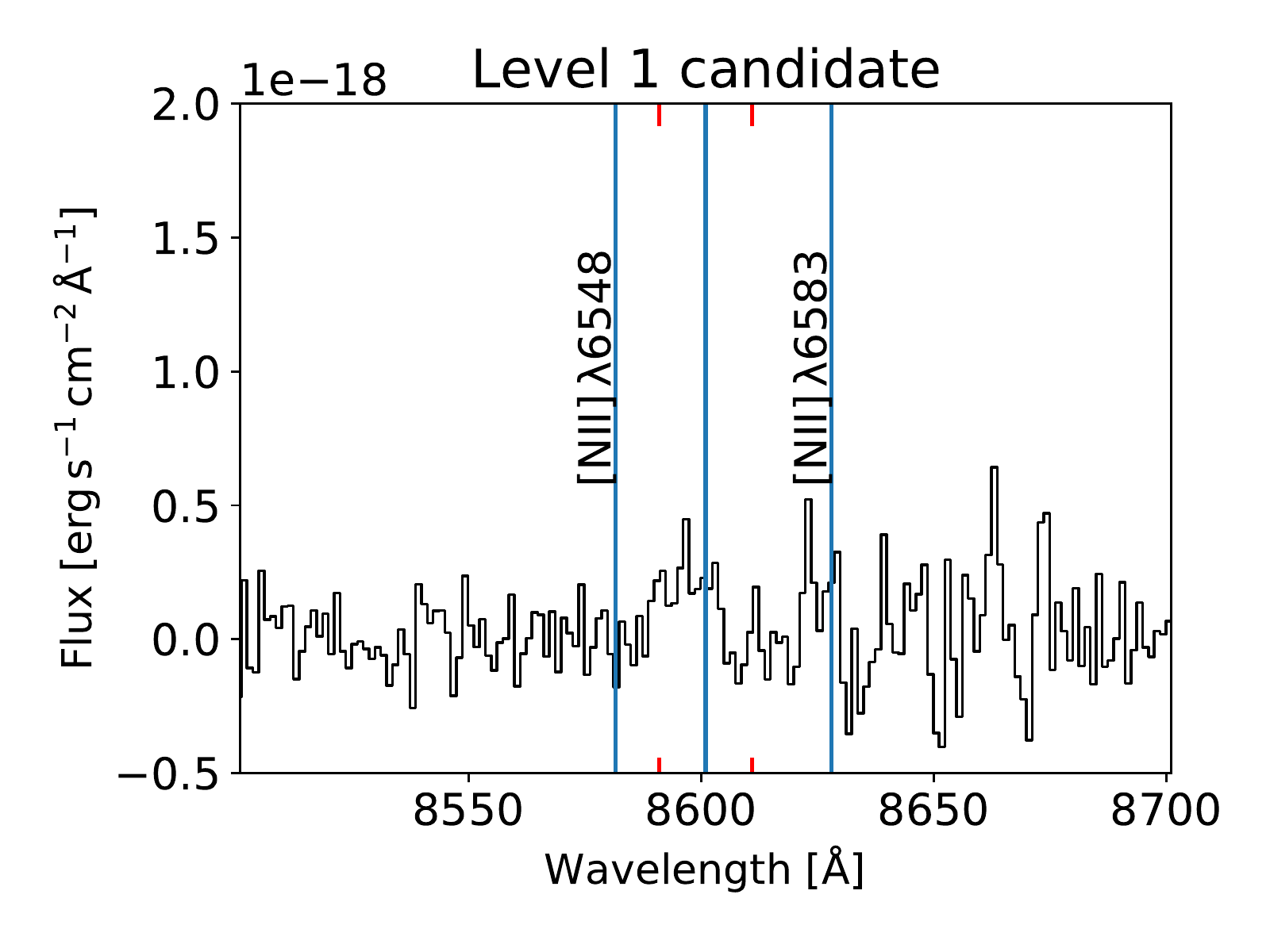}
\includegraphics[width=0.3\linewidth]{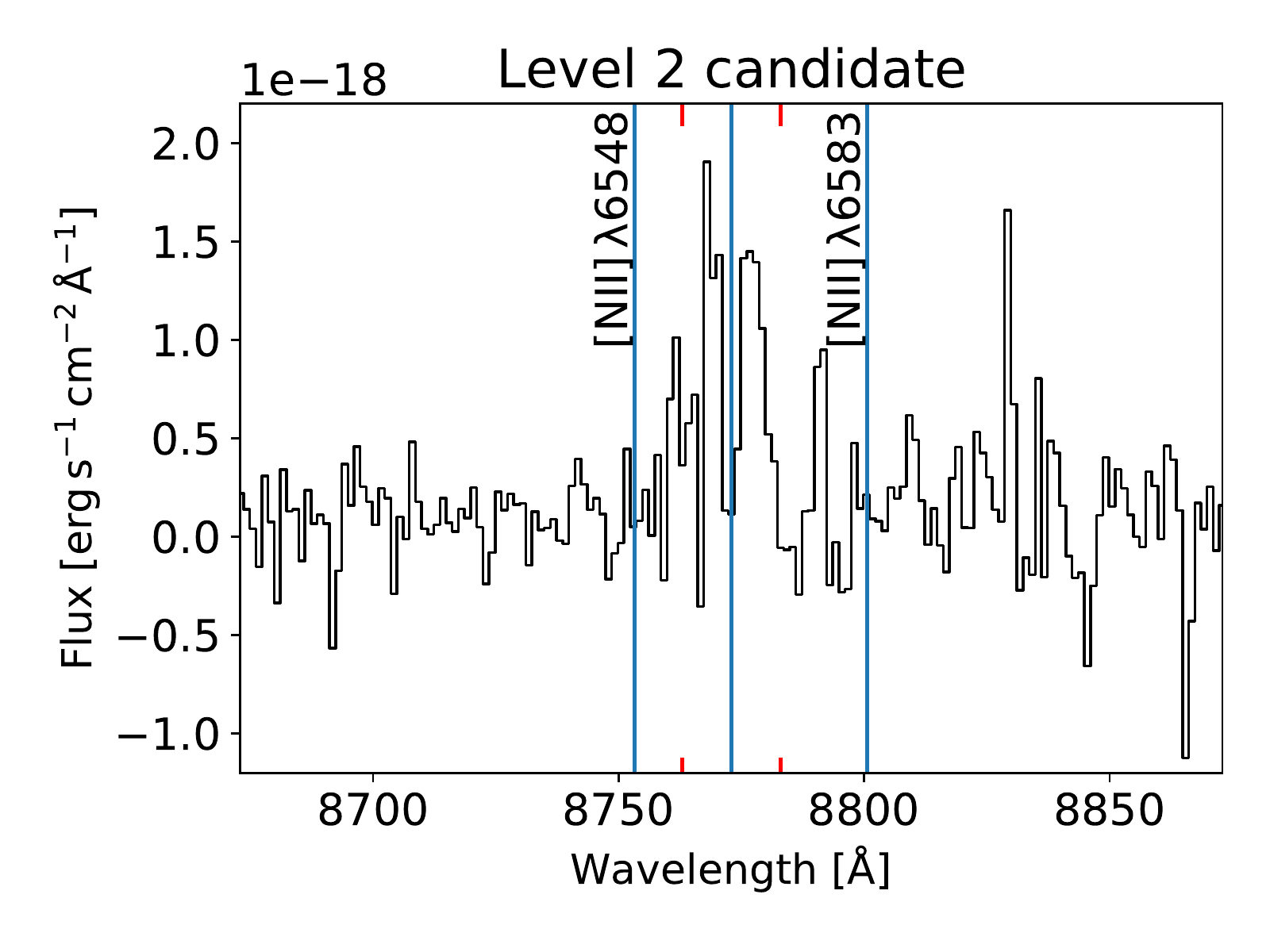}
\includegraphics[width=0.3\linewidth]{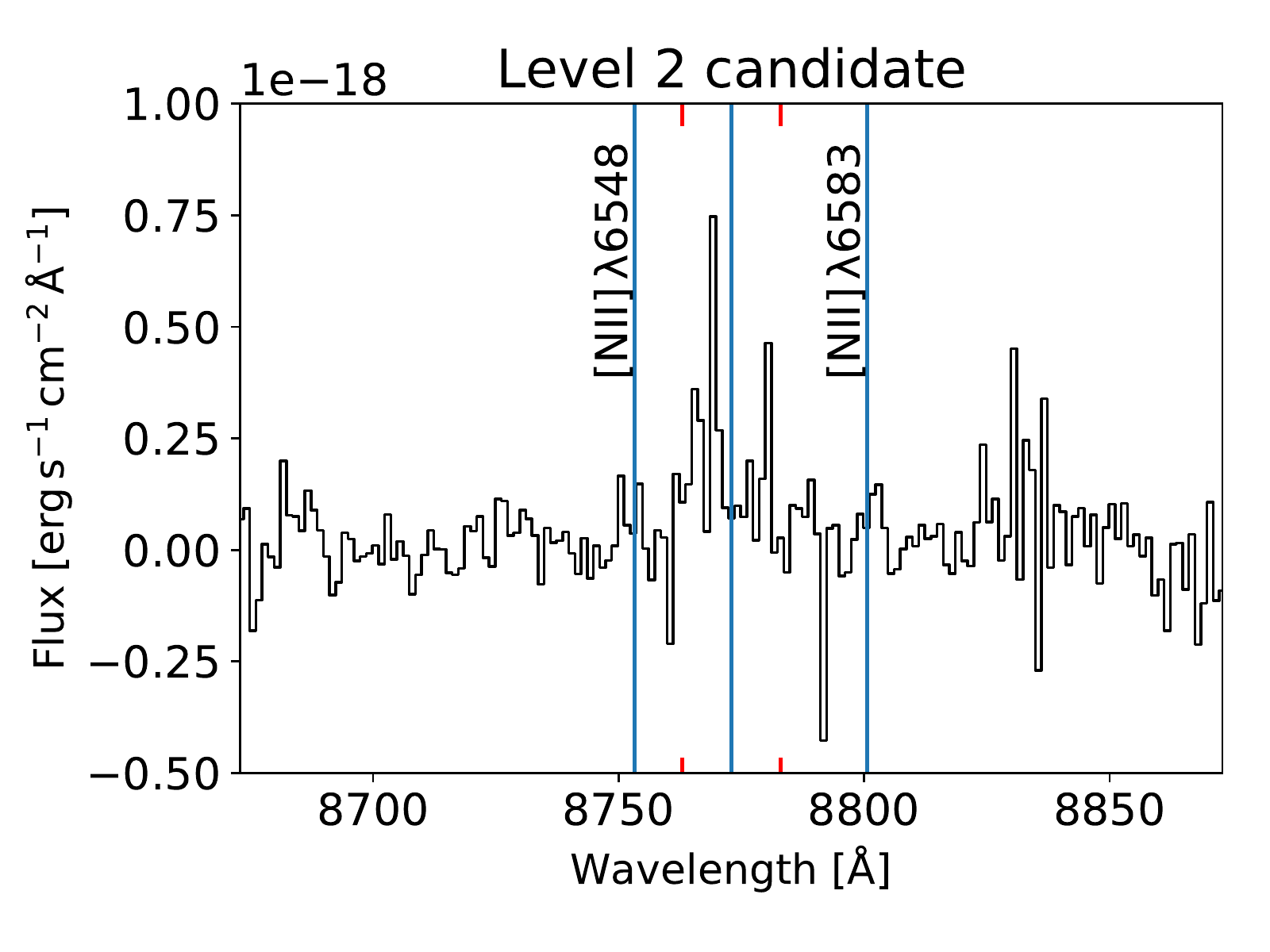}
\includegraphics[width=0.3\linewidth]{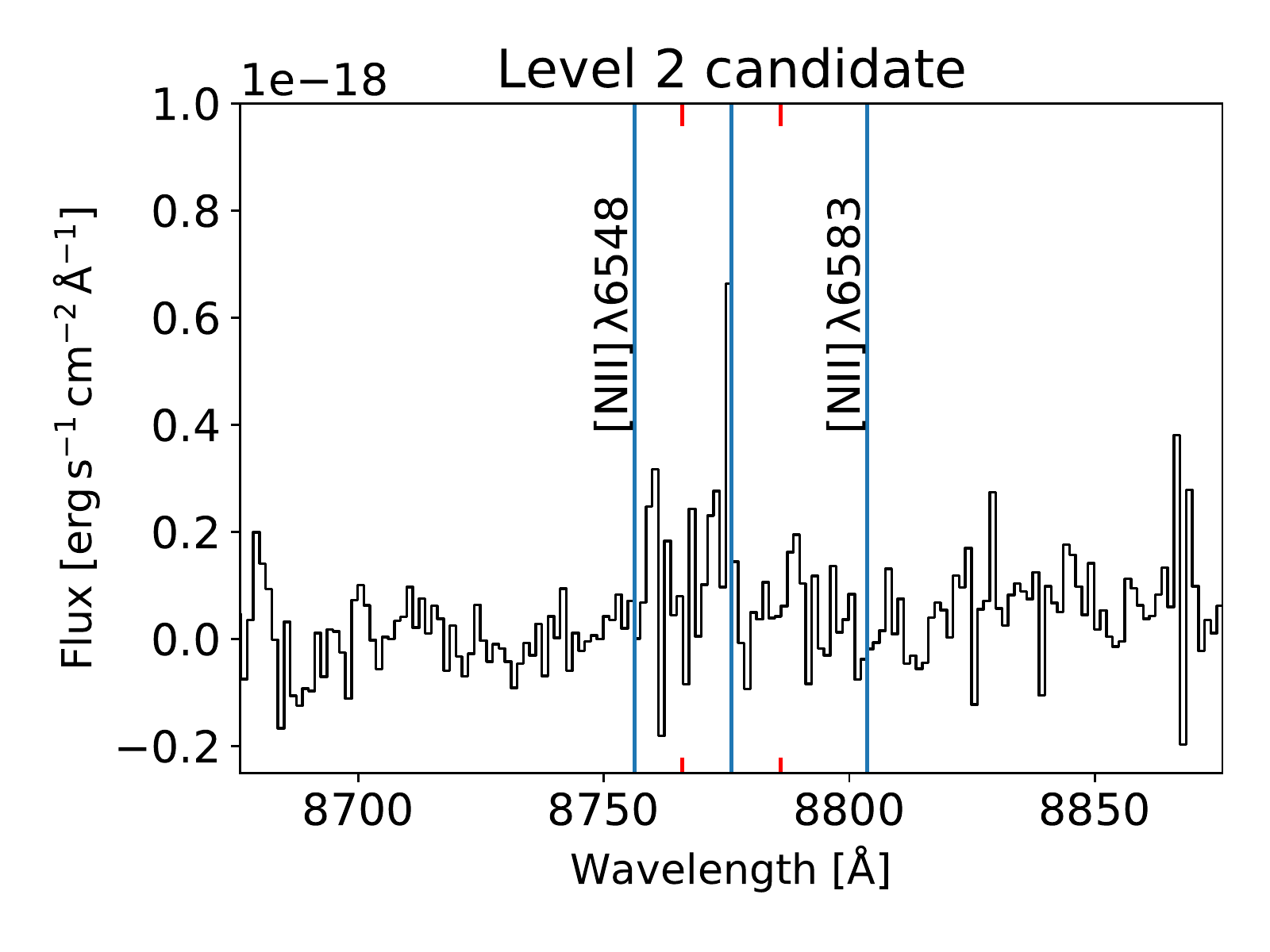}
\includegraphics[width=0.3\linewidth]{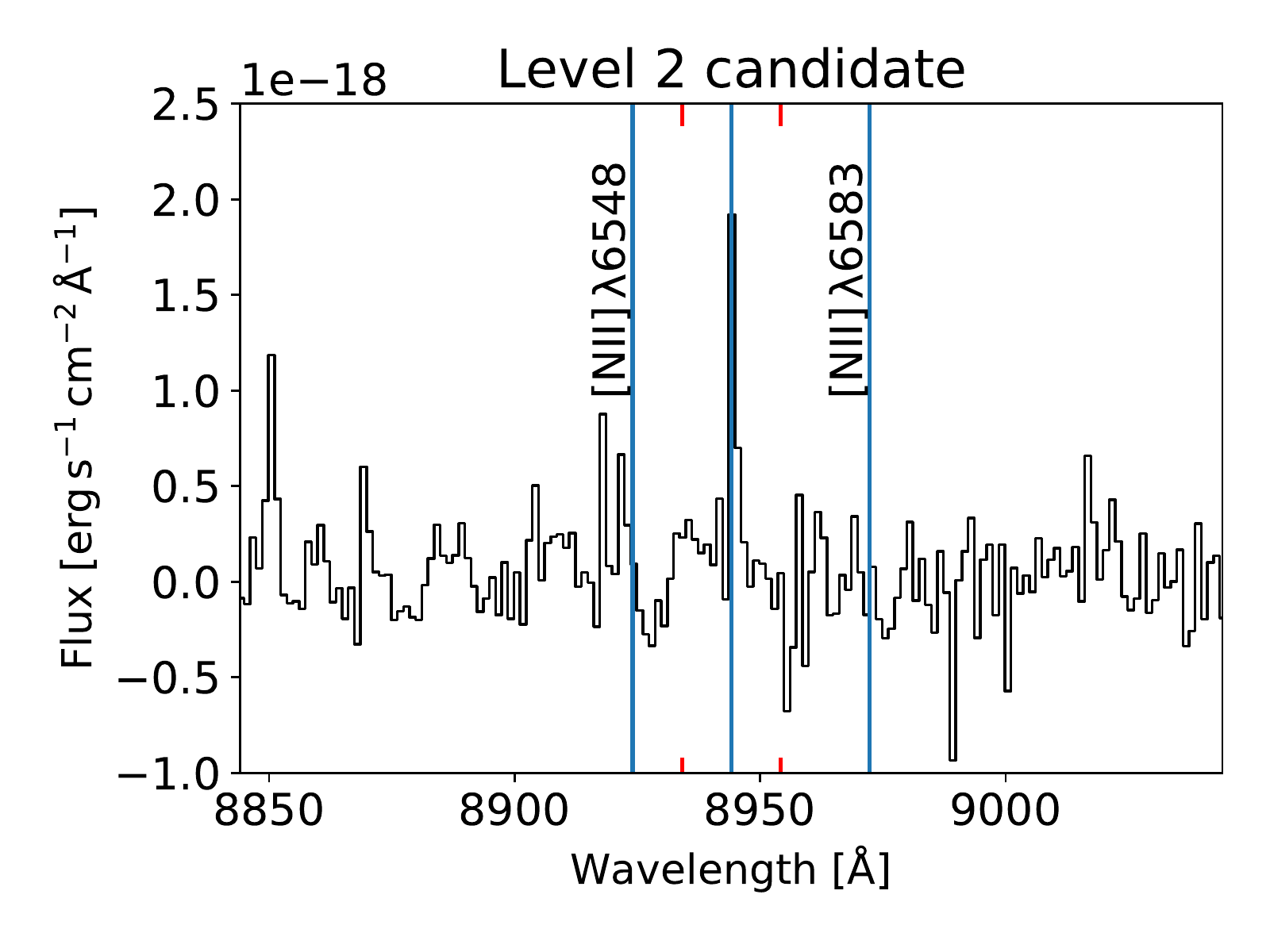}
\includegraphics[width=0.3\linewidth]{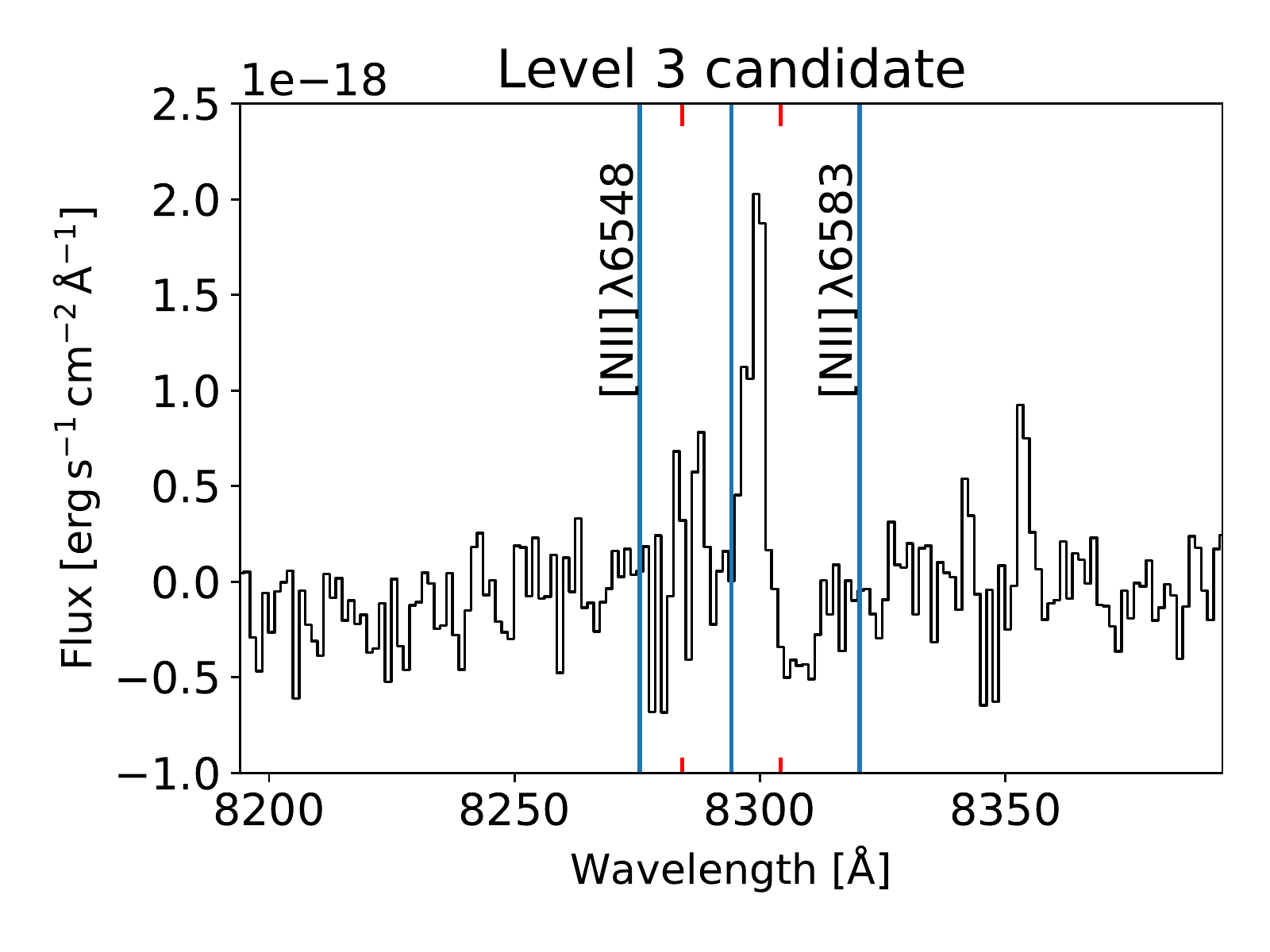}
\includegraphics[width=0.3\linewidth]{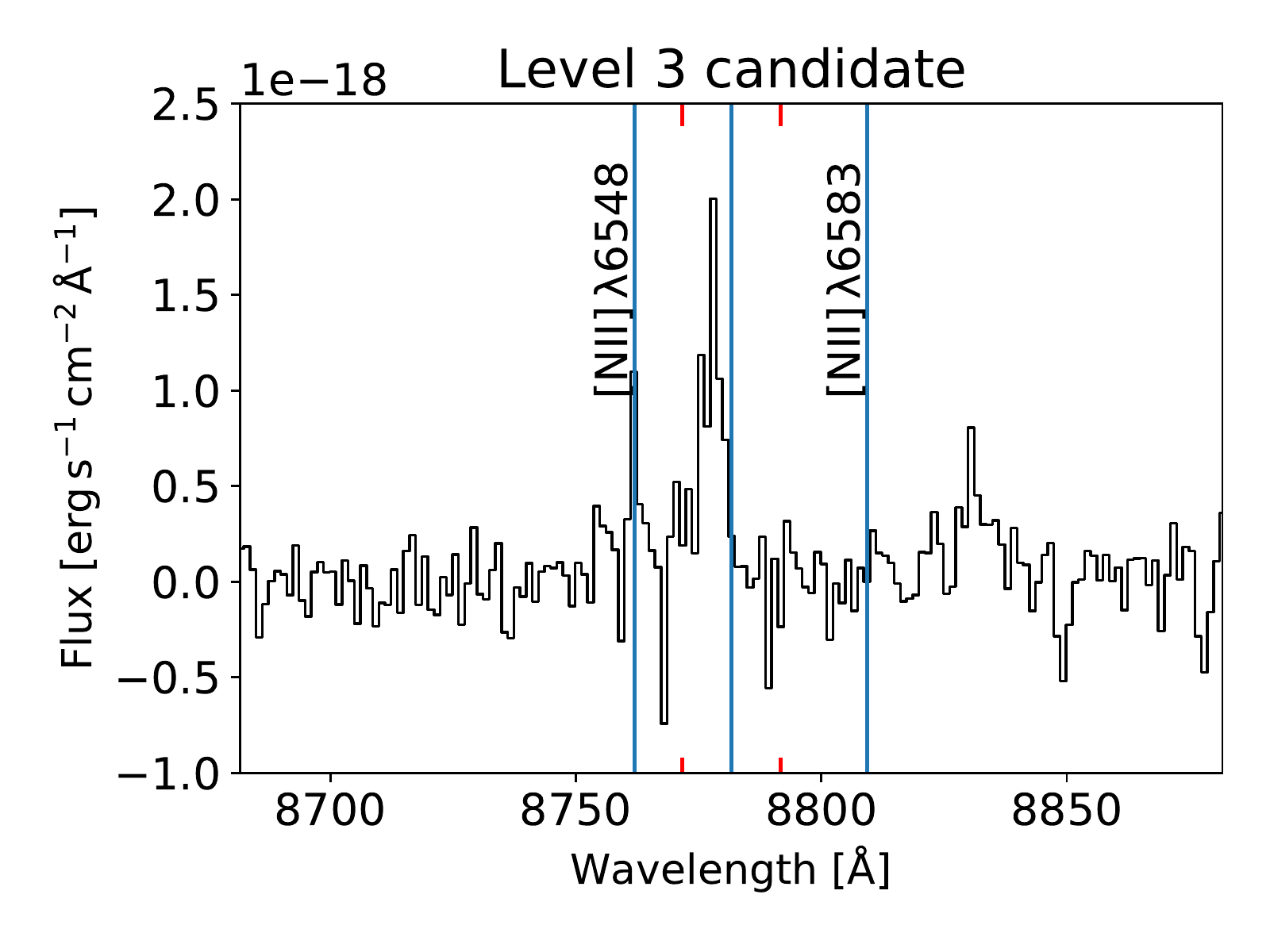}
\caption{H$\alpha$ line profiles for several of the candidates selected by inspection of the H$\alpha$ and broad band images. They illustrate the differences between levels 0, 1, 2,  and  3, that go from unreliable (0) to very reliable (3). There are examples of single peak and multiple peak H$\alpha$ profiles. The upper central and upper right plots illustrate a {\em top-hat} feature found in some of the candidates. %\comment{Joao: I do not see it.}\joao{replaced with better examples} 
These candidates are never classified as higher than level 1. The red markers on the top and bottom of the plots indicate the wavelength range covered by our narrow H$\alpha$ image (see Sect.~\ref{Ha,R,G images}). The vertical blue lines mark the expected position of H$\alpha$ and the two [N{\sc ii}] lines at the redshift of the central galaxy. Note that the vertical scale has to be multiplied by $10^{-18}$. 
}
\label{fig:Line_profiles}
\end{figure*}
%%%
%%%%%%%%%%%%%%%%
\subsection{Selection of H$\alpha$ emission clump candidates, assessment, and classification}\label{candidate_classification}
%\label{sec:candidates}

%\comment{This was part of Sect.~\ref{Ha,R,G images} plus the section on classification of candidates.}

We use the images described in Sect.~\ref{Ha,R,G images} to look for faint H$\alpha$ emitting structures while excluding sources that present a broad band contribution. Imposing  lack of broad band emission allows us to discard potential contaminants like obvious stars and background galaxies. Unfortunately, it does not exclude line emitter interlopers (like Ly$\alpha$ emitters). We discuss the problem of possible astronomical contaminants in Sect.~\ref{sec:origin}. 

The search for H$\alpha$ emitting candidates was conducted by visual inspection of the narrow and broad band images simultaneously (Fig.~\ref{fig:narrow_broad_bands} gives an example). While in principle there was the possibility of designing an automated algorithm for the search, it was discarded because the characteristics of the H$\alpha$ emission to be identified became known only a posteriori. Thus, we carried out a visual inspection of the H$\alpha$ and broad band images for each host galaxy, searching for emission in the H$\alpha$ image and selecting those candidates that did not have a counterpart in $r$ and $g$. Sizes were estimated visually, assuming a circular geometry for the candidates. The process typically required around 40\,--\,60\,minutes per galaxy amounting to around 20 8-hour days of visual inspection in order to cover the total 164 host galaxies within the redshift range. The procedure led to the identification of 621 potential candidates to be further inspected. %Details on these candidates are given in Sect.~\ref{sec:statistics}.

The criteria to select H$\alpha$ clumps are somewhat arbitrary, however this fact does not downgrade our study, which is confessedly exploratory. It aims at detecting so far undiscovered faint H$\alpha$ emission around galaxies. This exploratory selection will allow us to characterize the properties of these objects so that future searches can be automated using more unambiguous criteria. In addition, all the H$\alpha$ emission candidates go through additional screening before they are regarded as true signals and further statistical tests on the detections support their reliability. We classify them according to their degree of certainty through a two stage inspection, or two passes.  They are based on the visual inspection of the spectrum resulting from averaging all the spaxels of the candidate in the spectral regions where the emission lines H$\alpha$, H$\beta$, [O\textsc{iii}]$\lambda$5007, [O\textsc{iii}]$\lambda$4959, and [O\textsc{ii}]$\lambda$3727 are expected. The shape and signal around H$\alpha$ is the main driver to assign one out of 4 grades or levels, as detailed below:
\begin{itemize}
\item[-]Level 0: Immediately discarded as it cannot be a valid candidate. This generally happens when no H$\alpha$ line can be detected, which is the absolute minimum for a candidate to be considered.

\item[-]Level 1: Probably not real but retain the possibility of revisiting it. In general, this happens when the H$\alpha$ line is not very clear, difficult to separate from background, and there are no other lines present.

\item[-]Level 2: Probably real. H$\alpha$ line appears clearly and/or there may be suggestions of other lines as well, such as [O{\sc iii}]$\lambda$5007.  Suspected Ly$\alpha$ emitting interlopers are also included here, as they are candidates we definitely want to revisit.

\item[-]Level 3: Real. H$\alpha$ is clearly present along with possible presence of the other emission lines expected from emitting gas.
\end{itemize}
Figure~\ref{fig:Line_profiles} shows examples of profiles at various levels. From the original 621 candidates, we end up with 288 level 0, 167 level 1, 155 level 2, and 11 level 3.%\comment{revise to be consistent with Sect.~\ref{sec:statistics}}

Only candidates classified as levels 2 and 3 are selected for a second pass. 
The spectra of these candidates are further inspected to see whether or not the emission lines, particularly H$\alpha$, could be contaminated by sky line residuals. The inspection is carried out by over-plotting the telluric lines compiled by \citet{2003A&A...407.1157H} (wavelength and intensity) on the observed spectra.
If a candidate shows no clear presence of a sky line on the vicinity of the emission line, then the possibility of a sky artifact is minimal and the candidate is immediately accepted. Also accepted are candidates where the spectra present telluric lines near the emission line, but there is not serious threat of contamination since nearby much stronger telluric lines do not show any noticeable residual in the observed spectrum.
If the emission coincides with a strong sky line and we cannot discard serious contamination, the candidate is classified as {\em unclear} and discarded from the list of viable candidates to be considered for further analysis.
From the original 166 level 2 and 3 candidates, we ended up with 118 good candidates. We keep the level 2 and 3 classification along the manuscript but consider only the 118 candidates validated through the second inspection. Further details on the statistical properties are given in Sect.~\ref{sec:statistics}. %\comment{check that these and the numbers finally given in Sect.~\ref{sec:statistics} are the same.}
In particular, the S/N  at the maximum signal is typically larger than 3 and reaches up to 20. The same figure also holds for the S/N of the integrated line profile (i.e., for the flux). 

This second inspection was blind, without knowing what the previous classification (level 2 or 3) was, in order to minimize biases on our part. In general, the second pass was mainly focused on H$\alpha$, as having this line was essential for a candidate to be picked up. %While the inspection was carried out also for the other emission lines, it is possible for a candidate to be a good one as long as the H$\alpha$ line is accepted, even if the other lines are contaminated.

We note that our selection of {\em good} candidates is quite conservative. The overlapping of a telluric line with the putative H$\alpha$ signal does not imply the existence of contamination. The telluric line removal procedure ({\sc zap}) does an excellent job removing strong emission lines (see Sect.~\ref{sec:telluric} for details). %In addition, as we explain in Sect.~\ref{skylines}, other features of the detected emission clumps disfavor their telluric origin, even when they roughly coincide in wavelength with a skyline.   

%
%%%%%%%%%
%
\subsection{Discarding contamination due to data-cube construction artifacts and sky emission}\label{skylines}

After the identification of candidates, a concern is the possibility of getting false positives due to artifacts arising from the merging of data-cubes (Sect.~\ref{datacubes}). Limiting these artifacts was the rationale leading to merging entire data-cubes rather than separate images (Sect.~\ref{datacubes}). 
In addition to this concern, there is also the possibility of contamination by artifacts from the slicer-stacker that built the original cubes themselves \citep[][]{2014Msngr.157...13B,2019A&A...624A.141U}. We discard these two sources of artifacts by inspecting the spatial distribution of the selected candidates (Fig.~\ref{fig:spatial_distro}). They do not follow the boundaries between merged data-cubes nor do they portray a discernible grid-like distribution, an indication that the majority of the sources cannot be due to artifacts resulting from the piecing together of the individual spectra that form the data-cubes.
%\comment{Joao: I comment out the new figure with the sky line residuals. It simply does not fit in the line of the argument. Let's see whether it can be used elsewhere.}

When it comes to sky emission, we consider the possibility of contamination from sky line residuals that have survived the \textsc{zap} treatment (Sects.~\ref{sec:telluric} and \ref{candidate_classification}). 
%
%One may naively think that by averaging spectra from spaxels without sources, the telluric line residuals would be disclosed if present. However, this is not the case because of the way \textsc{zap} works. The sky correction in each spaxel has a common spectral shape but different amplitude, and shapes and amplitudes are computed so that they tend to cancel out to zero when averaging regions of the sky devoid of astronomical signal \citep[][and also Appendix~\ref{app:b}]{2016MNRAS.458.3210S}.
%
%Figure ~\ref{fig:super_residuals} shows this aspect of the sky averaging. it contains the average spectrum of the telluric lines between $\rm 7200\, \AA-7400\, \AA$, estimated from a circular region of 100 pixel ($\rm 20 \arcsec$) radius where we mask all emission higher than $\rm 3\sigma$. Notice that the average signal of the residuals is 1 to 2 orders of magnitude fainter than the emission spectrum of our selected candidates (see Figure \ref{fig:Line_profiles}). The small and wiggly residuals left, with their characteristic jumps from negative to positive flux values and vice-versa, are characteristic of the PCA-based telluric line removal used by MUSE-Wide (see Appendix~\ref{app:b}).
%
%We also compare our target emission lines to the shape expected to be presented by sky line residuals left by \textsc{zap}. We do this by carrying out a PCA decomposition of $\rm 10^{4}$ mock emission line spectra and compare the residuals left with the observed H$\alpha$ profiles. 
%
%
The shape of the observed H$\alpha$ profiles (Fig.~\ref{fig:Line_profiles}, to be discussed later on in Sect.~\ref{sec:line_shapes}) does not present the wiggles expected if it were due to residuals from \textsc{zap}, and we take this as one evidence for our H$\alpha$ detections not being artifacts due to telluric contamination (see App.~\ref{app:b} for further details).
%
%\begin{figure}
%\includegraphics[width=0.95\linewidth]{./figures/Stacked_sky_Figure_3.png}
%\caption{Average telluric line residuals in the 7200$\rm \AA$-7400$\rm \AA$ range in a region of the FOV devoid of astronomical sources. The red markers on the top part of the plot represent the wavelengths of the 4 stongest sky lines as determined by \cite{2003A&A...407.1157H} in this region of the spectrum. Notice the marked decrease in general flux intensity and the abrupt changes between positive and negative flux values, characteristic of the PCA-based telluric line removal used by MUSE-Wide (see Appendix~\ref{app:b}).}
%\label{fig:super_residuals}
%\end{figure}

%
A second argument against the signals being caused by telluric line contamination is the small size of the emission features. Their typical size is around $\rm 1\arcsec$ (see Sect.~\ref{sec:physical_properties} and Fig.~\ref{fig:histo_diameters_arcsec}). Most telluric emission comes from the mesopause \citep[see, e.g.][]{1921PGro...31....1V,2014A&A...567A..25N}, at a height of around 100\,km. Thus, for the observed emission to be produced in there, the physical size of the emitting gas would have a characteristic length scale of $\rm \sim 0.5\,m$, which is too small for any physically meaningful scale at this atmospheric layer. For example, the radial width of the mesopause is at least several km, and a typical wind of 10\,m\,s$^{-1}$ would move a 0.5\,m patch outside the MUSE FOV in just a few seconds. These order of magnitude arguments are fully consistent with the typical extension of the skyline emission measured to be around 1\arcmin\ \citep[e.g.,][]{2010arXiv1009.0554R,2010SPIE.7735E..6LR}, which is much too large to produce the small observed clumps.

A third sanity check was carried out comparing the observed emission line signals and the wavelength of intense telluric lines as compiled by \cite{2003A&A...407.1157H}. Very often there is not overlap, as we explain in Sect.~\ref{candidate_classification}. 

Finally, an extra argument against the telluric origin of the emission line signals comes from their spatial distribution. The observed galactocentric distances and azimuths are inconsistent with a random distribution of sources in the FOV. Rather, they cluster toward the major axis of the central galaxy (Sects.~\ref{sec:Azimuths} and ~\ref{sec:distances}). This fact disfavors any non-astronomical origin for them, including sky emission. 

%
%%%%%%%%%%%
\section{Results}\label{sec:results}

Here we present the properties of the 118 good H$\alpha$ emission candidates selected in Sect.~\ref{candidate_classification}. It starts by discussing general statistical properties (Sects.~\ref{sec:statistics}) and the shape of the line profiles (Sect.~\ref{sec:line_shapes}). Fluxes, sizes, and redshifts  are presented in Sect.~\ref{sec:physical_properties}. In Section ~\ref{sec:gas_mass}, we estimate the mass of emitting gas based on the H$\alpha$ luminosity and some assumptions on the physical properties of the gas. Sections ~\ref{sec:Azimuths} and ~\ref{sec:distances} analyze the spatial distributions of our candidates: their distance and orientation with respect to the major axis of the central galaxy.  Kolmogorov-Smirnov tests \citep[KS tests; see, e.g,][]{1986nras.book.....P} on various of these distributions support the reliability of the detection.

%
%%%%%%%%%%%
%
\subsection{Number of H$\alpha$ emitting blobs}\label{sec:statistics}

We have examined the outskirts of 164 galaxies with H$\alpha$ within the bandpass of \muse . This exercise resulted in 621 candidates being selected as possible H$\alpha$ emitting gas clouds, with an average of 3 to 4 candidates per galaxy.  After extracting 1D spectra for our candidates and applying the first-pass classification (Sect.~\ref{candidate_classification}), the number of valid candidates decreases considerably. Only 27\,\% (166 candidates) were classified as level 2 (probably real) or level 3 (secure) and were automatically selected for the second-pass classification to discard telluric line contamination. %Of these, 155 were classified as level 2 (25\,\% of original 621) and only 11 classified as level 3 (2\,\%). Of the remaining candidates, 167 (27\%) were classified as level 1 (probably not real, but with possibility of being revisited) and a large 46\% (288 candidates) were classified as level 0 (false positives, as no line could be clearly identified).
%
%As explained in Sect.~\ref{candidate_classification}, once the candidates are filtered through the first classification scheme, we apply a second revision to the 166 level 2 and 3 candidates in order to check whether or not the lines could be contaminated by skyline residuals (Sect.~\ref{candidate_classification}). 
Most of them passed the test (118 candidates), but  the rest were put aside to be conservative because there was a potential telluric contaminant coinciding with the emission feature. 
%\comment{is this too much of a repetition?. decide in the next round.}

%We make a rough estimation of the size of our candidates by assuming a circular geometry and guessing the radius of the clumps through visual inspection. Our candidates exhibit an average radius of 2.3 pixels (0.46\arcsec, assuming circular shape) with standard deviation of 4.3 (0.86\arcsec) pixels. Assuming the candidates stand at the same redshift as the central galaxies and taking the average redshift as reference, this results in our candidates having a typical size of 10.2\,kpc.
%
%There is no obvious indication that these candidates follow any kind of bias in spatial distribution (e.g., tracing the slicer-stacker bands or the merger bands in the data cubes) so, even if some of the signals could potentially be artificially created during the reduction process, this bias does not represent the main source of signal and should not influence the results (see also Sect.~\ref{skylines}).

Overall, the above numbers translate into roughly one secure candidate per host galaxy being found on average. Before applying the second inspection, around 87\,\% of the sample had between one and two candidates per host galaxy, with only a few outliers ($\sim 13\%$) having a higher count. After passing the two selections, a total of 56 galaxies (from the original 164, equivalent to $\rm \sim 34\%$) are shown to host clumps consistent with our selection criteria. Thus, a typical galaxy with clumps has 2 of them ($118/56\simeq 2.1$). 

%
%%%%%%%%%%%%%%%%%%
%
\subsection{Line profiles}\label{sec:line_shapes}
The spectra in Fig.~\ref{fig:Line_profiles} show examples of H$\alpha$ profiles, labeled according the classification from Sect.~\ref{candidate_classification}. From level 0 and 1, considered unlikely to be real, to levels 2 and 3, which we consider likely candidates. One remarkable characteristic is the appearance of line profiles presenting double peaks  (Fig.~\ref{fig:Line_profiles}, middle row, first panel).
A few of the candidates ($\rm \sim 2\,\%$ of the original total sample) also show no clear line on the H$\alpha$ range but instead present a kind of top-hat shape, causing the candidate to show up in the H$\alpha$ narrow band images. One such profile is shown in Fig.~\ref{fig:Line_profiles}, top right panel. 
Some of the double peaks present a boosted blue wing, which goes against what we expect from high redshift Ly$\alpha$ emitter interlopers (see Sect.~\ref{sec:origin}).

%

 %\comment{Joao: I do not see it. we should see together these profiles.}. 
\begin{figure}
\centering
\includegraphics[width=0.95\linewidth]{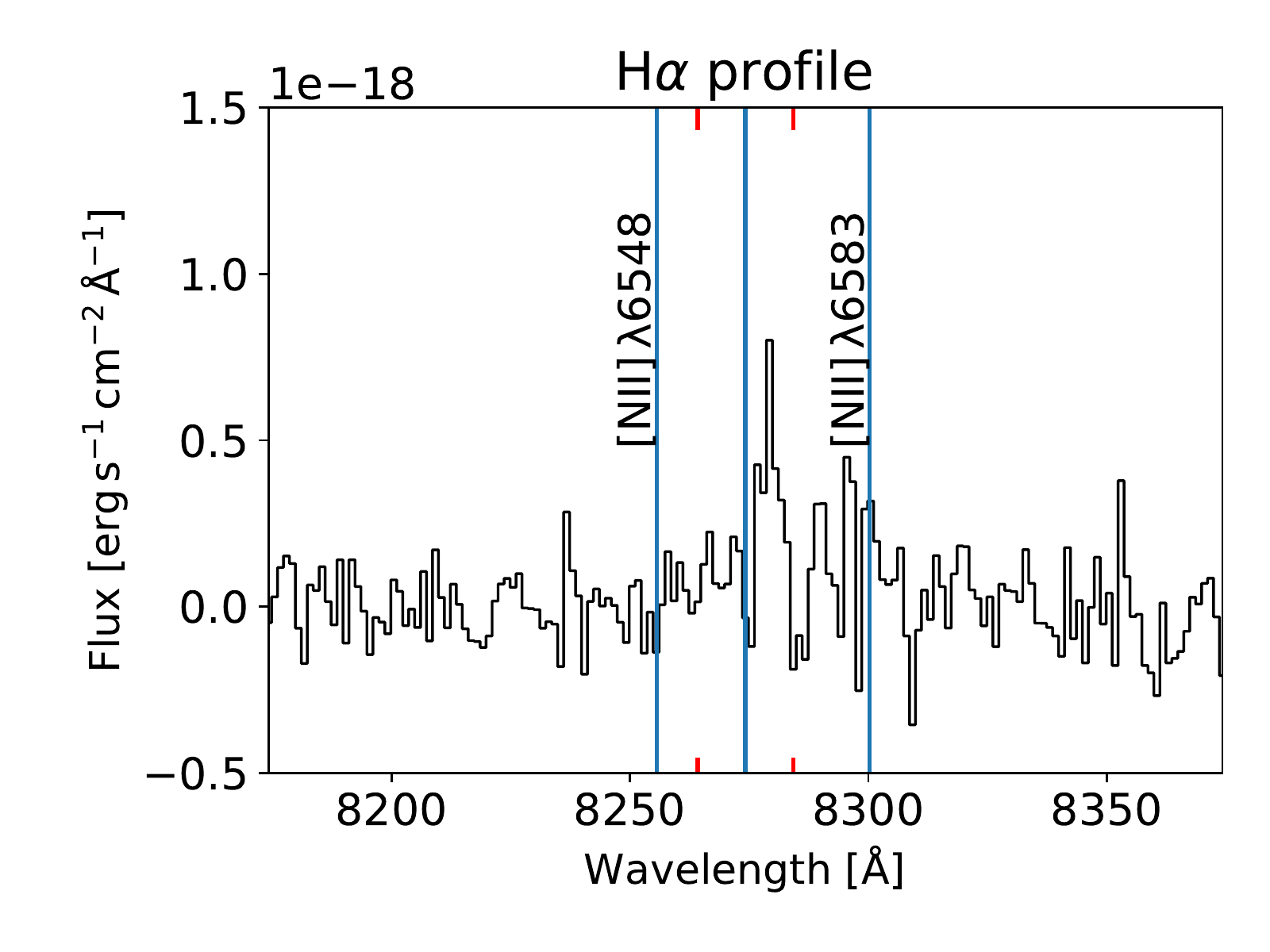}
\includegraphics[width=0.95\linewidth]{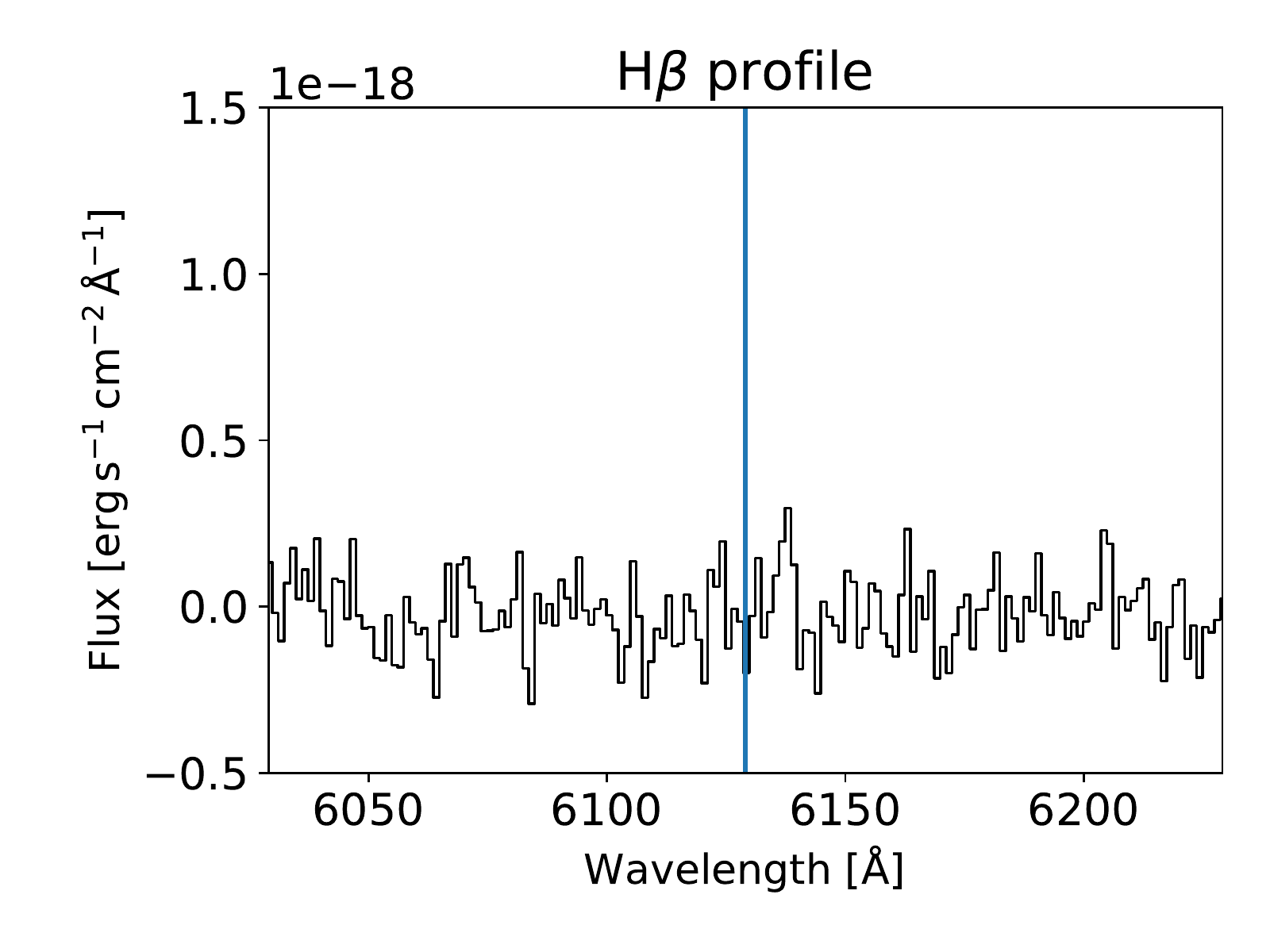}
\caption{Example of a candidate with hints of H$\beta$ emission. Both H$\alpha$ and H$\beta$ are redshifted  by ${\rm \sim 90\, km\,s^{-1}}$ with respect to their position in the central galaxy, which is marked by the unlabeled vertical lines. The ratio H$\beta$ to H$\alpha$ is close to $1/3$, as expected from Case~B recombination. 
}
\label{fig:profiles_Hb}
\end{figure}

\begin{figure}
\centering
\includegraphics[width=0.95\linewidth]{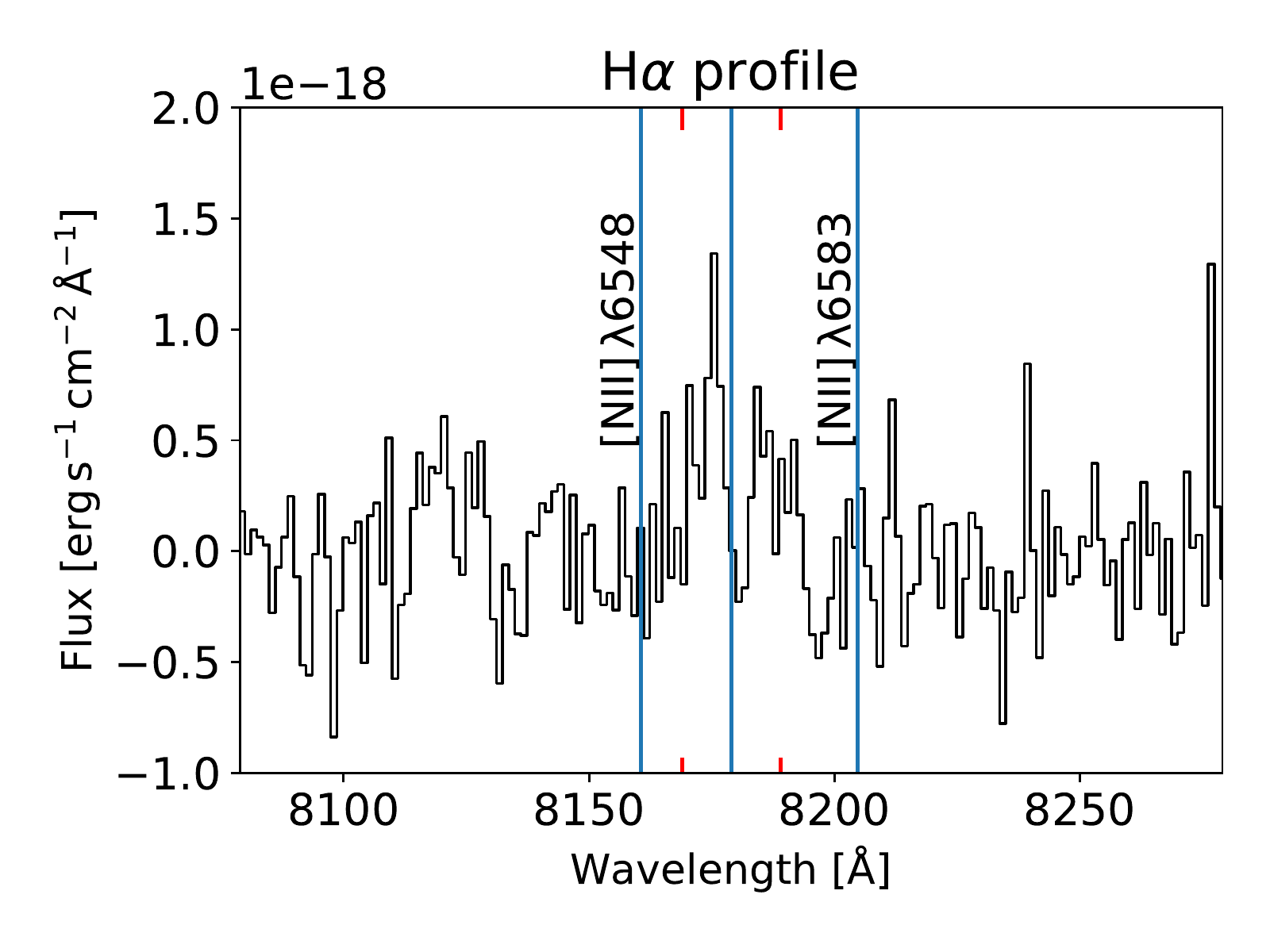}
\includegraphics[width=0.95\linewidth]{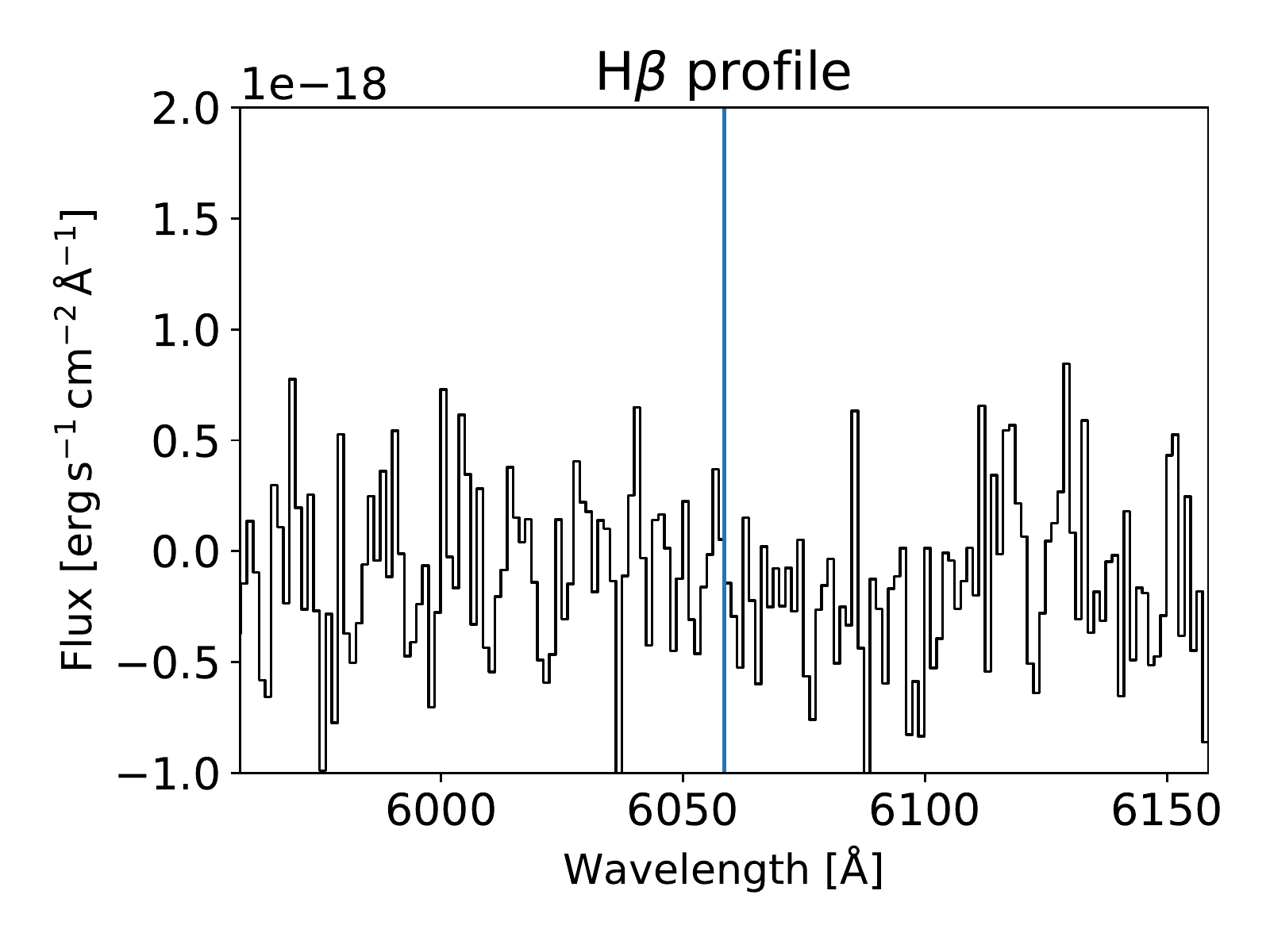}
\includegraphics[width=0.95\linewidth]{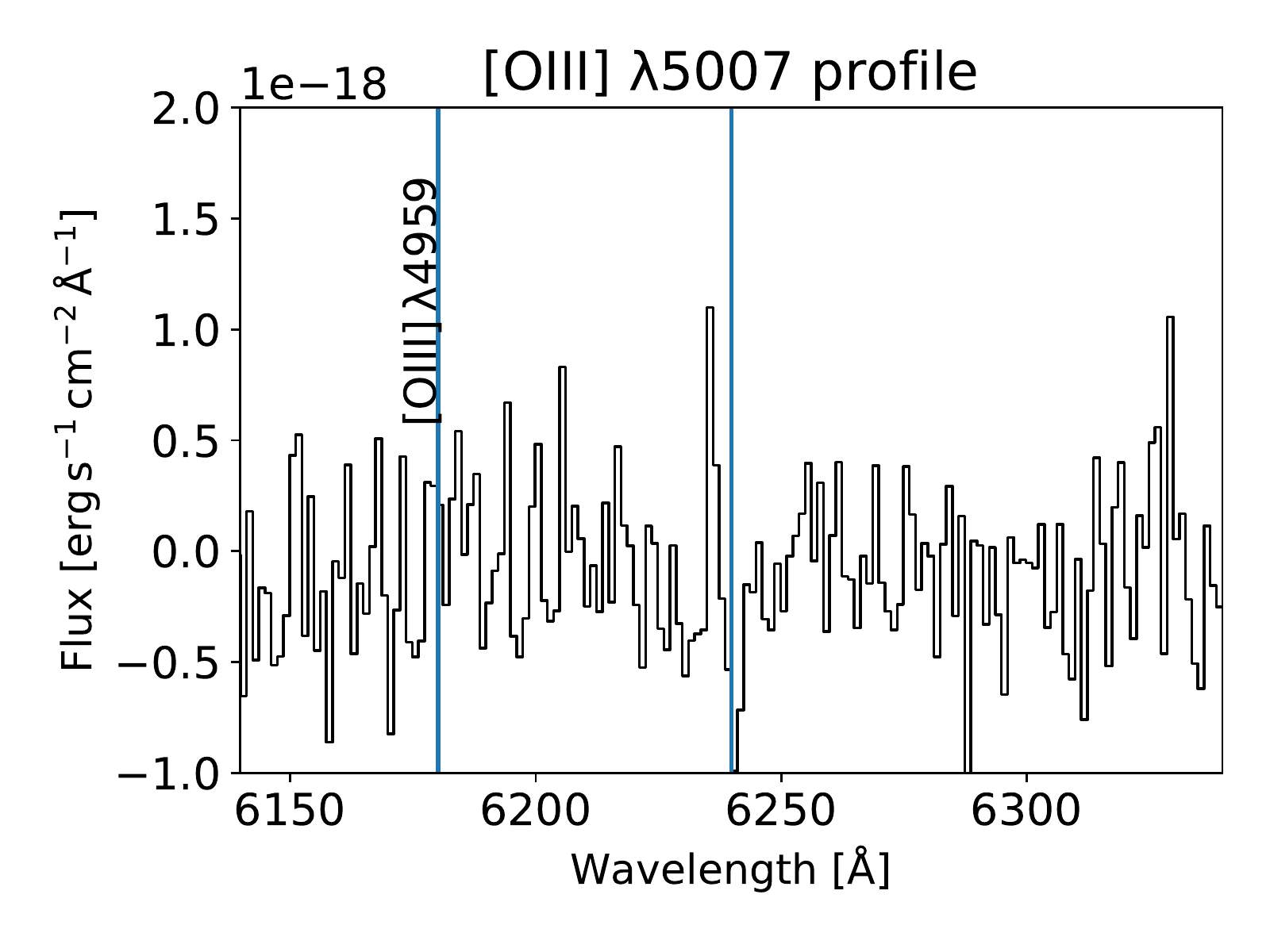}
\caption{Example of a candidate with [O{\sc iii}]$\lambda$5007 emission. The lines are clearly blueshifted with respect to the central galaxy (marked by unlabeled vertical lines). A possible H$\beta$ emission is at noise level, but complies with the expected $\rm \sim 1/3$ flux ratio with respect to H$\alpha$.
%Also seen emission for [OIII]5007, also blueshifted. It is possible that [OIII]4959 is also present but, if so, it is very much at noise level and we can see no clear evidence of it.
%\comment{Flux units 'erg/cm2/s' --$>$ 'erg/s/cm2'}
%\comment{For consistency with Fig. 3 and the upper panels, take of the lable [OIII]5007}
%\comment{[N{\sc ii}]$\lambda$6585 $\longrightarrow$ [N{\sc ii}]$\lambda$6583}
}
\label{fig:profiles_OIII}
\end{figure}
By selection, all candidates show signal at H$\alpha$. The spectra of some of them also suggest the presence of other characteristic emission lines. Figures~\ref{fig:profiles_Hb} and \ref{fig:profiles_OIII} show two candidates with possible emission in H$\beta$ and [O{\sc iii}]$\lambda$5007, respectively. %\comment{Joao: I do not think that Hb is evident at all in Fig.~\ref{fig:profiles_OIII}. Do we have another example?}\joao{I''ll look for another but the reason I chose this one was because it still showed hints of Hb while also showing "clear" OIII, all agreeing with the line shift of Ha.}

\begin{figure*}
\centering
\includegraphics[width=0.90\linewidth]{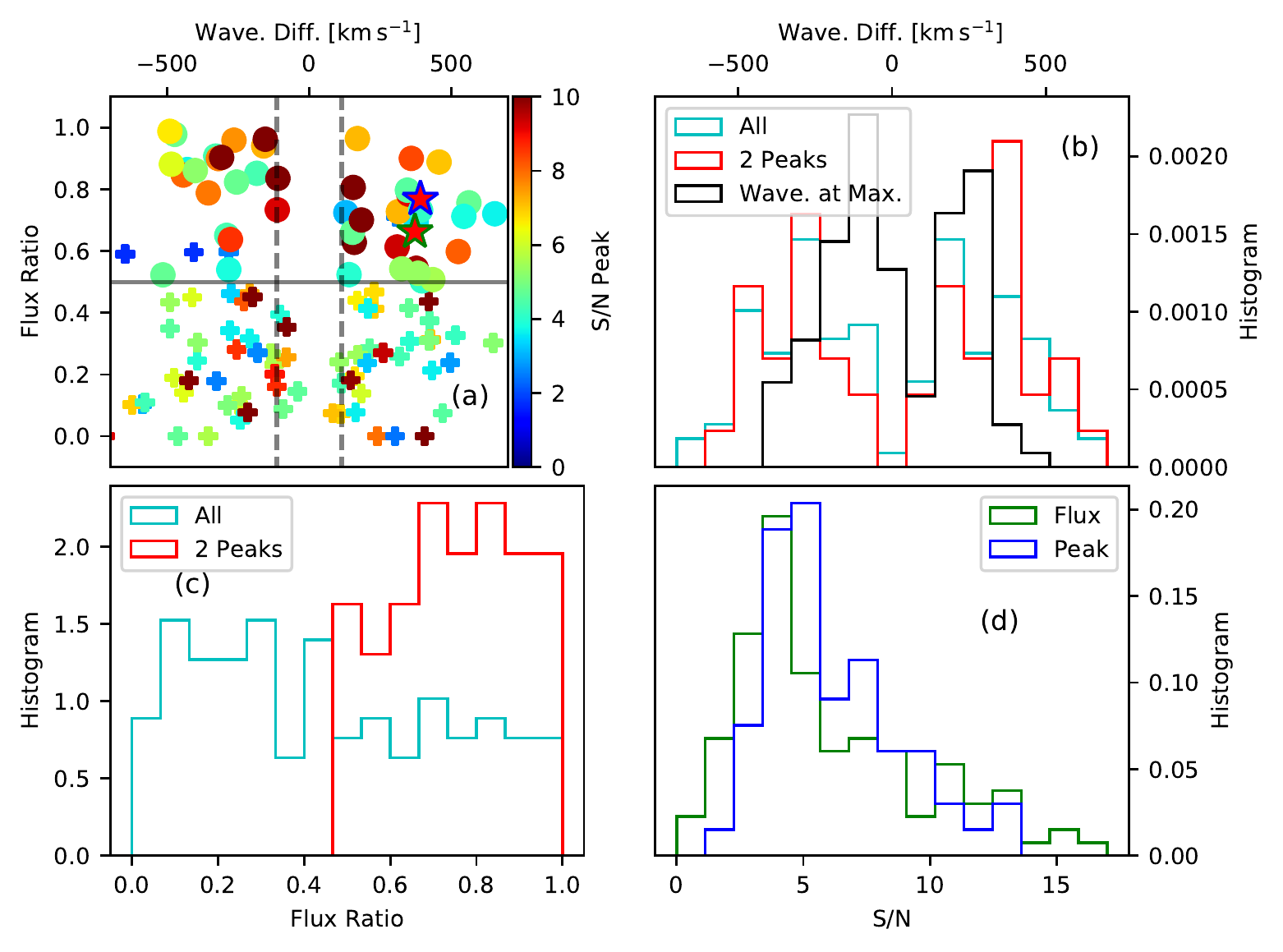}
\caption{
    Properties of the two Gaussians plus continuum fit to the observed H$\alpha$ emission line profiles.
(a) Scatter plot with the flux ratio versus the wavelength separation of the two Gaussians, color-coded with the signal to noise ratio (S/N) at the maximum emission. Double peak profiles (the bullet symbols) are defined as those clumps where the flux ratio is $> 0.5$, the peak separation is larger than the spectral resolution (marked by the two vertical dashed lines), and  ${\rm S/N} > 3$. Those clumps that do not meet these conditions are represented by plus symbols. The star symbols mark the location of the two types of stacked spectra.  
(b) Distribution of wavelength separation between the two Gaussians: for all clumps (the cyan line) and only those that meet the conditions to be regarded as double peak profiles (the red line). The black line does not show separation between peaks but difference between the wavelength of the peak emission and H$\alpha$.  Wavelength differences are expressed in km\,s$^{-1}$.
(c) Distribution of flux ratios of the two Gaussians: for all clumps (the cyan line) and only those that meet the conditions to be regarded as double peak profiles (the red line). 
(d) Distribution of S/N for the peak emission (the green line) and the wavelength integrated flux of the emission line (the blue line). All histograms are normalized to area equals one.
}
\label{fig:spectra2_e}
\end{figure*}

  We approach the problem of measuring the properties and shapes of the profiles by fitting two Gaussian functions plus a continuum to each individual H$\alpha$~profile. Provided the fits are good, the ratio between the fluxes of the two fitted Gaussians allows us to quantify the fraction of double peak  profiles. The top left panel in Fig.~\ref{fig:spectra2_e}a displays the diagnostic diagram used to carry out such classification. It represents a scatter plot with the flux ratio versus the wavelength separation of the two Gaussians, color-coded with the S/N at the maximum emission. The noise for the S/N is obtained as the standard deviation of the spectrum in a nearby continuum window (from 6500\,\AA\ to 6540\,\AA\ in rest-frame wavelengths). Double peak profiles are defined to be those where the secondary component over the main component flux ratio  is $> 0.5$, the peak separation is larger than the spectra resolution ($> 2.5{\rm \AA}\, \equiv \,115\, {\rm km\,s^{-1}}$; see Sect.~\ref{sec:data}), and ${\rm S/N} > 3$.
The double peak line profiles thus selected make up around 38\,\%\ of the candidates (45/118). The rest of our candidates (73/118) are regarded as single peak emission. The properties of the resulting two Gaussian fits are summarized in Fig.~\ref{fig:spectra2_e}b (distribution of wavelength separations),  Fig.~\ref{fig:spectra2_e}c (distribution of flux ratios).
The fits leading to Fig.~\ref{fig:spectra2_e} were carried out fitting a single Gaussian around the maximum emission, removing this Gaussian from the observed profiles, and then fitting a second Gaussian to the residuals.

Figure~\ref{fig:spectra2_e}d shows the distributions of S/N ratios at the peak emission and integrated along the line profile (i.e., the flux). As we explain above, the noise for this estimate was computed as the standard deviation of the spectrum in a nearby continuum window to the red of H$\alpha$. In the case of the integrated flux, the noise is found by propagating this error into the integral using the error propagation equation \citep[e.g.,][]{1971stph.book.....M}. We find that 97\,\%\ of profiles have ${\rm S/N} > 3$ in either peak or flux. % from spectra2 

As a byproduct to initialize the first Gaussian fit, we had to compute the centroid of the maximum emission within the H$\alpha$ detection band pass. The distribution  of wavelengths corresponding to this centroid is represented in  Fig.~\ref{fig:spectra2_e}b, the black line, and it provides further evidence for the detected signals to be real. Noise is expected to fake signals at random wavelengths. Thus, the wavelength of the peak emission produced by noise should follow a uniform distribution within the detection band pass. However, the observed distribution  presents two conspicuous peaks (Fig.~\ref{fig:spectra2_e}b, the black line). One can use the KS test \citep[e.g.,][]{1986nras.book.....P} to assess the null hypothesis that the wavelengths of the maximum emission are  drawn from the uniform distribution expected from noise. This exercise provides a {\em p-}value of only 0.026, thus discarding the null hypothesis with a probability of 97.4\,\%. Other similar KS tests described in Sects.~\ref{sec:Azimuths} and \ref{sec:distances} also discard the null hypothesis with even higher confidence.

%
%%%%%%%%%%%%%%%%%%%%
\begin{figure*}
\centering
\includegraphics[width=0.45\linewidth]{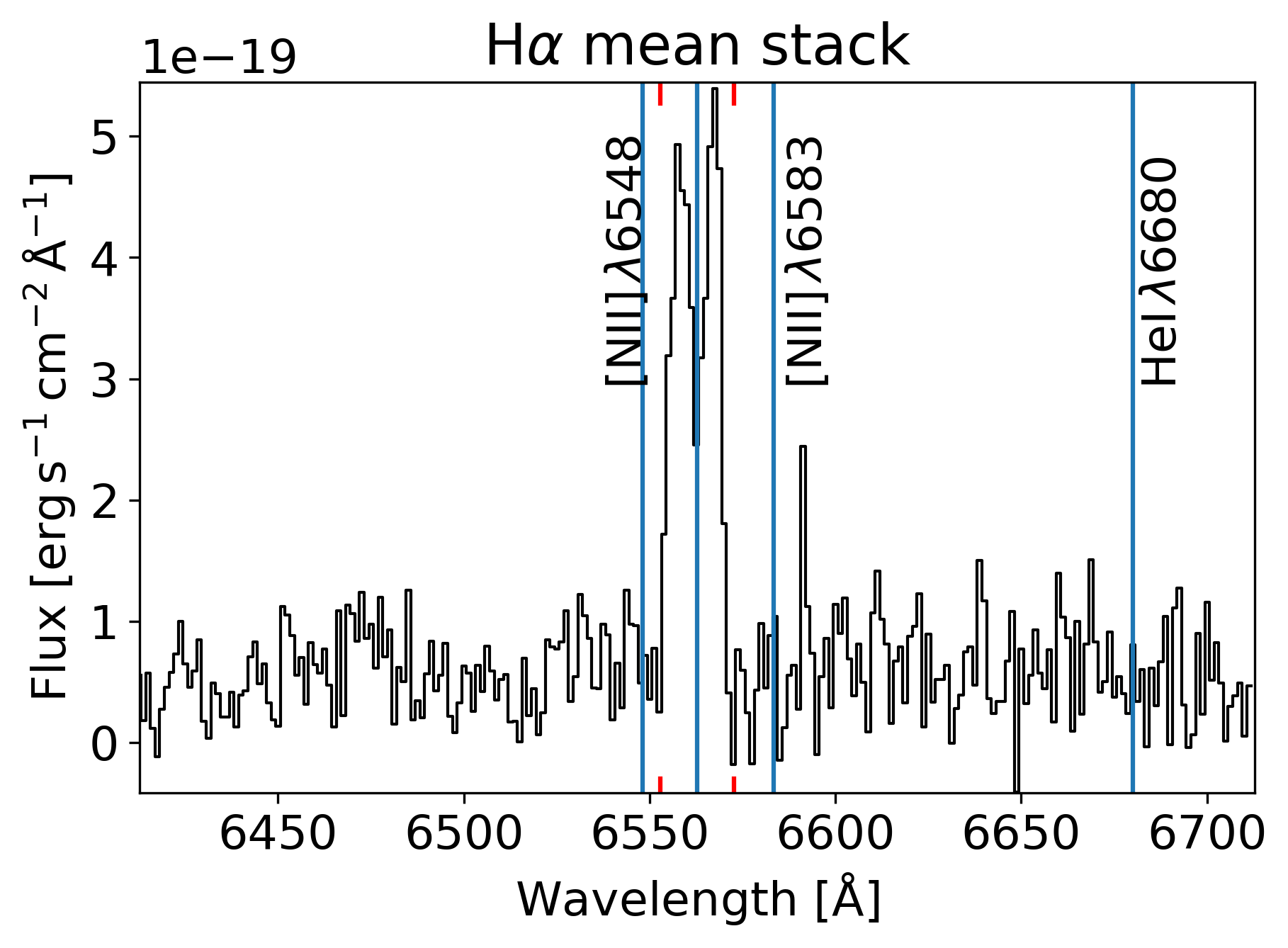}
\includegraphics[width=0.45\linewidth]{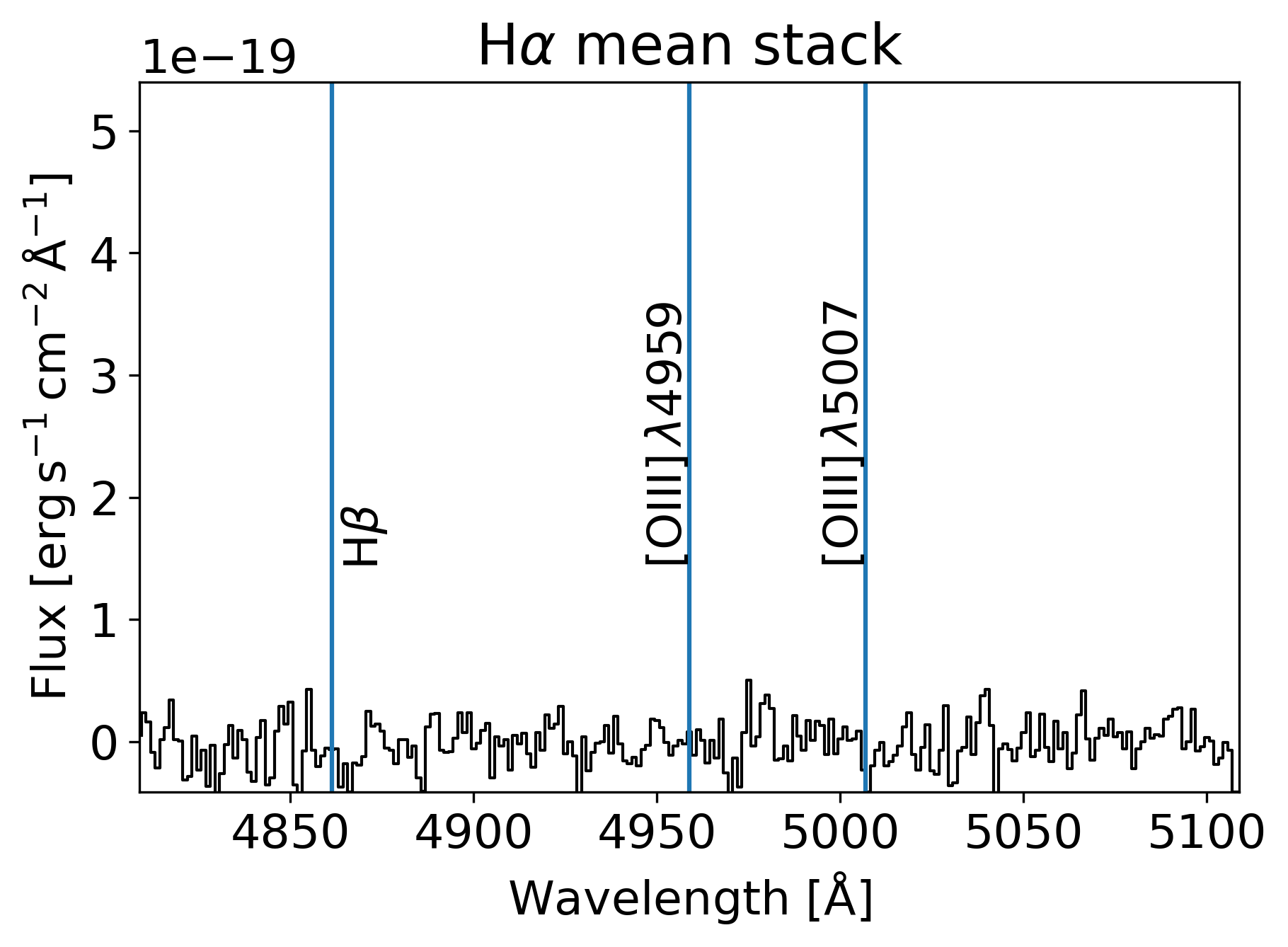}
\includegraphics[width=0.45\linewidth]{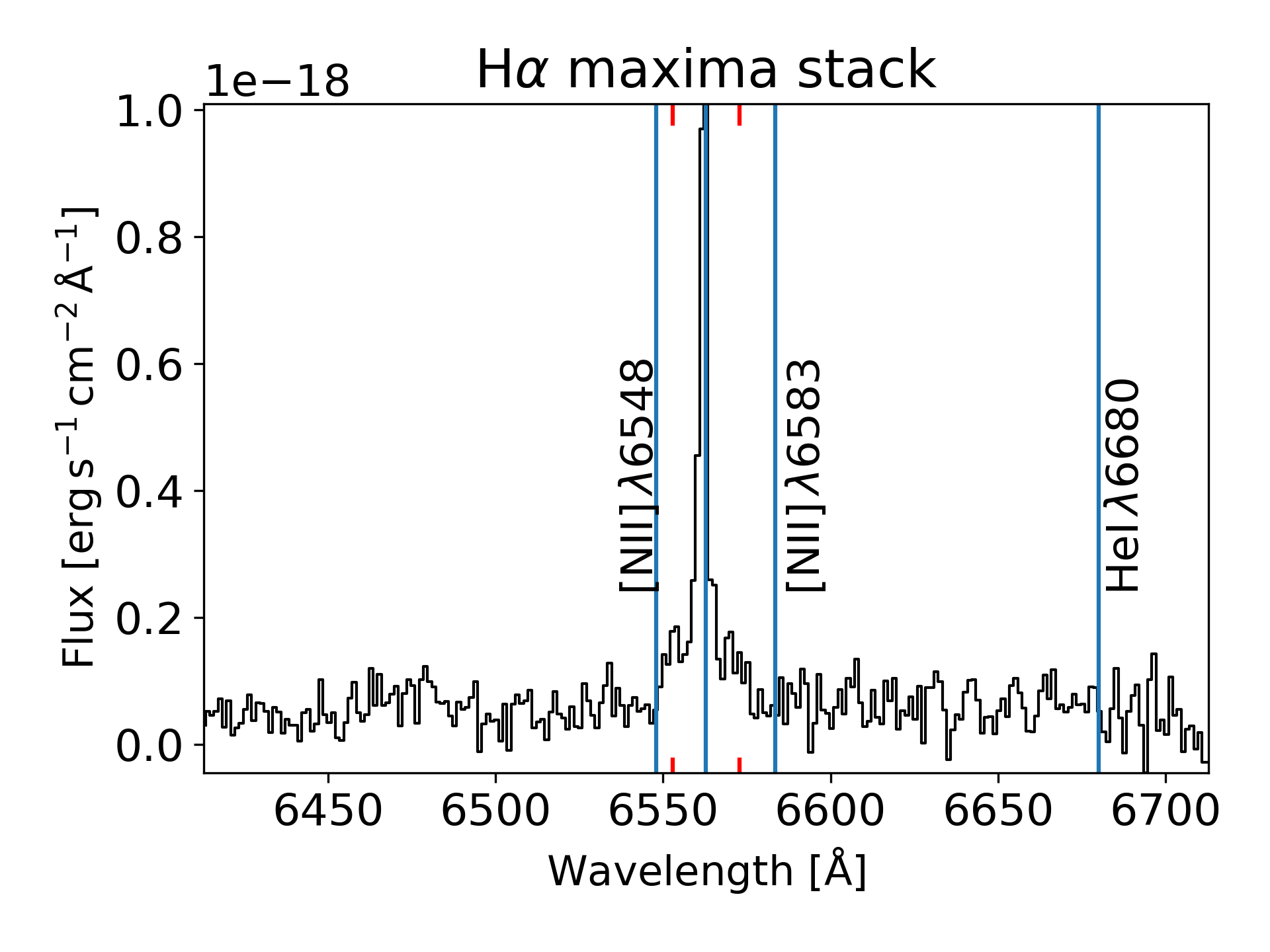}
\includegraphics[width=0.45\linewidth]{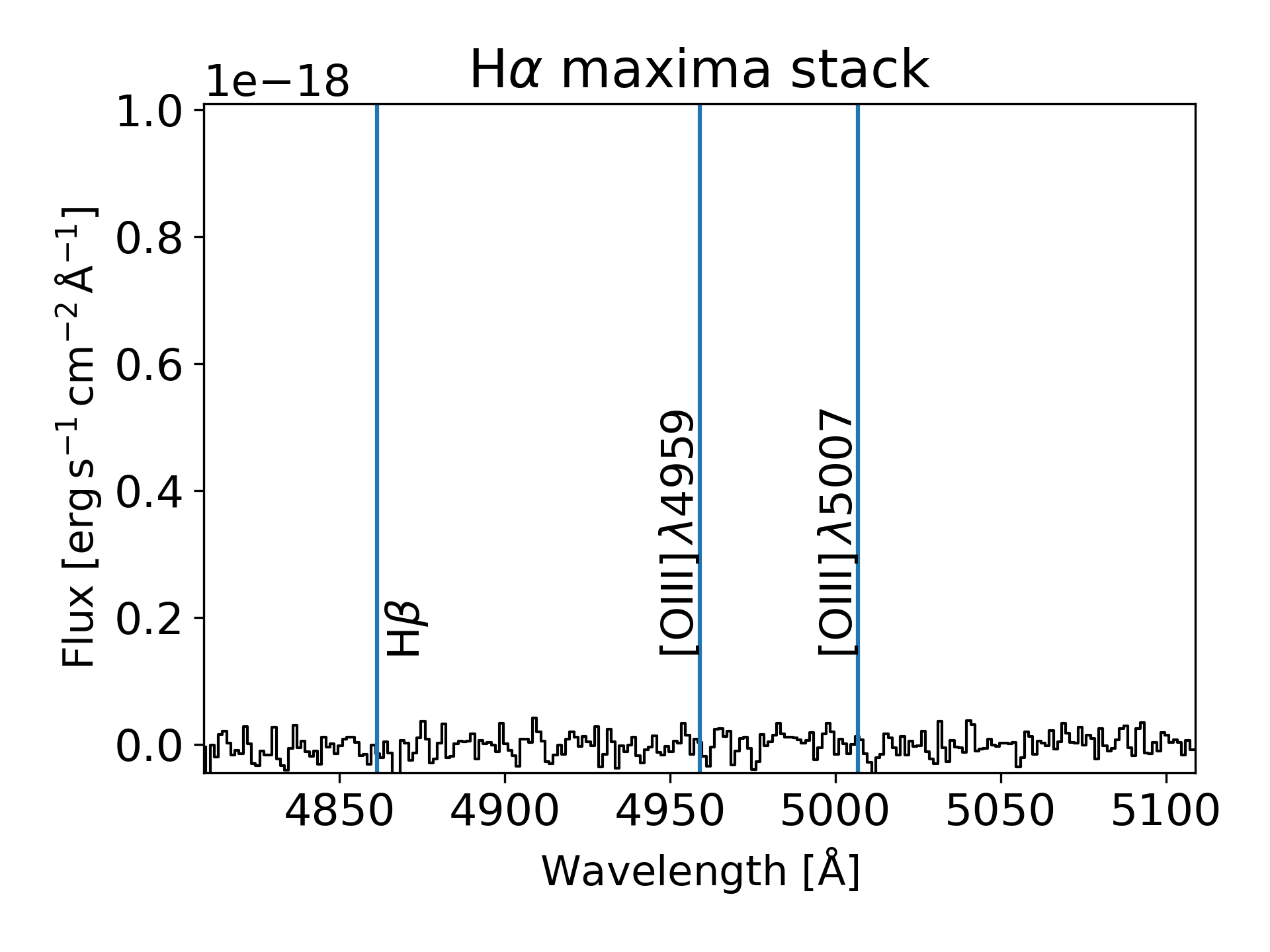}
\includegraphics[width=0.45\linewidth]{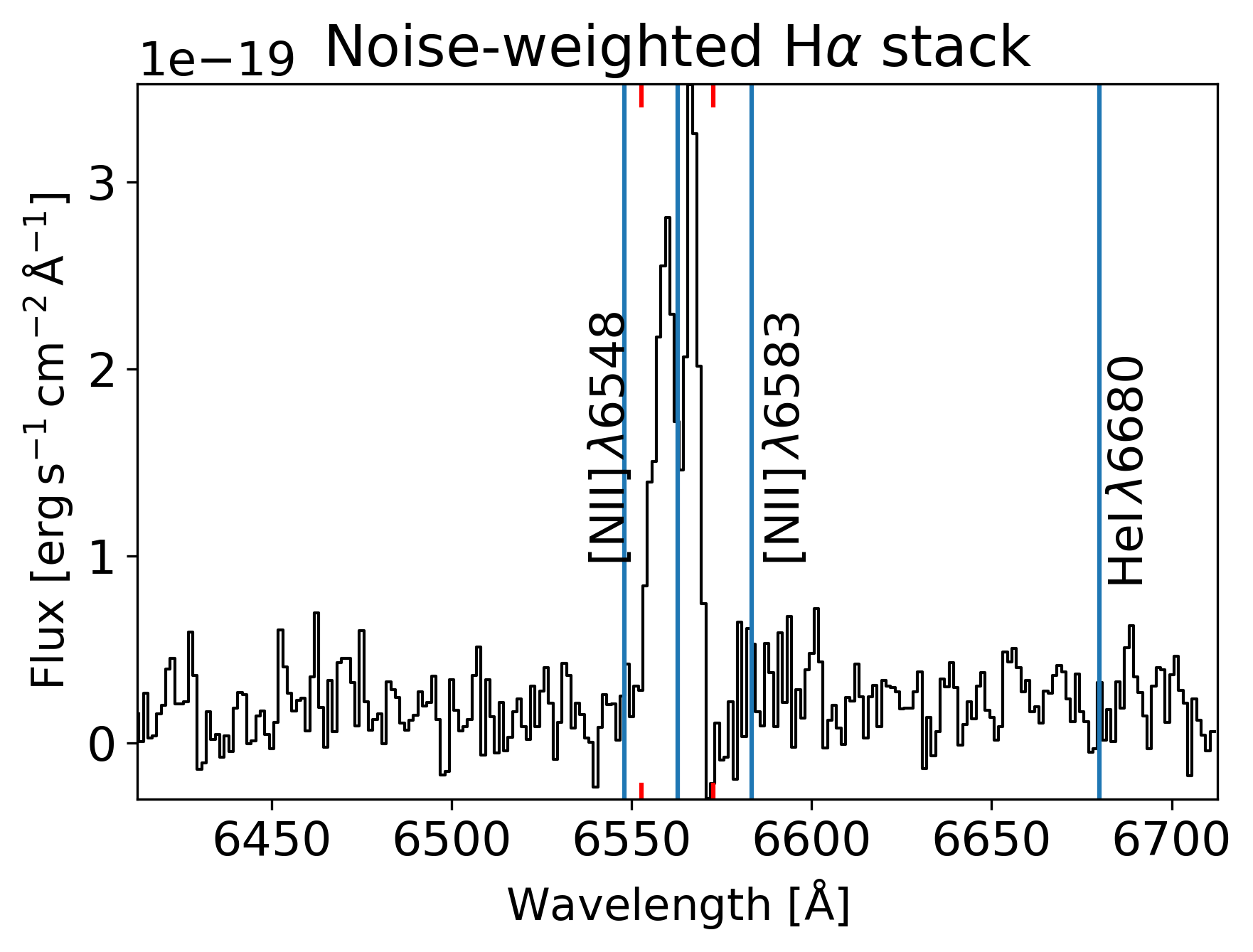}
\includegraphics[width=0.45\linewidth]{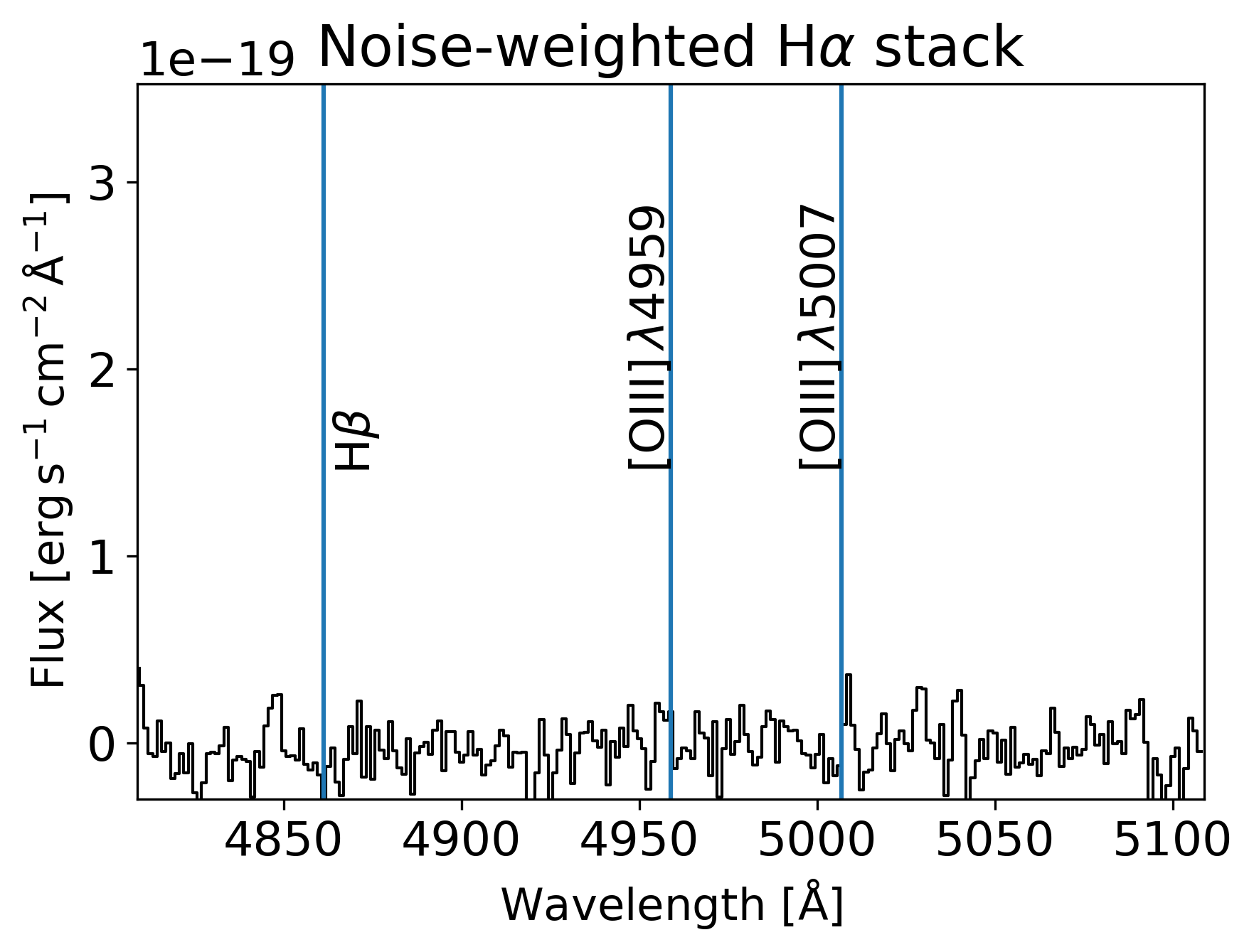}
\caption{Stacking all spectra of the good candidates. Spectra around H$\alpha$, H$\beta$, and [O{\sc iii}]$\lambda$5007 obtained by stacking assuming the redshift of the central galaxy (top row), the redshift of the maximum of the H$\alpha$ emission feature (central row), and using the redshift of the central galaxy but weighting each spectrum according to the noise level in the continuum next to H$\alpha$. The red ticks on the top and bottom edges of the H$\alpha$ panels mark the wavelength range used to search for H$\alpha$. Each row shares a common vertical scale.\\
%
%\comment{Joao: Ana Luisa suggests, and I am with her, showing H$\beta$ and [OIII]5007 in a single panel. Thus, this figura will only contain 3x2 panels. In order to have enough space on top, write ''H$\alpha$ maxima stack'}\joao{Done.}\\
%\comment{Joao:'Maxima H$\alpha$ stack' $\longrightarrow$ 'H$\alpha$ maxima stack'}\joao{Done.}
%
%\comment{Joao: some of the panels still have $\lambda\lambda$. Please, correct.}\joao{Corrected. I think I got them all now.}
%\comment{Joao: Labels on top the last column. ${\rm OIII(5007)\longrightarrow [O{\sc iii}]\lambda5007}$  }
%\joao{Replace the $\lambda \lambda$ with just $\lambda$. Remove wavelengths from H$\alpha$ and H$\beta$.}
%\comment{Are these stacks 'averages' or 'sum' of the individual spectra. Please, clarify.}
%\comment{Are the pix$^{-1}$ correct? Please, revise.}
%\joao{Revised. Plots now show the flux as determined by multiplying each individual spectrum by its respective collection area and then stacking. pix has been removed and $\rm \AA^{-1}$ added for correctness. All rows now have the same scale. Have been making the labels larger but the results aren't as obvious as I wanted, I suspect because Latex re-scales everything down to fit the page, so I will likely have to enlarge the font much more. Jorge once again proves to be a superior expert in Python. The line now allows for easy enlarging of the fonts, which I have done. Might need to go a bit further.}
}
\label{fig:Stacked_profiles}
\end{figure*}
In order to increase the signal-to-noise ratio in the emission lines, we stacked the spectra of the 118 candidates that passed the second inspection. The spectrum of each candidate has to be shifted in wavelength and re-sampled to a common laboratory wavelength scale, adopting the original MUSE sampling of 1.25\,\AA . We create three different stacks differing in the way the individual spectra are shifted and averaged.
(1) The first stack assumes the redshift of the candidate to be the same as the redshift of the central galaxy. This ignores proper motions and should give us a good general idea of the typical spectrum of a good candidate. All spectra are weighted equally. We apply a plain averaging for this stacking and do not normalize the spectra, which biases the final spectrum towards the strongest signals. 
(2) The second stacking assumes the redshift of the candidate to be set by the wavelength of the peak emission inside the wavelength range used for detection and classification (observed H$\alpha$ $\pm\,10\,$\AA ; Sect.~\ref{Ha,R,G images}). 
This procedure partly corrects for the relative proper motion of the H$\alpha$ emission clump with respect to the central galaxy, thus enhancing the resulting emission line signals. The weighting is the same as for (1).
(3) This stacking uses a redshift equals to that of the central galaxy but weights each individual spectrum according to the noise of the continuum next to H$\alpha$ (from 6500\,\AA\ to 6540\,\AA ). %This time the signals are enhanced with respect to (1) because the highest signal-to-noise ratio spectra over-contribute to the average.
Figure~\ref{fig:Stacked_profiles} shows the average spectrum obtained following these three approaches for H$\alpha$, H$\beta$, and [O{\sc iii}]$\lambda$5007. The stacked spectra remains noisy for all lines except H$\alpha$, which gets significantly enhanced. 
Note how the H$\alpha$ stacked spectra have a small but noticeable continuum.

The stacking using only the redshift of the central galaxy (Fig.~\ref{fig:Stacked_profiles}, top row) does not appear to show any feature at H$\beta$ or [O{\sc iii}]$\lambda$5007. While this could be readily interpreted as absence  of these lines in the majority of the stacked spectra, we note that the noise-weighted stack shows a dip with a small emission core at the position [O{\sc iii}]$\rm \lambda$5007 (Fig.~\ref{fig:Stacked_profiles}, bottom right panel).
%The decrease of the spectra may be due to some sort of self-absorption and re-emission, however this process is expected to be difficult to happen, due to the high gas density necessary for it to occur. We discuss this possibility in Sect.~\ref{sec:origin}.
%\comment{I do not see the dip. This has to be discussed.}
%\joao{Revisit once Figure 5 is updated with more examples, as references to particular line/rows will have to be corrected.}
%\comment{where is the staking of the 'Contaminated' spectra? Include a figure with them.} \joao{No longer applicable, following the decisions on the classification scheme.}
%\comment{the above description has to be completed once proper figures are available.}
%\comment{A discussion on how dust reddening could make H$\beta$ so small remains to be included.}

The ratio of H$\alpha$ to H$\beta$ fluxes is expected to be around 3 if it is produced by recombination in a fully ionized medium at some $10^4\,$K. We measure the ratio to be larger than approximately 6 (Fig.~\ref{fig:Stacked_profiles}). Thus, should the lack of H$\beta$ would be due to dust reddening, the $V$ band extinction\footnote{Assuming a \citet{1989ApJ...345..245C} law with a MW-like extinction.} has to be $A_V > 2.2$, which is quite significant. This high extinction would also explain the absence of [O{\sc iii}]$\lambda$5007. Alternatively, one could also reproduce the large observed flux ratio if part the emission is due to collisional excitation rather than to recombination \citep[e.g.,][]{2015RMxAA..51..231R}. These conditions may be met in shocks \citep[e.g.,][]{2012MNRAS.422.2357T}, which also produce emission where [O{\sc iii}]$\lambda$5007 is negligibly small compared to H$\alpha$ \citep[e.g.,][]{2019RMxAA..55..377A}.
The  dense environment existing  in the broad line region of AGNs also disfavors the collisionally excited [O{\sc iii}]$\lambda\lambda$4959,5007 lines with respect to the Balmer lines \citep[e.g., ][]{1984QJRAS..25....1O,2008ARA&A..46..475H}.

To further characterize the emission, we also fit two Gaussians plus a continuum to the H$\alpha$ stacks showing double peaks (Fig~\ref{fig:Stacked_profiles}, top and bottom left panels). In both cases, we find the two Gaussians to be separated by  8\,--\,9\,\AA, equivalent to a velocity of $\rm 350-400 \, km \, s^{-1}$, and implying a proper motion with respect to the central galaxy of the order of 150\,--\,200\,$\rm km \, s^{-1}$.  The FWHM of the Gaussians is 6.9\,\AA\ (315\,km\,s$^{-1}$) for the plain stack and 4.6\,\AA\ (210\,km\,s$^{-1}$) for the noise-weighted stack. Despite the uncertainties, they are much larger than the spectral resolution of \muse\ (2.5\,\AA), implying the existence of a large velocity dispersion within the emitting gas. The continua are $7.5\times 10^{-20}$ and  $\rm 9.2 \times 10^{-21}\,erg\,s^{-1}\,cm^{-2}\,\AA$ for  the plain stack and the noise-weighted stack, respectively. 
%\joao{The lowering of the continuum for the last stack is probably an artificial result due to the noise-weighting applied, where the more noisy spectra will necessarily contribute less to the overall result (see Fig. \ref{fig:Stacked_profiles}). \comment{Joao: I do not quite get the argument}}

The line [N{\sc ii}]$\lambda$6583 is never detected. The flux ratio between [N{\sc ii}]$\lambda$6583 and H$\alpha$ is a proxy for the metallicity of the gas \citep[][]{2002MNRAS.330...69D}, thus, the non-detection can be transformed to an upper limit on the metallicity. Using the calibration by \citet{2004MNRAS.348L..59P} and the same flux ratio limit employed above (i.e., [N{\sc ii}]$\lambda$6583/H$\alpha < 1/6$), we get ${\rm 12+\log(O/H) < 8.5}$. This limit is not very restrictive, though.  Using the solar metallicity measured by \citet{2009ARA&A..47..481A}, it corresponds to around 60\,\% of the solar metallicity.

\begin{figure} %%%
\centering
\includegraphics[width=0.9\linewidth]{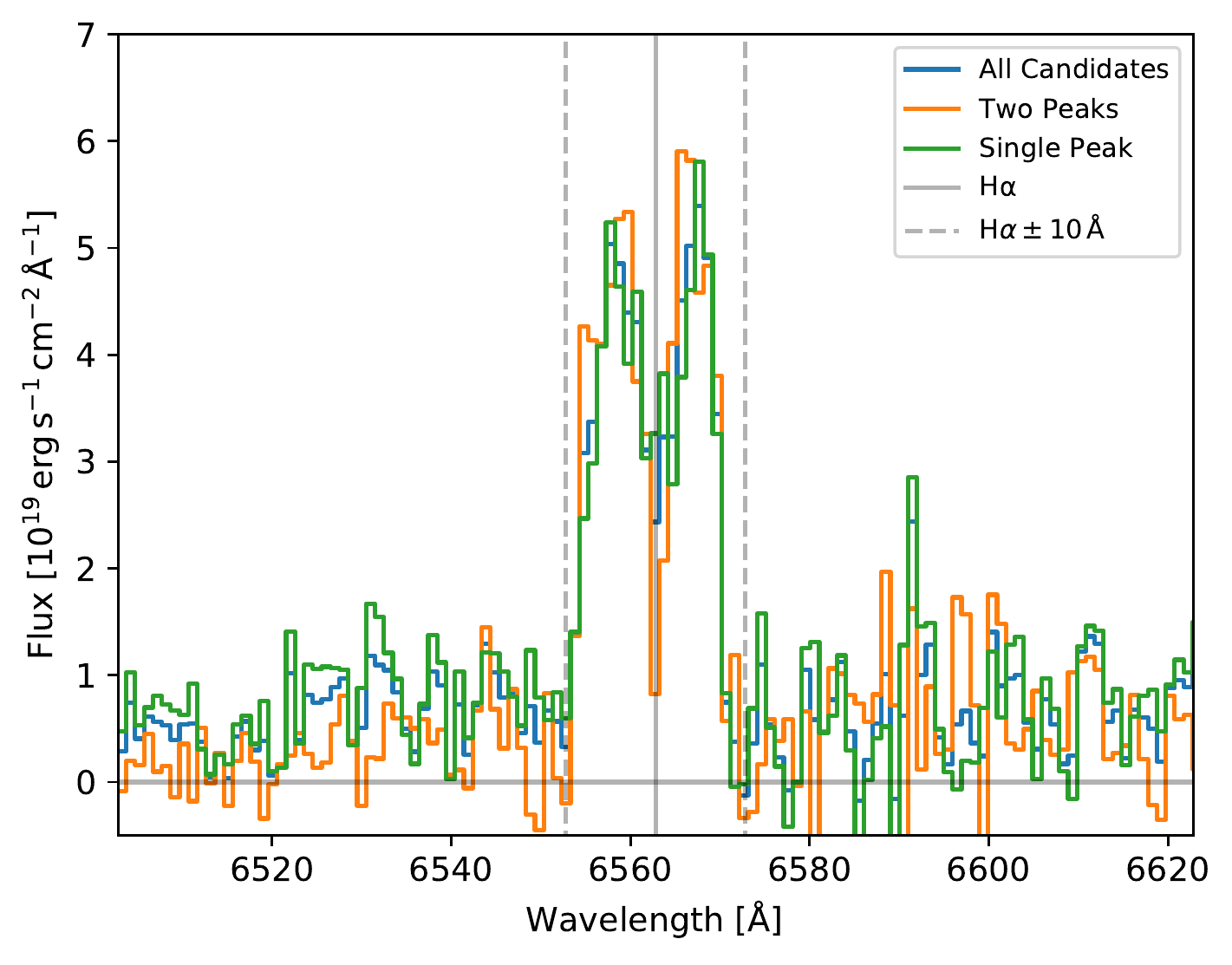}
\caption{
Stacked H$\alpha$ spectra considering all 118 candidates (blue line), only spectra classified as double peaks (orange line) and only spectrasified as single peak (green line). Note how the mean spectra of single peaks also shows two peaks. The H$\alpha$ wavelength and the wavelength window used for  detection are indicated as explained in the inset. The averages are plain means  as those corresponding to the top panel in Fig.~\ref{fig:Stacked_profiles}.
}
\label{fig:separate_stacks}
\end{figure}

One final remark is in order. Some of the single peak profiles may be double peaks with one of the two peaks missing. This is suggested by a number of issues. Most of the profiles contributing to the stacked spectra are single peak and yet the stacks show double peaks. The distribution of wavelengths at the maximum emission of all profiles show two peaks (Fig.~\ref{fig:spectra2_e}b, the black solid line). Finally, the stacking of only single peak spectra also results in a double peak profile (see Fig.~\ref{fig:separate_stacks}). 
%

%
%%%%%%%%%%%%%%%%%%%%
%
\subsection{Fluxes, luminosities, redshifts, and sizes of the H$\alpha$ emitting clumps}\label{sec:physical_properties}

\begin{figure} %%%
\centering
\includegraphics[width=0.9\linewidth]{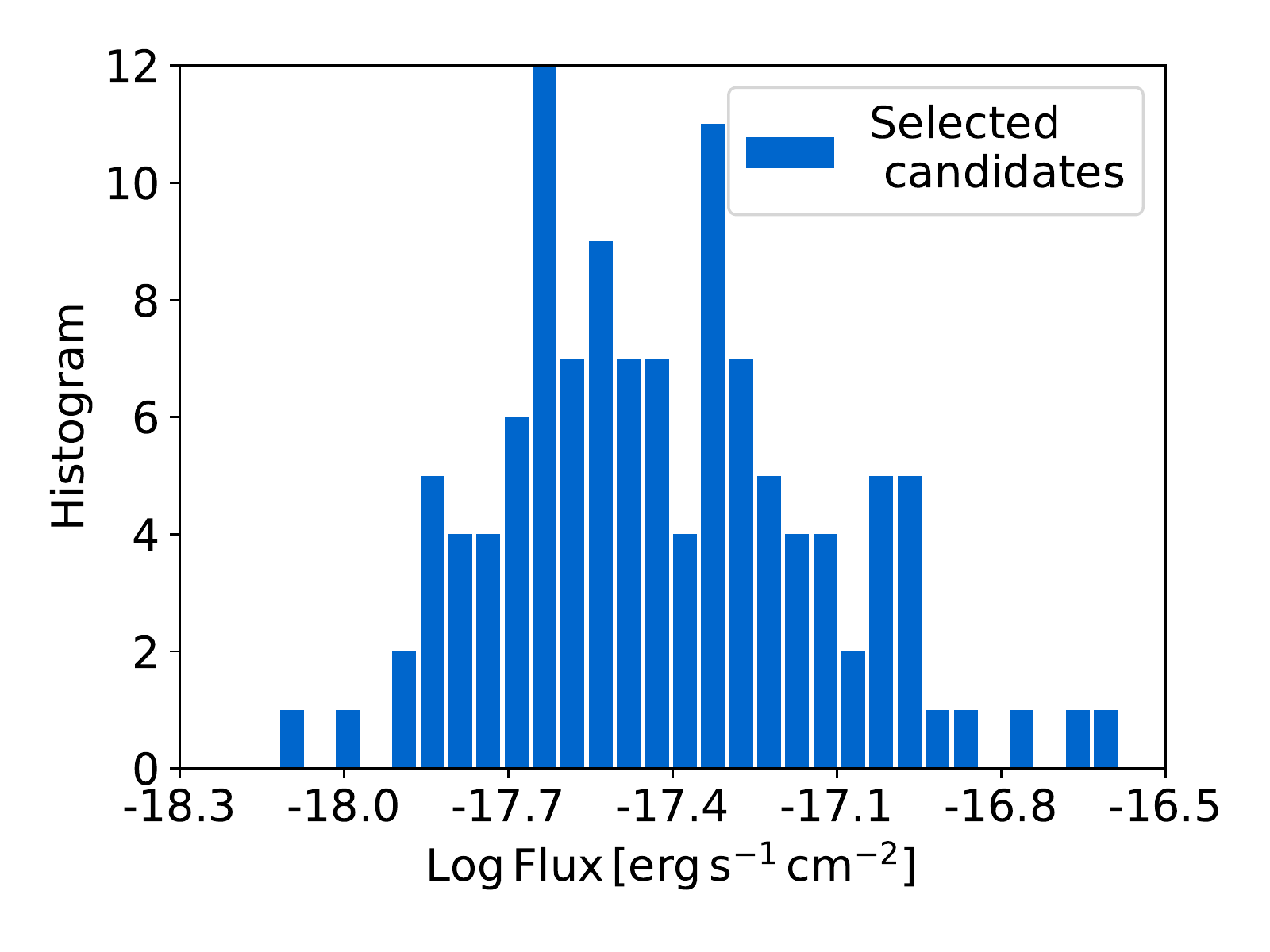}
\includegraphics[width=0.9\linewidth]{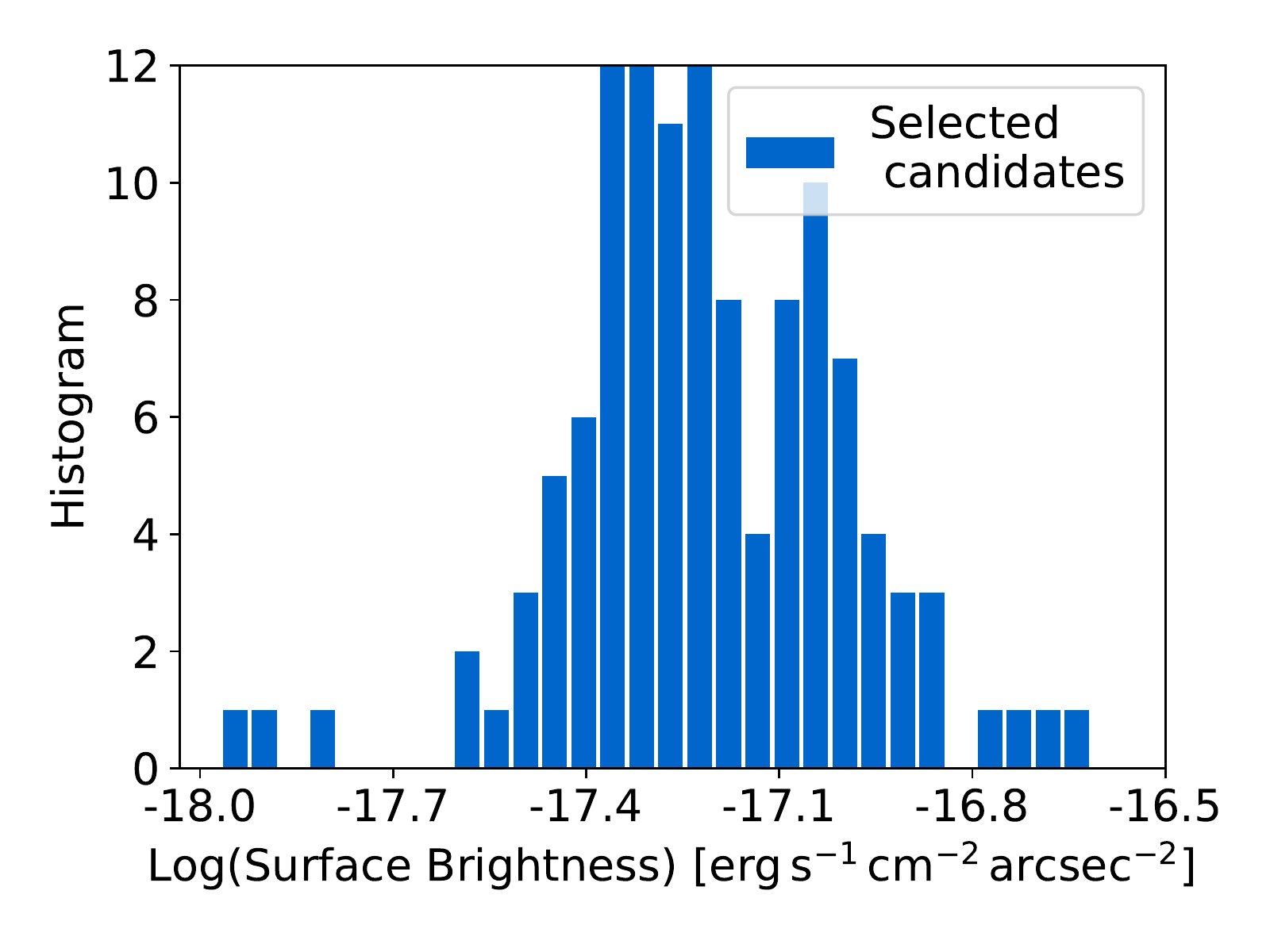}
\caption{Distribution of log flux (top panel) and log surface brightness  (bottom panel), given in cgs units as indicated in the labels. Our candidates have a log flux of $\rm -17.3 \pm 0.3$ and a log surface brightness of $\rm -17.2 \pm 0.2$.
%\comment{${\rm Log_{10}}$ --$>$ $\log$.}
%\comment{${\rm erg\,cm^{-2}\,s^{-1}}$ --$>$ ${\rm erg\,s^{-1}\,cm^{-2}}$}
%\comment{VERY extrange. Horizontal Ticmarks are sometimes separated by 0.2 dex while other times the saparation is 0.3. I do not understand what happends unless you put the labels by hand. Happens in both panels. Please, correct.}\joao{It's a problem with the decimal places shown. I will fix it.}
}
\label{fig:Gaussian_Fits_Histogram}
\end{figure}
\begin{figure} %%%%%
\centering
\includegraphics[width=0.9\linewidth]{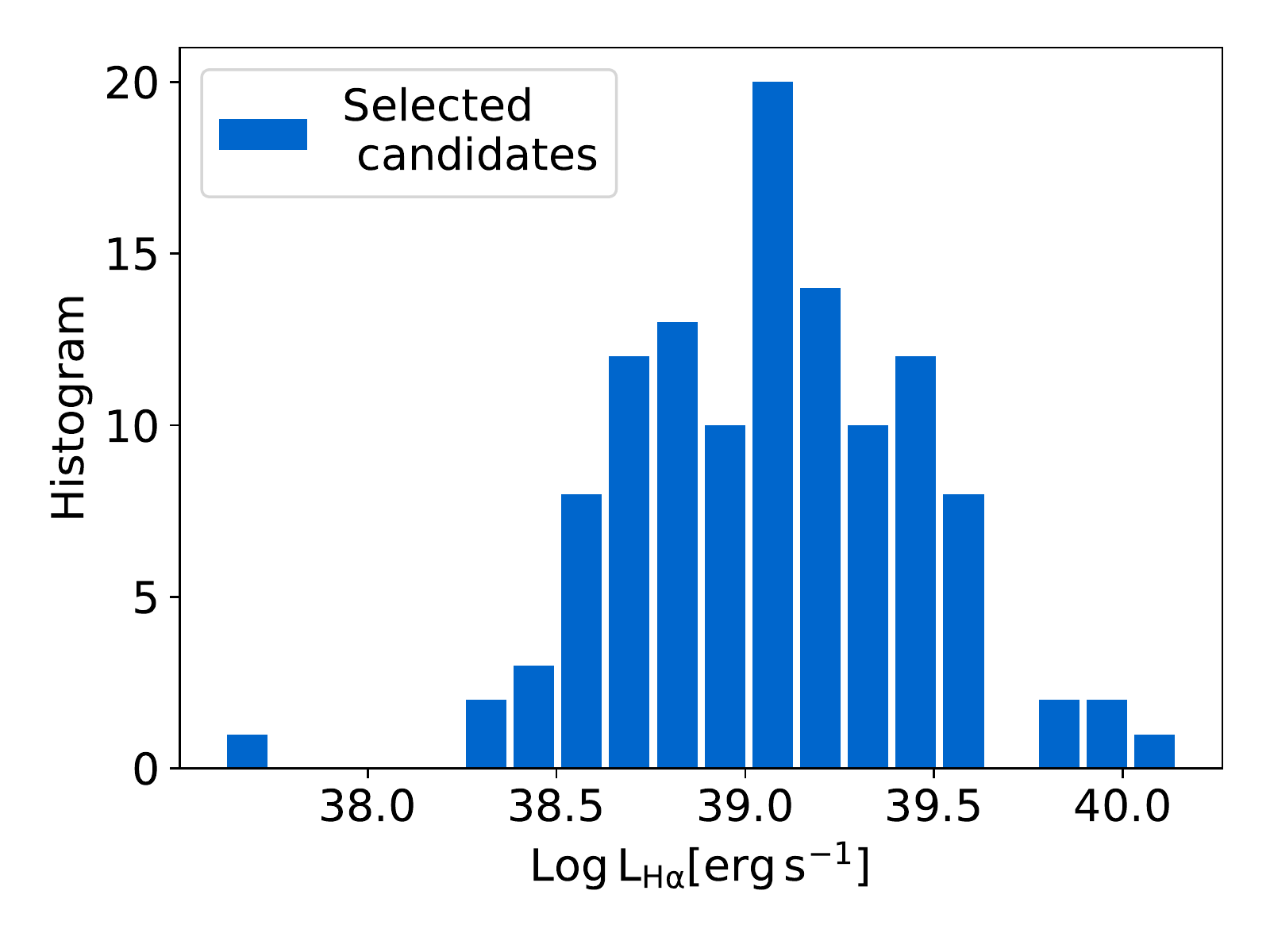}
\caption{Distribution of  H$\alpha$ luminosity for the candidates that passed both inspections. The detected clumps are relatively faint with a mean luminosity around $\rm 10^{39.2} \, erg \, s^{-1}$ and an upper limit at around $\rm 10^{40} \, erg \, s^{-1}$. 
%\comment{${\rm Log_{10}}$ --$>$ $\log$.}
%\comment{Lum$_{H\alpha}$ --$>$ $L_{\rm H\alpha}$}
}
\label{fig:luminosity}
\end{figure}
We estimate the flux of the H$\alpha$ emission features as the total flux of the two Gaussian fit described in Sect.~\ref{sec:line_shapes}.
Figure~\ref{fig:Gaussian_Fits_Histogram} shows the distribution of observed fluxes (upper panel) and surface brightness (lower panel).  
The final selected H$\alpha$ candidates present a $\log$ flux distribution with mean and standard deviation of $\rm -17.3 \pm 0.3$, with the flux given in cgs units ($\rm erg \, s^{-1} \, cm^{-2}$). The distribution of $\log$ surface brightness has mean and standard deviation $\rm -17.2 \pm 0.2$, in units of $\rm \, erg \, s^{-1} \, cm^{-2} \, arcsec^{-2}$.

Figure~\ref{fig:luminosity} contains the distribution of H$\alpha$ luminosities, derived from the observed fluxes ($F_{\rm H\alpha}$) as, 
\begin{equation}
    L_{\rm H\alpha} = 4 \pi D_L^2 F_{\rm H\alpha},
    \label{eq:lumha}
\end{equation}
where $D_L$ is the luminosity distance, estimated from the redshift of the central galaxy and the cosmological parameters given in Sect.~\ref{sec:intro}.

\begin{figure} %%%
\centering
\includegraphics[width=0.9\linewidth]{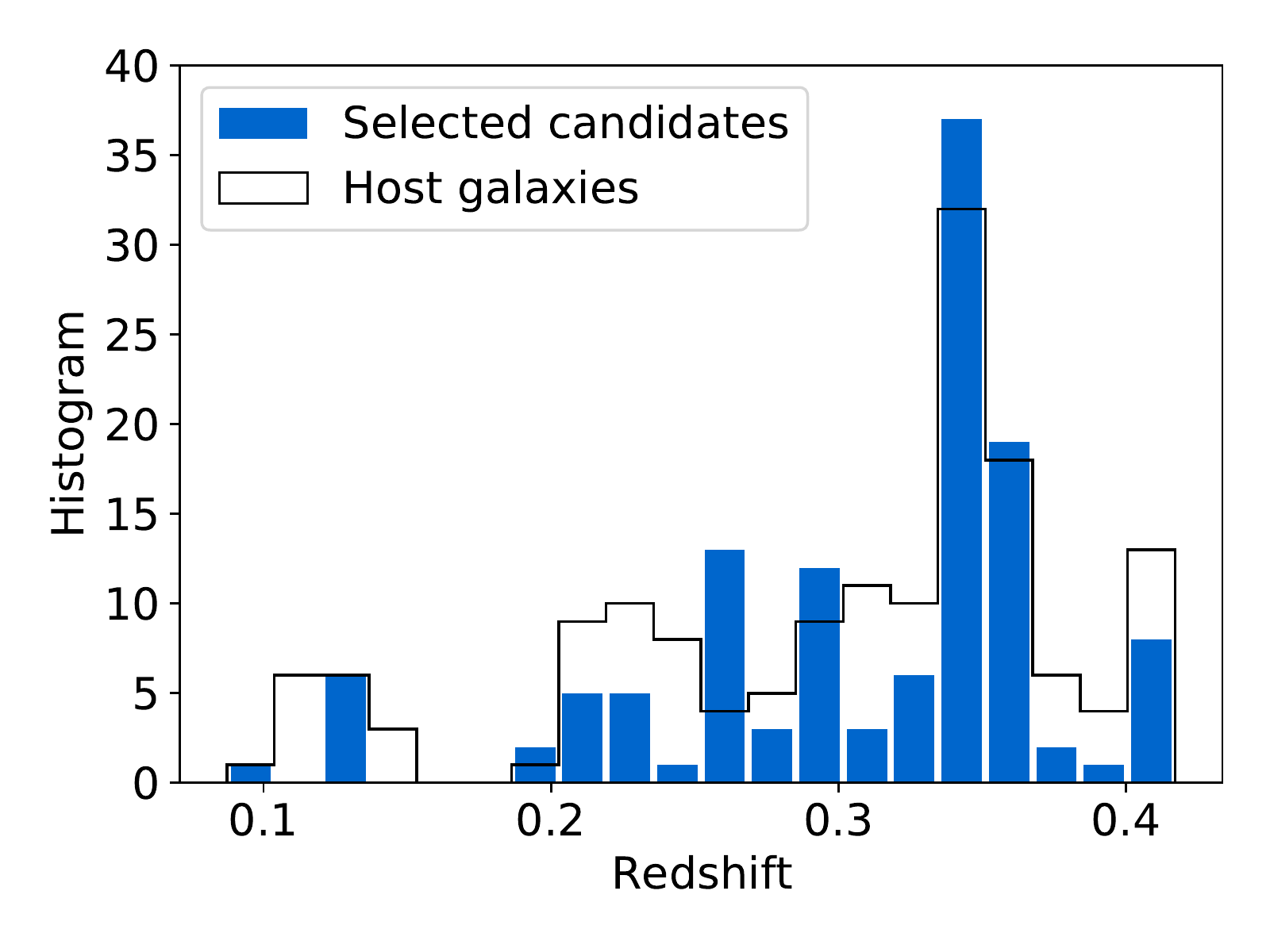}
\caption{Distribution of redshift for the selected candidates. The average redshift is $\rm 0.31\pm0.07$, with the error bar giving the standard deviation. The black line represents the distribution of redshifts of the galaxies originally selected for inspection, and is included here for reference.
}
\label{fig:histo_redshifts}
\end{figure}
Figure~\ref{fig:histo_redshifts} shows the distributions of redshift of the H$\alpha$ clumps. The candidates have an average redshift of $0.31 \pm 0.07$, with the error giving the standard deviation of the distribution. The distribution mimics the distribution of redshifts of the central galaxies with a preference for redshifts around 0.35 (see the black solid line in Fig.~\ref{fig:histo_redshifts}). 

\begin{figure} %%%%%
\centering
\includegraphics[width=0.9\linewidth]{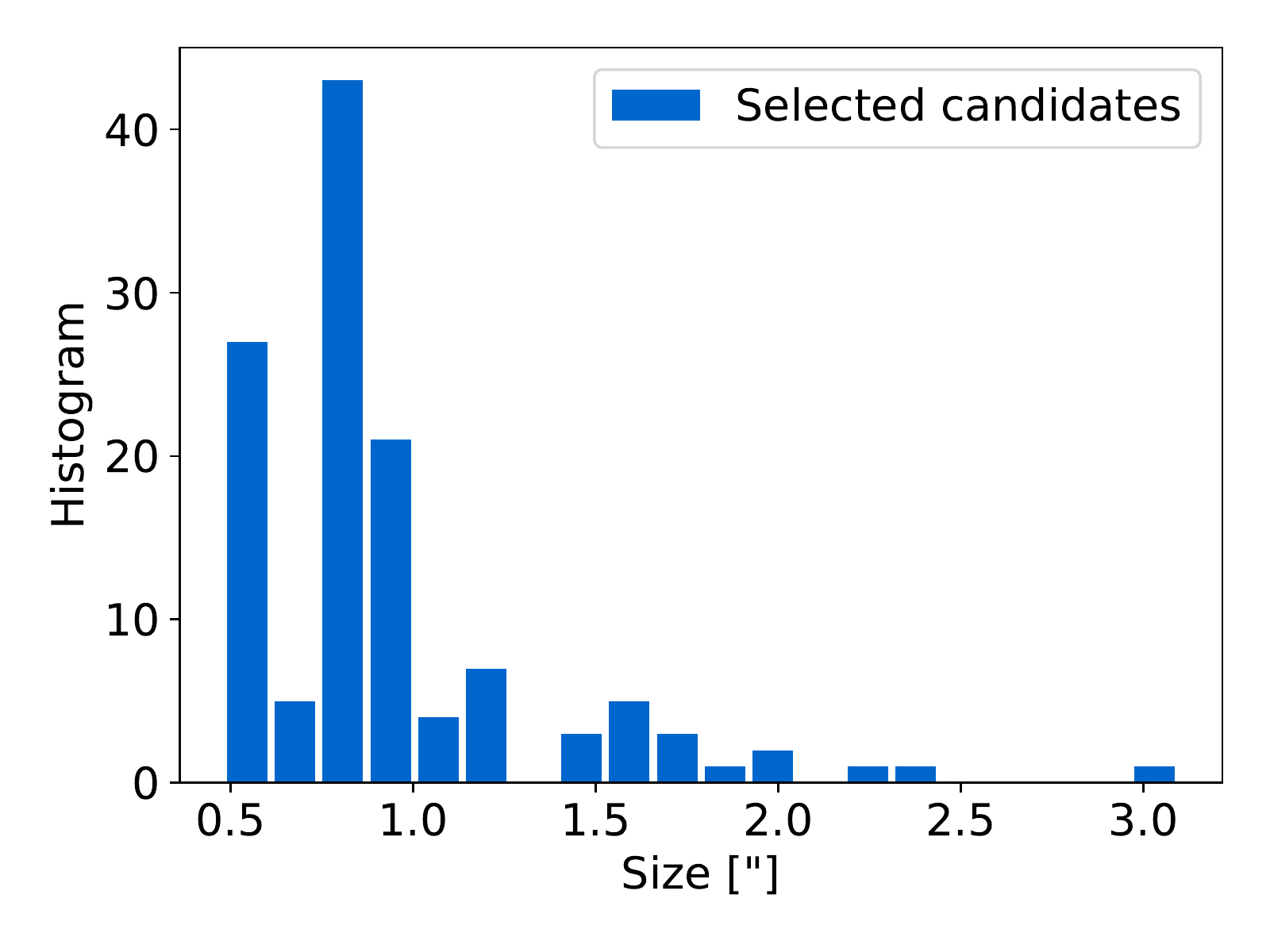}
\includegraphics[width=0.9\linewidth]{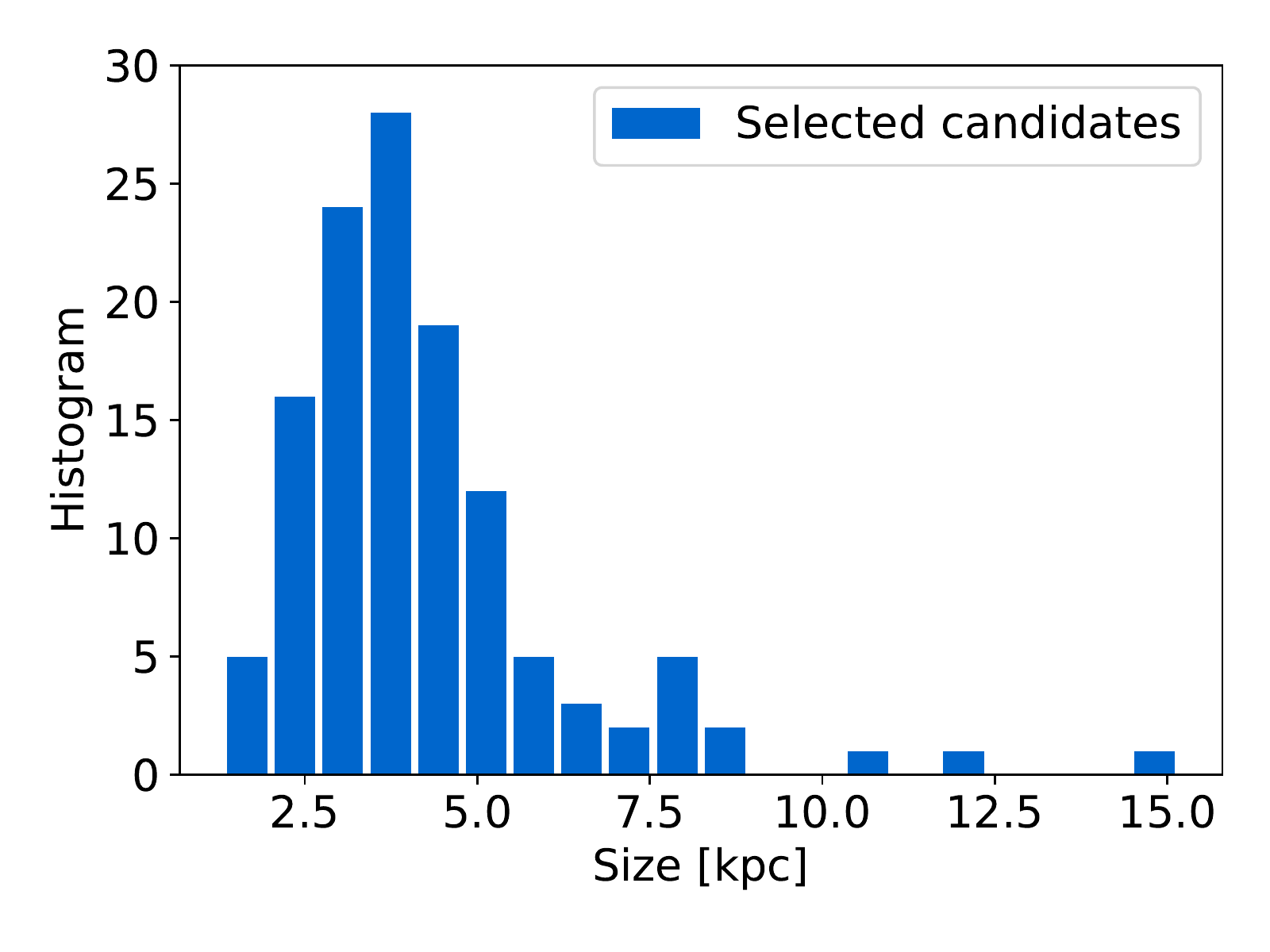}
\caption{Distribution of the apparent diameters (in arcsec; top panel) and physical diameters (in kpc; bottom panel). The apparent diameters are estimated from the circle that encompasses the totality of the emitting clump, even when the shape of the structure is not completely circular.}
\label{fig:histo_diameters_arcsec}
\end{figure}
Figure~\ref{fig:histo_diameters_arcsec}  shows the distribution of projected sizes and intrinsic sizes of the H$\alpha$ clumps. Apparent diameters are estimated from the circle that covers the full emitting clump (Fig.~\ref{fig:narrow_broad_bands}), even when the shape of the structure is not truly circular. The physical size follows from the apparent size assuming the angular size distance inferred from the redshift. %\comment{this time is the angular size distance rather than the luminosity distance. Please check}  
The average apparent diameter is $0.9\arcsec \pm 0.4\arcsec$, with the error giving the standard deviation of the distribution (Fig.~\ref{fig:histo_diameters_arcsec}, top panel). These apparent diameters correspond to physical diameters of $4.3 \pm 2.0$ kpc (Fig.~\ref{fig:histo_diameters_arcsec}, bottom panel).
%
%The sizes do not change much when considering only level 3 candidates, but they do become slightly smaller ($0.84\arcsec \pm 0.35\arcsec$).
%
The spatial resolution of \musew\ is around 1\arcsec\ \citep[see Sect.~3.1 and Table~1 in][]{2019A&A...624A.141U}, therefore, except for the larger H$\alpha$ clumps, the measured sizes seem to be set by the \muse\ spatial resolution. Consequently, most of the emitting structures are spatially unresolved. Clumps significantly smaller than the resolution (Fig.~\ref{fig:histo_diameters_arcsec}, top panel) can be explained by noise, which prevents us from detecting the full extend of the underlying H$\alpha$ emitting structure.  

%
%%%%%%%%%%%%%%%%%%%%%%%%%%
\begin{figure}
\centering
\includegraphics[width=0.9\linewidth]{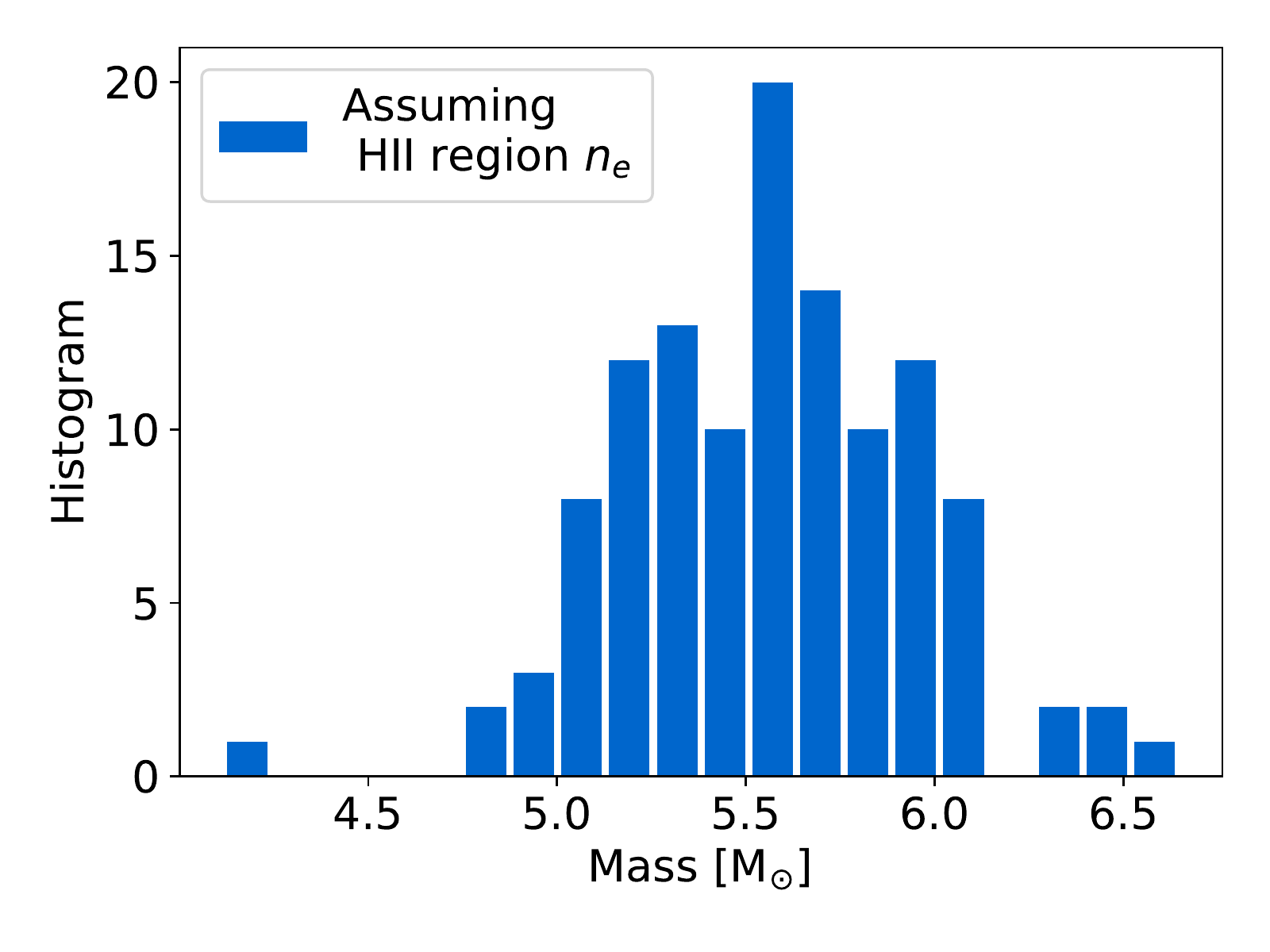}
\caption{Gas masses of the H$\alpha$ emitting clumps. The masses peak around $\rm \sim 10^{5.5}\,  M_{\odot}$, although the value depends on the assumed electron density (10\,${\rm cm}^{-3}$ in this case; see Eq.~[\ref{eq:gas_mass}]). If, rather than a value for galactic HII regions, we use a value characteristic of the CGM or IGM, all masses increase by a factor of $10^3$.
}
\label{fig:gas_mass}
\end{figure}
%
%%%%%%%%%%%%%%%%%%%%%%%%%%%%%%%
%
\subsection{Mass of the  {\rm H$\rm \alpha$} emitting gas}\label{sec:gas_mass}

We estimate the mass of the {H$\alpha$ emitting} gas following the approach described in  
\citeauthor{2017ApJ...834..181O}~(\citeyear{2017ApJ...834..181O}; see also \citeauthor{2015A&A...580A.102C}~\citeyear{2015A&A...580A.102C}). The emission is assumed to be produced by recombination of hydrogen in a fully ionized plasma (i.e., in an HII region) having the solar composition. Under these conditions, the mass of emitting gas, $M_g$, can be expressed as,
\begin{equation}
    M_{g}=3.15 \times 10^3\,{\rm M}_\odot ~t_4 \, \frac{L_{{\rm H}\alpha}/10^{38}\,{\rm erg \, s^{-1}}}{n_e / 10^2\,{\rm cm^{-3}}},
    \label{eq:gas_mass}
\end{equation}
where $t_4$ is the temperature of the gas in units of $10^4$\,K, $L_{{\rm H}\alpha}$ stands for the total H$\alpha$ luminosity (Eq.~[\ref{eq:lumha}]), and $n_e$ represents the electron density. The interested reader is directed to the work by \citet{2017ApJ...834..181O} for further details. Assuming values of $n_e$ and $t_4$  typical of H{\sc ii} regions ($10 \, {\rm cm}^{-3}$ and $t_4 = 1$, respectively; see, e.g., \citeauthor{1974agn..book.....O}~\citeyear{1974agn..book.....O}), and the luminosities derived in Sect.~\ref{sec:physical_properties}, we obtain gas masses ranging from $\rm 10^4$ to  $10^{6.5}\,{\rm M}_{\odot}$, with a mean mass of $\rm 10^{5.7} \, {\rm M}_{\odot}$. The full distribution of values is displayed in Fig.~\ref{fig:gas_mass}. If rather than $n_e$ for galactic H{\sc ii} regions we assume a value characteristic of the CGM or IGM \citep[say, $10^{-2}\,{\rm cm}^{-3}$; e.g.,][]{2014A&ARv..22...71S} 
all masses have to be increased by a factor of the order of $10^3$, becoming from $\rm 10^7$ to $10^{9.5} \, {\rm M}_{\odot}$, with mean mass of $10^{8.8} \, {\rm M}_{\odot}$.

With HII region $n_e$, the
masses are significantly smaller than the stellar mass of the central galaxy (Sect.~\ref{sec:central_galax}), with a typical mass ratio of the order of a thousandth (on average, $\log[M_g/M_\star] \simeq -3.1$). If, on the other hand, CGM -- IGM $n_e$ are assumed, then $M_g$ and $M_\star$ are comparable.

%
%%%%%%%%%%%%
%
\subsection{Azimuthal distribution of the {\rm H}$\alpha$ emission with respect to the central galaxy}\label{sec:Azimuths}

\begin{figure}
\centering
\includegraphics[width=1.1\linewidth]{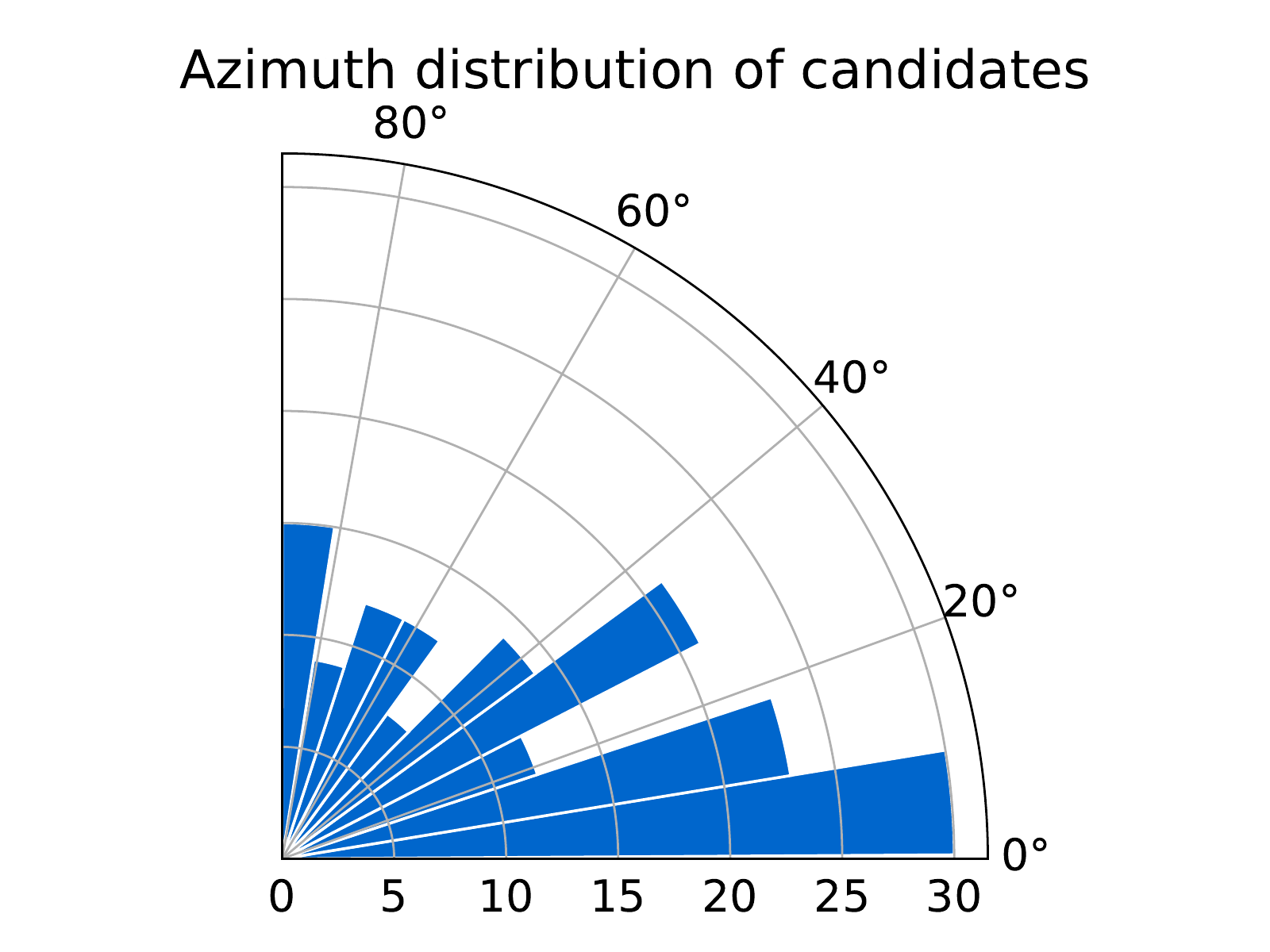}
\caption{Distribution of the relative azimuth of the candidates with respect to the major axis of the corresponding central galaxy. Angles are folded to appear in the 1st quadrant. The favoured orientation seems to match the major axis, suggesting an alignment with the galaxy plane.
%\jorge{I would take out the label on vertical axis. It is not clear the meaning and does not help to understand the histogram.}
}
\label{fig:histo_azimuths}
\end{figure}

As we explain in Sect.~\ref{sec:intro}, the gas from cosmological accretion is expected to prefer the plane of the disk of the central galaxy. On the contrary, outflows from the galaxy are channeled along the direction perpendicular to the plane of the disk. In order to investigate the existence of any preference, and to find additional arguments for the association of the emission with the central galaxy, we estimate the azimuthal distribution of the H$\alpha$ clumps with respect to the central galaxy.
The orientation of the major axis of the host galaxies is taken from the catalog by \cite{2012ApJS..203...24V}, which provides the position angle (PA) of the galaxy with respect to the celestial north.
The PA of our candidates is measured taking the galaxy as pivot and measuring the angle counter-clockwise from the north. The difference of PAs gives the relative orientation, which we summarize in Fig.~\ref{fig:histo_azimuths} as a polar-plot histogram.

The candidates prefer relative azimuths in the range of 0\,--\,30$^\circ$, although with considerable dispersion (Fig.~\ref{fig:histo_azimuths}). The H$\alpha$ emitting clumps seems to be roughly aligned with the plane of the galaxy, which  has two important implications. Firstly, the scenario where the emitting gas is pristine and coming from cosmological accretion is favored. Secondly, the clumps are not randomly distributed in azimuth. The clumps {\em know of} the existence of a central galaxy, which adds on to the argument that the observed emission lines are neither random artifacts nor created by interlopers.

In order to quantify the second argument, we carried out a KS test to evaluate the null hypothesis that the clumps are randomly distributed in the sky and so their azimuth is drawn from a uniform distribution. The resulting {\em p-}value, i.e., the probability of getting the distribution in Fig.~\ref{fig:histo_azimuths} if the signals were randomly distributed in the sky, is only 0.002. Thus we can discard the null hypothesis with 99.8\,\% confidence. Note that under the assumption that artifacts are expected to be randomly distributed in the sky, we are discarding with 99.8\,\% confidence that the signals are fake. 

%
%%%%%%%%%%%%%%%%%%%%%
%
\subsection{Distribution of distances from the  H$\alpha$ emission to the central galaxy}\label{sec:distances}
\begin{figure} %%%%%
\centering
\includegraphics[width=0.9\linewidth]{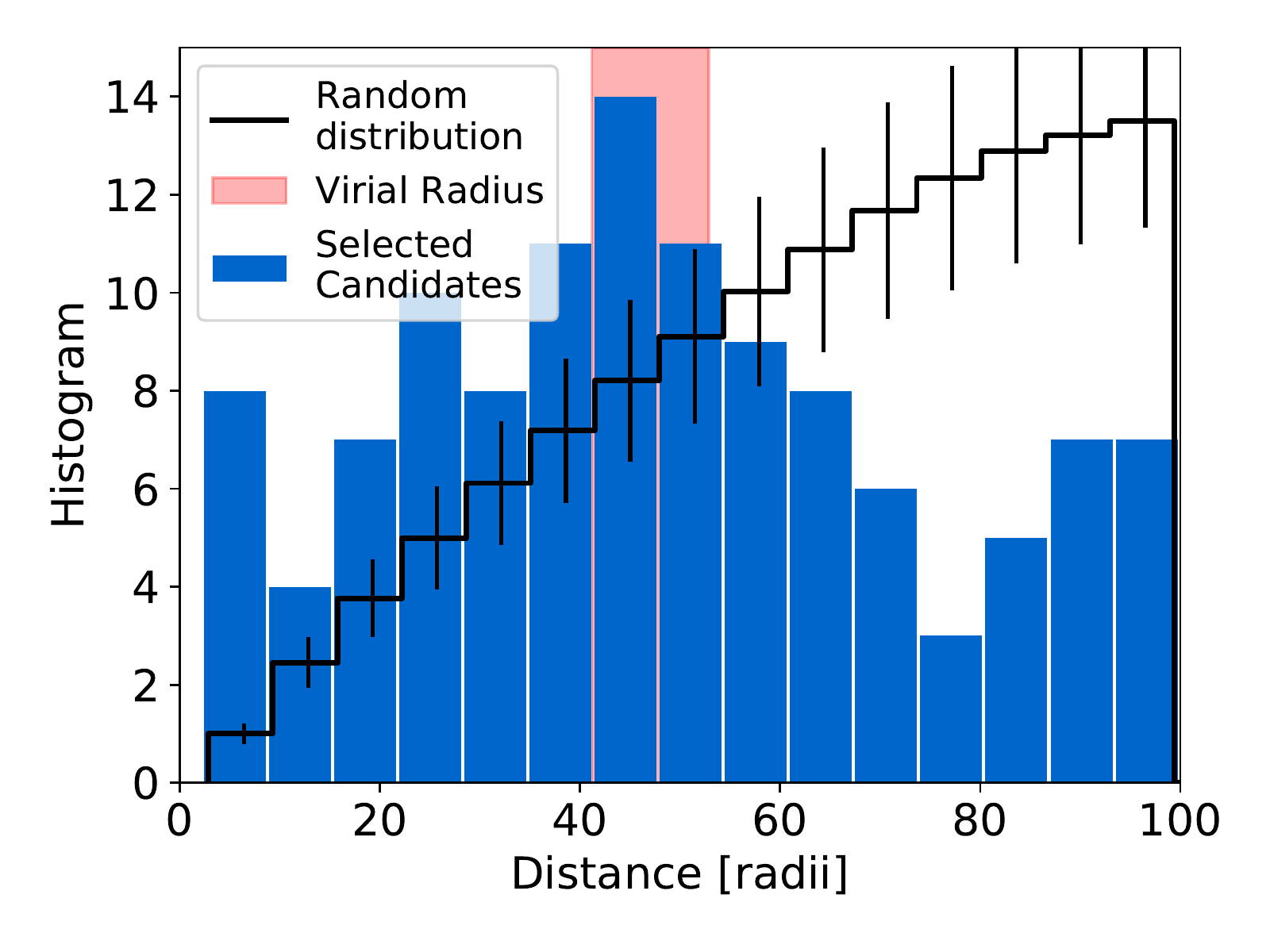}
\caption{
Distribution of distances between the H$\alpha$ emission and their central galaxy. Distances are represented in multiples of the galaxy radius adopted in the work (i.e., radius containing 80\,\%\ of the light at $\sim 1.5\,\mu{\rm m}$). Most clumps seem to be at $\rm \sim 50$ galactic radii, with a secondary peak at just under 100 radii. The black line with error bars represents the distribution expected if the emitting clumps where randomly distributed throughout the \musew\ footprint, and it has been normalized to the total number of candidates. The red band approximately indicates where the transition between CGM and IGM is expected to be.
}
\label{fig:histo_distances}
\end{figure}
The histogram in Fig.~\ref{fig:histo_distances} shows the distribution of distances from the line-emitting clumps to their respective central galaxy. In order to plot together different galaxies, distances are normalized to the radius of the galaxy (Sect.~\ref{sec:area}). The distribution has a mean and standard deviation of $\rm 53 \pm 25$ radii,  with a drop toward 100 radii, which corresponds to the radius of the searching area.
Most of our candidates seem to be located at a distance of 40\,--\,50 radii from their respective host galaxies, with the number of objects decreasing for lower and higher distances. 

The drop of the histogram at $> 50$ radii approximately coincides with the expected virial radius of the central galaxy, i.e., the boundary between the CGM and the IGM. We estimate the virial radius to be between 42 and 54 times the radius adopted for the galaxies (Sect.~\ref{sec:data})\footnote{Through abundance matching, \citet{2013ApJ...764L..31K} shows the virial radius to be approximately 70\,--\,90 times the half-mass radius of a galaxy. The typical ratio between the half-light radius and the radius containing 80\,\% of the light, used in this work, is of the order of 0.6 in our galaxies. Thus, 70 -- 90 times the half-light radius corresponds to 42\,--\,54 times our 80\,\% light radius.}. This range is indicated as a red shaded area in Fig.~\ref{fig:histo_distances}.   

The question arises as to whether the observed distribution of distances may be due to a random distribution of sources in the FOV. If the emitting clumps were randomly distributed, their number should scale with the available searching area at each distance from the source. This area, averaged over the 164 scrutinized galaxies and normalized to the total number of detected emission clumps, is shown in Fig.~\ref{fig:histo_distances} as the solid line with error bars. The error bars represent the scatter of this scaled searching area as inferred from the scatter among the areas of individual galaxies. The available area should increase linearly with distance if the searching area was a perfect disk. However, the outermost parts of the searching area often extend beyond the boundary of the \musew\ footprint, due to host galaxies being close to the edge. The effect of this can be seen in the model histogram of Fig.~\ref{fig:histo_distances} (the black solid line), which grows linearly up to approximately 50 radii, and then flattens up at larger radii.  
Since the observed distribution of distances is so different from the expected distribution of random sources, we conclude that the detected emission clumps tend to cluster around the central galaxy. There is a significant excess of sources for distances $< 50$ radii and a dearth above this distance. The divide occurs at roughly the virial radius of the central galaxy (the red stripe in Fig.~\ref{fig:histo_distances}).

As we did for the distribution of azimuths in Sect.~\ref{sec:Azimuths}, a KS test was carried out to evaluate the null hypothesis that the clumps are randomly distributed in the sky and so their true number density is described to the black line in Fig.~\ref{fig:histo_distances}. This time the resulting {\em p-}value is of the order of $10^{-10}$, fully discarding the null hypothesis. As a side-effect of these results, the fact that the observed clumps know about the existence of a central galaxy provides further evidence that they are neither artifacts nor created by randomly occurring interlopers.

\section{Cross-match with catalogs of X-ray and radio sources}\label{sec:Xrays_radio_match}

In order to further characterize the selected H$\alpha$ emitting clumps that have passed both our inspections, we cross-match their position with catalogs of X-ray and radio sources. By selection, they should not overlap with bright continuum sources (although see the follow up work mentioned at the end of Sect.~\ref{sec:conclusions}).
%\comment{Although wait for Ana Luisa's results}. 

Following \cite{2019A&A...624A.141U}, we cross-match our candidates with the Chandra CDFS-7Ms observations. The large total integration time allows this survey to reach depths in the soft (0.5--2\,keV), hard (2--7\,keV), and full (0.5--7\,keV) Chandra bands of  $ 6.4 \times 10^{-18}$, $2.7 \times 10^{-17}$,  and $\rm 1.9 \times 10^{-17}\, erg\,s^{-1} \, cm^{-2}$,  respectively. %\comment{Joao: could you please re-order the sigmas? I did but the deepest should be the full ... but it is not ...}\joao{The sensitivity is better in the soft band because the background level in that energy band is lower than in the other two.} 
We use 3 times the X-ray positional accuracy from \cite{2017ApJS..228....2L} as a limit for the uncertainty in the matching radius. %\cite{2019A&A...624A.141U} already did a match of \musew\ sources with this X-ray catalog, however, our match is not redundant because they looked for associations with galaxies whereas we are looking for counterparts to the faint emission clumps.

We find no good candidates separated from an X-ray source by less than three times the positional accuracy in \cite{2017ApJS..228....2L}.  Two level 1 candidates are at around 1\,arcsec, but these are classified as probably not real H$\alpha$ signal by our screening (Sect.~\ref{candidate_classification}). Moreover,
these X-ray sources seem to be AGNs  %\joao{ (check if previously defined. Checked. This is the first time.)}
with redshifts of 0.5 and 1 \citep{2017ApJS..228....2L}, thus outside the redshift range we targeted. 

We also stacked X-ray images corresponding to all the good candidates and those with double peak emission lines (App.~\ref{sec:xray_stacking} gives details). The procedure does not yield any detection, but it poses and upper limit for the average X-ray luminosity between $2.1 \times 10^{39}$ and  $\rm 4.6 \times 10^{39} \, erg \, s^{-1}$ for the Chandra full band ($\rm 0.5-7.0 \, keV$).
%while stacking the full final sample of 120 selected level 2 and 3 candidates.} 
%\comment{Joao: I take out the comment on X-ray binaries because this is not the place. Accretion disks are discussed in length in Sect. 6} \joao{Understood.}
%This is not enough to detect objects like high-mass x-ray binaries, which normally reach only luminosities of $\rm 10^{37-38} \, erg \, s^{-1}$, during outburst phases \citep[][]{KRETSCHMAR2019101546, 2001A&A...377..161O}}%. \comment{joao: band? galaxy set?}\comment{The limit seems to me too low. These are the luminosities of X-ray binaries. Can we detect X-ray binaries at redshift 0.3?. Please, revise.} \joao{I will check, but I was under the impression X-ray binaries were at most $10^{35}$ which would still be 2 orders of magnitude below our detection. HMXR binaries can reach $10^{37}$. Will recheck the stack but wanted to finish the application for Heidelberg today, if possible.}  }

We also cross-match our catalog of H$\alpha$ clumps  with the Very Large Array (VLA) 1.4\,GHz survey for the Extended Chandra Deep Field South \citep[E-CDFS;][]{2008ApJS..179...71K, 2013ApJS..205...13M}. The second data release goes down to an average depth of $\rm \sim 7 \mu Jy$. Following \cite{2019A&A...624A.141U}, we use a fixed 1\arcsec\ search radius for matching, as the resolution of the radio images and of  \musew\ are comparable. We find no match with any radio source. %, which could be interpreted as a sign that there is no appreciable star formation or AGN activity in the vicinity of our gas clump candidates.

%%%%%%%%%%%
%
% Table with the pros and cons of the various explanations 
\begin{deluxetable*}{lcccccccc}[h]
%\tablenum{1}
\tablecaption{Possible physical origin of the detected signals \label{tab:procons}}
\tablewidth{0pt}
\tablehead{
\colhead{Mechanism} & 
\colhead{Surface} & 
\colhead{Number} &
\colhead{Double} &
\colhead{Central} &
\colhead{H$\beta$/H$\alpha$}&
\colhead{No} &
\colhead{Size}&
\colhead{Spatial}\\[-0.2cm]% to fit space between lines
&\colhead{Brightness}&
\colhead{Density} &
\colhead{Peak} &
\colhead{Drop} &
&\colhead{Continuum} & 
&
\colhead{Distribution}\\
~~~~~~~~~~~~~~~(1)&(2)&(3)&(4)&(5)&(6)&(7)&(8)&(9)
}
%\decimalcolnumbers
\startdata
Accretion disks (IMBH)&\cmark&\cmark&\cmark&\cmark&\qmark&\cmark&\cmark&\cmark\\
Expanding bubbles (SN)&\cmark&\cmark&\cmark&\cmark&\qmark&\cmark&\cmark&\xmark\\
Cosmological gas& \cmark &\cmark&\qmark&\qmark&\xmark&\cmark&\cmark&\cmark\\
Planetary nebulae&\xmark&\cmark&\cmark&\cmark&\qmark&\cmark&\cmark&\qmark\\
X-ray binaries&\xmark&\cmark&\cmark&\cmark&\qmark&\cmark&\cmark&\qmark\\
Shocks &\qmark& \cmark &\cmark& \xmark&\qmark&\cmark&\qmark&\cmark\\ 
Galaxy outflows&\cmark& \cmark &\cmark& \xmark&\cmark&\cmark&\qmark&\xmark\\ 
Tidal disruption events&\qmark&\xmark&\cmark&\cmark&\qmark&\cmark&\cmark&\xmark\\
Interlopers&\cmark & \xmark & \qmark & \xmark&\cmark&\cmark&\qmark&\xmark\\
Jets&\qmark&\qmark&\cmark&\cmark&\qmark&\qmark&\qmark&\xmark\\
%YSOs&\xmark&\cmark&\qmark&\qmark&\cmark&\cmark&\xmark\\
\enddata
\tablecomments{The symbols \cmark , \xmark , and \qmark\ stand for {\em yes}, {\em no}, and {\em don't known}, respectively. 
(1) Potential explanation discussed in Sect.~\ref{sec:origin}.
(2) Does the physical mechanism produce surface brightness signals in the observed range?
(3) Does it predict the observed number density of objects? 
(4) Does it produce double peak spectra?
(5) Does the drop in-between peaks appear at the redshift of the central galaxy?
(6) Does it account for a flux ratio H$\beta$/H$\alpha\, < 1/6$? 
(7) Does it lack continuum emission as observed?
(8) Are the expected structures spatially unresolved?
(9) Does it reproduce the radial and azimuthal distributions? 
}
\end{deluxetable*}

%
%
%%%%%%%%%%%%%%%%%
%
\section{Possible astrophysical origin of the line emission signals}\label{sec:origin}

As we argue in previous sections, the emission line signals, even if weak and unusual, have to be of astronomical origin. One can think of various processes to produced them, but none of the ones we came up with is completely free from trouble. This section puts forward several possibilities, discusses their viability, and highlights which ones are more likely. The degree of agreement with observations of the various physical mechanisms is summarized in Table~\ref{tab:procons}, where they are ordered according to the number of observables they may be able to satisfy.

%
%%%%%%%
%
\smallskip\noindent$\bullet$ {\em Contamination from emission line background sources (interlopers).} Emission line sources in the background may leak into the band-pass used to select the H$\alpha$ emission. Ly$\alpha$ is commonly the strongest UV line and so the usual suspect to contaminate. Given the Ly$\alpha$ luminosity function at the appropriate redshift, in App.~\ref{app:a} we work out the expected number of Ly$\alpha$ emitters that may be detected in the FOV around the observed local galaxies. This number is shown to be $\lesssim$0.03 contaminants per galaxy, which is much too small to explain the observed numbers (Sect.~\ref{sec:statistics}). Other properties of the detected emission also disfavor the Ly$\alpha$ contamination. Even if Ly$\alpha$ sometimes show double peaks, this shape is exceptional and should not appear in a supposed Ly$\alpha$ average profile. Moreover, double peak Ly$\alpha$ profiles tend to have a blue lobe significantly smaller than the red lobe \citep[e.g.,][]{2021MNRAS.505.1382M}, a feature which is not present in the observed profiles (Sect.~\ref{sec:line_shapes}; Figs.~\ref{fig:Line_profiles} and \ref{fig:Stacked_profiles}). Finally, the drop in intensity between the two peaks (Figs.~\ref{fig:Line_profiles}, \ref{fig:profiles_OIII}, and \ref{fig:Stacked_profiles}) coincides with H$\alpha$ at the redshift of the central galaxy, which discards the contribution to this feature of sources unrelated to the local galaxy.  
Other line emitters (C{\sc iii}], [O{\sc ii}], and H$\beta$) are also considered and discarded in App.~\ref{app:a}. In particular, the double [O{\sc ii}]$\lambda\lambda$3727,3729 has a separation insufficient to account for the split that we observe (Sect.~\ref{sec:line_shapes}).

%Another indirect argument also disfavors emission caused by high redshift interlopers. Some of the clumps are spatially resolved (Fig.~\ref{fig:histo_diameters_arcsec}). Should the emission would be due to Ly$\alpha$ at redshift $\sim 5$ (observed wavelength around 7000\,\AA ), a 2\arcsec\ resolved structure would have to be larger than 15~kpc. This size is too large even for a massive galaxy at redshift~5 \citep[e.g.,][]{2014ApJ...788...28V}.

Finally, Sects.~\ref{sec:line_shapes}, \ref{sec:Azimuths}, and \ref{sec:distances} conclude that the observed line emission {\em knows} of the existence of the central galaxy, which would be extremely weird if the typical signals were produced by sources at redshift around 5. A casual coincidence can be discarded with high confidence.

%
%%%%%%
%
\smallskip\noindent$\bullet$ {\em Cosmological gas accretion.} The level of H$\alpha$ emission to be expected from processes associated with cosmological gas accretion is in the range from $10^{-17}$ to $10^{-21}~\cgsunits$ (App.~\ref{app:predictions}). Since this range is very wide, it does not pose a strong constraint, but the magnitude of the detected surface densities is consistent with this possibility (Fig.~\ref{fig:Gaussian_Fits_Histogram}, top panel). We do not claim this to be the origin of the signals since (a) it is unclear whether gas accretion can produce the double peak profiles that we often observe and (b) there are hints of significant dust obscuration (H$\beta$ is fainter than expected without reddening; Sect.~\ref{sec:line_shapes}), which disagrees with the belief that cosmological gas is metal poor (Sect.~\ref{sec:intro}). A precise determination of metallicities is needed to explore this issue because, at present, the upper limit we can set is too loose (Sect.~\ref{sec:line_shapes}). 

A property consistent with cosmological gas accretion is the preference for the signals to appear aligned with major axis of the central galaxy, and consequently, in the plane of the galaxy (Sect.~\ref{sec:Azimuths}). The distribution with distance from the central galaxy (approximately uniform within the virial radius and with a drop toward the IGM; Sect.~\ref{sec:distances}) is also tenable if the emission traces cosmological gas in the process of being accreted.   

Conceptually, double-peak profiles can be produced by self-absorption in the emitting gas cloud. However, H$\alpha$ self-absorption is very unexpected. It should be tracing extremely dense gas pockets, in conditions similar to those existing in the central regions of AGNs \citep[$t_4=1$, $n_e\sim 10^9\, {\rm cm}^{-1}$; e.g.,][]{1979ApJ...233L..91K,1981ApJ...243..390C}, 
so that electron collisions are able to maintain a significant fraction of the neutral H in the $n=2$ excited state \citep[e.g.,][]{1980ApJ...241L.137S}.

Shocks involving cosmological gas are to be expected when gas accreted from the IGM meets the gas pre-existing in the halo of the central galaxy (Sect.~\ref{sec:intro}). These shocks can create double-peak line profiles. However, one of the emission components should be at rest with respect to the central galaxy, a feature that we do not observe.  We discuss this issue further in the next item on {\em shocks}.
%The next item, on , discusses this issue in more detail. \joao{I would just say:  To keep the references to shocks at a minimum, in a section where we don't really address them, but this is a minor point and can be ignored.}

%
%%%%%%%%%%%%
\smallskip\noindent$\bullet$ {\em Shocks.} They are expected in the CGM, where gas accreted from the IGM encounters hot gas already existing in the CGM, or when outflows driven by stellar winds, SNe, or AGNs collide with halo gas (see Sect.~\ref{sec:intro}). Since shocks involve gas having two distinct velocities with the transition occurring through a sharp interface, they are expected to produce double peak emission lines \citep[e.g.,][]{2005MNRAS.358.1195S,2004A&A...420..423F}. Indeed, these double peaks are observed in shocks running through the atmospheres of Mira variables \citep[][]{1983A&A...128..384G} or in solar flares \citep[e.g.,][]{2015ApJ...813..125K}. The expected difference of speeds between the CGM and the infalling gas can be pretty large, of the order of several hundred km\,s$^{-1}$ \citep[e.g.,][]{2020MNRAS.499..597B}, which is fully consistent with the line splitting we detect (Sect.~\ref{sec:line_shapes}). The spatial distribution to be expected is uncertain but if shocks occur where gas from the IGM meets CGM gas, then they should be concentrated toward the plane of the disk of the central galaxy, as observed.  

So far shocks seem to be consistent with all observed features, however, there is an additional property difficulty to reconcile with observations: one of the two peaks of the H$\alpha$ profile is created by gas at rest with respect to the central galaxy, however,  the two observed peaks are generally split with respect to the velocity set by the central galaxy.  

\smallskip\noindent$\bullet$ {\em Accretion disks around compact objects.} 
Compact low-luminous massive objects are expected to lurk the CGM of galaxies. From lone low-mass stars \citep[e.g.,][]{2020ARA&A..58..205H}, to BHs arising from stellar evolution \citep[e.g.,][]{2013MNRAS.430.1538F}, including remnants of Pop\,{\sc iii} stars \citep[e.g.,][]{2001ApJ...551L..27M,2018MNRAS.478.2541F}, and primordial BHs \citep[e.g.,][]{1974MNRAS.168..399C,2015PhRvD..92b3524C}. The number density of these objects is assumed to be high; for example, there are claims that primordial BHs account for all the dark matter in the Universe \citep[][]{2015PhRvD..92b3524C}. Even if this is not the case \citep[e.g.,][]{2020ARNPS..70..355C}, the expectations clearly overwhelm the density required to account for all the emission signals that we observe, of the order of one clump per central galaxy (Sect.~\ref{sec:statistics}). When these compact objects are surrounded by accretion disks, they should emit in H$\alpha$, with the rotation of the disk giving rise to two-horn H$\alpha$ profiles when observed with the appropriate viewing angle \citep{1969AcA....19..155S}. This kind of double peak H$\alpha$ emission is observed in X-ray binaries  \citep[][]{2007ApJ...660.1398G,2013A&A...559A..87Z,2018MNRAS.481.4372C,2020MNRAS.496.3615M} and cataclysmic variables \citep[e.g.,][]{2011A&A...526A..84Z}, with peak separations of up to hundreds of km\,s$^{-1}$ \citep[][]{2013A&A...559A..87Z}.
At a completely different mass scale,  double peak H$\alpha$ emission is sometimes observed in the broad line region of AGN, where it is also supposed to trace a rotating disk \citep[e.g.,][]{1994ApJS...90....1E,2008ARA&A..46..475H}.

The double peak H$\alpha$ emission observed around stellar-mass compact objects comes together with continuum emission \citep[e.g.,][]{2010ApJ...724..379M,2019A&A...622A.173Z}. There are hints of continuum emission in our stacks of observed spectra (Fig.~\ref{fig:Stacked_profiles}), but the question arises at to whether this very faint continuum is or not consistent with an accretion disk around a compact object. A number of arguments show that the observed lack of continuum emission does not discard the accretion disk scenario. Firstly, large H$\alpha$ equivalent widths (EWs) are sometimes observed in X-ray binaries.  H$\alpha$ EWs larger than 100\,\AA\ are not rare during quiescence \citep[][]{2009MNRAS.393.1608F,2018MNRAS.481.4372C}, with maxima reaching 2000\,\AA\ \citep{2016Natur.534...75M}, which implies line to continum flux ratios enough to account for our observations (Fig.~\ref{fig:Stacked_profiles}). %Note that double peak H$\alpha$ profiles are particularly prominent during quiescence. %
Secondly, H$\alpha$ photons are to be produced by recombination of H atoms photo-ionized by the accretion disk \citep[e.g.,][]{2015MNRAS.450.3331M}. Under particular circumstances, the nebular emission produced by a photo-ionizing source can emit an H$\alpha$ line with very little underlying continuum. For example, young starbursts produce H$\alpha$ with EW in excess of $10^3$\,\AA\ \citep{1999ApJS..123....3L}, and galaxy-integrated spectra with EW of hundreds of \AA\ are not uncommon \citep[e.g.,][]{2011ApJ...743...77M}. The main physical ingredient for the photo-ionization source to produce little continuum is presenting a hard spectrum with an ionization flux greatly exceeding the flux in the optical. Accretion disks can easily match this requirement since they can be extremely hot with spectra peaking in the far UV and X-ray.    

The small size of the observed sources, often spatially unresolved (Fig.~\ref{fig:histo_diameters_arcsec}), also fit in the accretion disk scenario. As for the spatial distribution (Figs.~\ref{fig:histo_azimuths} and \ref{fig:histo_distances}), it can be seamlessly accommodated within this explanation as well. BHs are expected to be present everywhere in the galaxy halo, however, they need gas to build the accretion disk. Getting gas around a lone BH is certainly very unlikely, but the chances are higher where the gas concentrates, i.e., near the galaxy and close to its disk. 

The H$\alpha$ luminosity of accretion disks around stellar-mass compact objects is observed to be between $10^{32}$ and $10^{35}$~erg\,s$^{-1}$. This estimate has been taken from the prototypical low-mass X-ray binary V404 Cygni \citep[][]{2018MNRAS.481.2646M}, with the range embracing from the quiescent to the active phases. Even during outbursts, when the brightness is largest, the characteristic luminosities are orders of magnitude smaller than the ones we observe (Fig.~\ref{fig:luminosity}). Therefore, stellar-mass compact objects, such as the ones resulting from stellar evolution described above, cannot be responsible for the observed H$\alpha$ emission. However, the bolometric luminosity of an accretion disk scales with the mass accretion rate, which itself scales with the mass of the central object  \citep[e.g.,][]{2006jebh.book....1F,2010arXiv1005.5279S,2014SSRv..183..163K}. If the bolometric luminosity and the H$\alpha$ luminosity scale with each other (a reasonable  ansatz), our signals could be produced by compact objects with mass between $10^3$ and $10^6\,{\rm M}_\odot$, i.e., in the realm of the IMBHs (Intermediate-mass BHs).\footnote{We note that the mass of emitting gas estimated in Sect.~\ref{sec:gas_mass}, betweem $10^{3}$ and $10^{6}$~M$_\odot$, also discards stellar-mass BHs and seems to be consistent with the IMBH interpretation.}
Rogue IMBHs in galaxies were theoretically predicted by \citet{2001ApJ...551L..27M} to result from direct gravitational collapse during early phases of the Universe. These objects keep growing through mergers as part of the hierarchical formation of galaxies. Even though they are generated at large mass over-densities, i.e., at the center of proto-galaxies, gravitational recoil during BH mergers can kick them out to appear in the outskirts of galaxies, far from their cradle \citep[][and references therein]{2019BAAS...51c.175B}. Numerical simulations \citep[e.g.,][]{2009MNRAS.395..781O,2011MNRAS.414.1127M,2014ApJ...780..187R} predict that tens to several thousands of IMBHs should exist in a galaxy like the MW. IMBHs can also be formed by mergers and associations of primordial BHs, a channel that outnumbers the production via  direct gravitational collapse \citep[e.g.,][]{2019arXiv190608217C,2020ARNPS..70..355C}.

Accretion disks around stellar-mass compact objects emit most of the radiation in X-rays. The luminosity of X-ray binaries spans from $10^{31}$ to $10^{34}$~erg\,s$^{-1}$ during quiescence and from $10^{35}$ to $10^{39}$~erg\,s$^{-1}$ during outbursts \citep[e.g.,][]{2009MNRAS.393.1608F,2018MNRAS.481.4372C}. As we explain above, the H$\alpha$ EW is largest during quiescence, therefore, for the accretion disk scenario to explain our H$\alpha$ luminosities the boosting factor has to be around $10^6$. A naive scaling of the X-ray luminosities during quiescence with this factor implies that the IMBH accretion disks should have an X-ray luminosity $<10^{40}$~erg\,s$^{-1}$, which renders a flux density at the typical redshift of our sources ($\sim 0.3$; Fig.~\ref{fig:histo_redshifts}) of  $<3\times 10^{-17}$~erg\,s$^{-1}$\,cm$^{-2}$. This luminosity and flux density are upper limits, not only because they correspond to up-scaling the largest observed X-ray luminosities  but because an increase of the central object mass drops the temperature of the accretion disk\footnote{The inner radius of the disk grows with increasing central mass thus dropping the disk temperature. This drop is quite large and cannot be compensated by the increase of temperate arising from the growth of the central mass; see, e.g., \citet[][]{2010arXiv1005.5279S}.}, which moves the bulk emission from X-rays to the far UV. An additional comment is in order: some of the IMBH candidates in literature correspond to X-ray sources with luminosities larger than $10^{40}$~erg\,s$^{-1}$ \citep{2017IJMPD..2630021M}, thus, larger than the upper limit used above. This, once again, is consistent with the scenario. These large X-ray luminosities may be the IMBH equivalent to the stellar-mass BHs emission during extreme outbursts.

We finally note that the X-ray luminosities and fluxes predicted above are close to but below the detection threshold of the current X-ray surveys explored in Sect.~\ref{sec:Xrays_radio_match}.

\smallskip\noindent$\bullet$ {\em Outflows and galactic fountains.} 
The process of star formation and the AGN activity expel gas from galaxy disks. This ejected gas ends up in the CGM and, depending on the depth of the gravitational well, also in the IGM. Part of this material rains back to the galaxy in the so-called {\em galactic fountains} \citep[e.g.,][]{2014IAUS..298..228F}. In principle, the outflows can be distinguished from cosmological accretion because they are not metal-poor and because they are channeled by the galaxy disk to break out in the direction perpendicular to the plane of the disk \citep[e.g.,][]{2014A&ARv..22...71S}. The existence of outflows with these properties has been inferred in various ways, for instance, as metal-rich absorption on background sources \citep[e.g.,][]{2006MNRAS.372..369P,2011MNRAS.413.2481F}, or as a bimodality in the azimuthal angle and metallicity of the absorbing gas around galaxies \citep[e.g.,][]{2012ApJ...760L...7K,2013ApJ...770..138L}.

Galaxy driven outflows are able to account for the number and the surface brightness of our H$\alpha$ emissions. However, three other properties of the observed clumps are not well reproduced within this scenario. Firstly, the observed clumps are concentrated toward the plane of the galaxy (Fig.~\ref{fig:histo_azimuths}) rather than where the outflows should be, in the direction perpendicular to this plane. Secondly, within the virial radius, the observed clumps are not centrally concentrated but spread out. Thirdly, outflows are often supersonic \citep[e.g.,][]{2017ApJ...834..181O} and can collide with the gas in the CGM, thus creating a shock front that emits double-peak line profiles. However, contrarily to observation, one of the two peaks has to be at rest with respect to the central galaxy. Double peak line emission are indeed observed in outflows near the disks of luminous IR galaxies, with one of the two components at the systemic velocity of the galaxy \citep{1990ApJS...74..833H}.

Outflows and the subsequent galactic fountains involve metal rich gas, where the existence of dust grains is to be expected. Thus, the observed large H$\alpha$ flux compared with H$\beta$ could, in principle, be explained as an effect of dust reddening in outflows from galaxies (see Sect.~\ref{sec:line_shapes}).

%
%%%%%%%%%%%%%%%%%%%%%%%%%%%
\smallskip\noindent$\bullet$ {\em Expanding Bubbles.} 
Expanding gas shells give rise to top-hat line shapes \citep[e.g.,][]{1996ApJ...456..264T}, which do not have a central dip. Such line shape is at variance with the double peaks that we often observe, however, if the shell is dusty then photons at the center of the line are preferentially absorbed within the shell, resulting in two-horn profiles \citep[e.g., appendix in][]{2017ApJ...834..181O}. In this case, the blue lobe tends to be largest. 

This type of shell producing double peak lines could be associated with planetary nebulae (PNe), i.e.,  the ionized circumstellar gas ejected during the late phases of evolution of solar-mass stars. It produces emission lines with little continuum since the ionization is provided by a very hot central star. However, the luminosity of one of these shells is not sufficient to account to the typical $3\times 10^{38}$\,erg\,s$^{-1}$ H$\alpha$ luminosity that we observe \citep[PN luminosities span from $10^{34}$ to $5\times 10^{35}$\,erg\,s$^{-1}$; ][]{2010PASA...27..149C}. 
In addition, PNe emit more in [O{\sc iii}]$\lambda$5007 than in H$\alpha$, a feature we do not see (Fig.~\ref{fig:Stacked_profiles}). %0909.4356.pdf  
Thus, PNe can be discarded as sources of the observed signals. An alternative shortcut to reach the same conclusion is noting that the mass of emitting gas (Fig.~\ref{fig:gas_mass}), all uncertainties notwithstanding, is much larger than the envelope that could be ejected from a low-mass star.  

Expanding bubbles driven by SN explosions do not have these drawbacks. SN remnants tend to have a H$\alpha$ flux significantly larger than [O{\sc iii}]$\lambda$5007 \citep[e.g.,][]{2018RNAAS...2...32D}. The H$\alpha$ luminosity at the explosion is much larger than the ones we observe, however, $\sim 10^3$ days later from the outset, the luminosity drops to the observed values \citep[e.g.,][]{2016MNRAS.456.3296B}. Thus, the luminosity itself is not an issue, but the fact that we do not observe SN remnants at their brightest point questions the whole scenario. Why do we observe SNe only after a particular time lag from the explosion?
Double peak H$\alpha$ profiles are common in SNe and they present a continuum consistent with observations \citep[e.g.,][]{2016MNRAS.456.3296B}. On the other hand, SN remnants are product of the stellar evolution and so should mimic the spatial distribution of the stars in the central galaxy. This fundamental characteristic of SNe is in disagreement with the spatial distribution of observed H$\alpha$ signals which, in addition to be spread out over large areas, they also appear in the CGM where the formation of the stars that go off SN is unlikely (Fig.~\ref{fig:histo_distances}).   

\smallskip\noindent$\bullet$ {\em Tidal disruption events (TDEs).} 
When a star passes sufficiently close to a super-massive BH (SMBH), it can be stretch out to disruption by tidal forces \citep[e.g.,][]{1982Natur.296..211C}. A fraction of the original mass is captured to form a transient accretion disk around the BH. The emission of the system resembles the one from accretion disks discussed above, in particular, double peak H$\alpha$ lines are produced \citep[][]{2020MNRAS.498.4119S, 2020ApJ...903...31H} with a maximum luminosity significantly larger than the ones that we observe ($>10^{41} \rm \, erg \, s^{-1}$; \citeauthor{2019ApJ...883..111H}~\citeyear{2019ApJ...883..111H}, \citeauthor{2020MNRAS.498.4119S}~\citeyear{2020MNRAS.498.4119S}). The luminosity and spectrum are variable so it is unclear whether the double peak line shape and the right H$\alpha$ luminosity occur simultaneously during the transient. Similarly, it is unclear whether a scaled down version of the observed TDEs, with smaller central BH masses, can pull apart a star and give rise to an accretion disk with the characteristics we observe.
The main handicap for the TDEs to explain our observations is the fact that the two required ingredients, a star plus a SMBH, cluster toward the center of the gravitational potential. Thus an encounter is far more likely to occur at the center of the galaxies, which is at variance with the spatial distribution of the H$\alpha$ signals (Fig.~\ref{fig:histo_distances}). In addition, it is also a short-lived event ($\lesssim 1$\,yr), therefore,  hard to reconcile with the relatively high number of the observed signals. 

%
%%%%%%%%%%
\smallskip\noindent$\bullet$ {\em Jets.} Astrophysical jets are quite common and they involve bipolar outflows which, once  spatially averaged, can give rise to double peak line profiles like the ones that we often detect.  They originate in accretion disks around compact central objects such as BHs of all masses, neutron stars, X-ray binaries, T Tauri stars, Herbig–Haro objects, or evolved post-AGB stars. Jets around stellar mass objects can be discarded with the argument that the H$\alpha$ emitting gas seems to involve between $10^3$ to $10^6\,{\rm M_\odot}$ (Fig.~\ref{fig:gas_mass}). This narrows down the possibilities to IMBHs, and many of the restrictions invoked above apply. In addition, the unresolved character of many of the detected sources (Fig.~\ref{fig:histo_diameters_arcsec}) may be at variance with the large scales expected in jets around massive BHs. Jets can carry away a significant part of the energy released during the accretion process and so, from the point of view of the energetics, they are similar to accretion disks \citep[e.g.,][]{2021AN....342..727R}.

%\tableofcontents

%%%%%%%%%
\section{Conclusions}\label{sec:conclusions}
Aiming at detecting cosmological gas being accreted onto galaxies of the local Universe (Sect.~\ref{sec:intro}), we examined the outskirts of all the 164 galaxies in the \musew\ footprint with H$\alpha$ within the bandpass of \muse\ (redshift $< 0.42$). After a first visual inspection of H$\alpha$ maps around all galaxies, an exhaustive screening of the putative faint emission line signals led us to identify what seems to be 118 H$\alpha$ emitting gas clouds (Sects.~\ref{candidate_classification} and \ref{sec:statistics}).
The signals are very faint, with a typical surface brightness of ${\rm 10^{-17.2\pm 0.2}\,erg\,s^{-1}\,cm^{-2}\,arcsec^{-2}}$ and a typical flux of ${\rm 10^{-17.3\pm 0.3}\,erg\,s^{-1}\,cm^{-2}}$ (Fig.~\ref{fig:Gaussian_Fits_Histogram}). 
We discarded that they are created by instrumental artifacts or telluric line residuals (Sects.~\ref{skylines}, \ref{sec:line_shapes}, \ref{sec:Azimuths}, and \ref{sec:distances}, and App.~\ref{app:b}). Neither are they high redshift interlopers (Sect.~\ref{sec:origin}, and App.~\ref{app:a}).
KS tests on the spatial and spectral distribution of the signals discard with high confidence ($> 97$\,\%)  that the signals are fake.

Assuming that the emission line signals are produced by gas clumps around the central galaxies, they have the following properties:

\begin{itemize}
\item [-] The H$\alpha$ line profile often shows a double peak with the drop in intensity at the rest-frame of the central galaxy (Figs.~\ref{fig:Line_profiles}, \ref{fig:profiles_OIII}, and \ref{fig:Stacked_profiles}). The typical peak-to-peak velocity is of the order of $\pm 200$\,km\,s$^{-1}$.
  The double peak line profiles make up almost 38\,\%\ of all candidates (Sect.~\ref{sec:line_shapes}). Some of the single peak profiles may be double peaks with one of the two peaks missing (Fig.~\ref{fig:separate_stacks}).

\item[-] Only a few of the H$\alpha$ emitting clumps also show  H$\beta$ (Fig.~\ref{fig:profiles_Hb}) and/or [O{\sc iii}]$\lambda$5007 (Fig.~\ref{fig:profiles_OIII}). These emission lines are virtually absent in the average stacked spectra (Fig.~\ref{fig:Stacked_profiles}), so that their fluxes have to be smaller than around 1/6 of the flux in H$\alpha$ (Sect.~\ref{sec:line_shapes}).  
\item[-] Using the non-detection of [N{\sc ii}]$\lambda$6583, we set a loose upper limit on the metallicity of the emitting gas, ${\rm 12+\log(O/H) < 8.5}$, which roughly corresponds to 60\,\% of the solar metallicity (Sect.~\ref{sec:line_shapes}).

\item[-] Most of the detected line emission clumps are at the resolution limit of \musew , around 1\arcsec. We measure diameters of ${\rm 0.9\pm 0.4\,arcsec}$ (Fig.~\ref{fig:Gaussian_Fits_Histogram}, top panel), which at the redshift of the central galaxies ($0.31\pm 0.07$; Fig.~\ref{fig:histo_redshifts}) correspond to ${\rm 4.3\pm 2.0\, kpc}$ (Fig.~\ref{fig:Gaussian_Fits_Histogram}, bottom panel).
\item[-] Using the distance of the central galaxy, the inferred H$\alpha$ luminosities are around $\rm 10^{39} \, erg \, s^{-1}$ and they seldom exceed $\rm 10^{40} \, erg \, s^{-1}$ (Fig.~\ref{fig:luminosity}).  
\item[-] The mass of emitting gas ($M_g$) has been estimated under the assumption that the emission is produced by H recombination in a fully ionized medium. With a temperature of ${\rm 10^4\,K}$ and $n_e={\rm 10\,cm^{-3}}$, $M_g$ goes from $10^4$ to ${\rm 10^{6.5}\,M_\odot}$ (Fig.~\ref{fig:gas_mass}). These masses are significantly smaller than the stellar mass of the central galaxy ($M_\star$; Sect.~\ref{sec:central_galax}), with an average  mass ratio of the order of $\log(M_g/M_\star) \simeq -3.1$. $M_g$ scales as $n_e^{-1}$ so for densities closer to those of the CGM and IGM, $M_g/M_\star\sim 1$.
\item[-] The signals are not isotropically distributed with respect to the central galaxy. The azimuth of the signals tend to be aligned with the major axis of the corresponding central galaxy (Fig.~\ref{fig:histo_azimuths}).
\item[-] The distances from the emitting clumps to the central galaxies are not randomly distributed (Sect.~\ref{sec:distances}). There is an excess of sources for distances $< 50$ galaxy radii and a dearth above this value (Fig.~\ref{fig:histo_distances}). The critical distance roughly corresponds to the virial radius of the central galaxies, which defines the transition between CGM and IGM. 
\item[-] We find no match between the position of the H$\alpha$ clumps and catalogues of X-ray sources \citep[][]{2017ApJS..228....2L} or radio sources \citep[][]{2008ApJS..179...71K, 2013ApJS..205...13M}; see Sect.~\ref{sec:Xrays_radio_match}. 
\end{itemize}

We explore several physical mechanisms to explain these observations including the cosmological gas accretion that motivated the study (Sect.~\ref{sec:origin}).
%
% we explore various physical mechanisms that may potentially explain the properties listed above. None of possibilities is fully satisfactory, including the cosmological gas accretion that motivated our work. 
%\comment{This may be changed depending on the way the agremment with IMBHs will be treated. See what happens tomorrow.}
The degree of agreement with observations provided by each one of them is summarized in Table~\ref{tab:procons}. According to number of observables that they satisfy, the main possibilities are:
\begin{itemize}
\item[-] Accretions disks around rogue IMBHs. They may provide the observed surface brightness, number density, double peak profiles centered at the redshift of the galaxy, lack of continuum emission, unresolved sizes, and correct spatial distribution. IMBHs may also be able to reproduce the observed small H$\beta$/H$\alpha$ flux ratio.
\item[-] Expanding bubbles caused by SN explosions. They account for all the observables mentioned in the previous item except for the spatial distribution (SN trace stellar sites, whereas the emission is spread over much larger distances to reach out the IGM). It also remains unclear whether they can account for H$\beta$/H$\alpha$.
\item[-] The cosmological gas accretion naturally accounts for the spatial distribution of the emitting clumps. They can also cope with their typical luminosities and sizes. However, it has difficulty to explain the double peak in H$\alpha$. In addition, the problem to explain H$\beta$/H$\alpha$ is particularly severe in this scenario since the cosmological gas is metal-poor, and so, unable to produce significant dust reddening. 
\item[-] Shocks driven either by cosmological gas accretion or by galaxy winds. In this case, the main obstacle is explaining double peak H$\alpha$ profiles with the intensity drop at the rest-frame of the central galaxy. Shocks with CGM gas can create double peak line profiles, but with one of the peaks at the systemic velocity. In addition, the spatial distribution is at odds with galaxy driven winds, while shocks involving metal-poor cosmological gas has difficulty to account for H$\beta$/H$\alpha$. 
\end{itemize}
Thus, from all the possibilities, accretion disks around rogue IMBHs satisfy the observational constraints best.

Note that the key discriminant among the posed physical scenarios is the need to reproduce the double peak emission centered at the redshift of the galaxy. Even if this feature is quite common (38\,\% or more; Sect.~\ref{sec:line_shapes}), most of the observed spectra present a single peak and so are not subject to this constraint. Thus, they can be produced by mechanisms other than the IMBHs, including the cosmological gas accretion we were after (Sect.~\ref{sec:intro}). The cause of the detected emission could be assorted, and follow up studies will be required to assign them individually.

%To sum up, we cannot offer a clear explanation for the physical origin of the signals yet. \comment{jorge: rephrase if needed} The cause could be \jorge{a combination} of several of the possibilities invoked above. Follow up studies will allow us to further probe these sources and also to extend the search to additional fields other than \musew.

%A final comment is in order.

The criteria we have employed to select H$\alpha$ clumps are rather fuzzy, however, this fact does not undermine our study, which is confessedly exploratory. It was aimed at detecting so far unknown faint H$\alpha$ emission around galaxies. Its detection, presented in this paper, will allow us to characterize its properties so that future searches can be automated, and so, based on less ambiguous criteria.
As an additional followup, we are in the process of cross-matching the positions of the H$\alpha$ emissions with existing ancillary data,  in particular, with deep Hubble Space Telescope multi-band images of the region \citep[][]{2019ApJS..244...16W}. Thus, several faint optical continuum counterparts of the emitting clumps have been found and their physical properties will be reported soon (Gonz\'alez-Mor\'an et al. 2022, in preparation).

\begin{acknowledgments}
Thanks are due to various colleagues that helped us along the way:
Ana Monreal Ibero, for discussions and references on may different aspects of the work.
Teo Mu\~noz Darias, for discussion on the interpretation of the double peak H$\alpha$ signals based on stellar mass accretion disks. He also pointed out the possibility of TDEs.
Mar Mezcua and Lucio Mayer, for discussions and references on the IMBH hypothesis.
Peter Weilbacher, for clarifications on how the removal of telluric lines is carried out, both using the standard approach and ZAP.  
Antonio Cabrera, for discussion and references on the size and timescales of the emitting patches if telluric.
Ignacio Trujillo, for various specific suggestions on cross matching with external catalogs and on the parameters to be elucidated.
Tanya Urrutia, for clarifications regarding the \muse\ data-cubes and their manipulation.
Graham Berriman, for help with Montage.
We are particularly thankful to an anonymous referee for helping us quantify the reliability of the detections.
We acknowledge support from the Spanish Ministry of Science and Innovation, project  PID2019-107408GB-C43 (ESTALLIDOS), and from Gobierno de Canarias through EU FEDER funding, project PID2020010050.
This research made use of Montage, which is funded by the National Science Foundation under Grant Number ACI-1440620, and was previously funded by the National Aeronautics and Space Administration's Earth Science Technology Office, Computation Technologies Project, under Cooperative Agreement Number NCC5-626 between NASA and the California Institute of Technology.

\end{acknowledgments}


\vspace{5mm}
\facilities{MUSE@VLT \citep{2014Msngr.157...13B}} %,  HST(STIS), Swift(XRT and UVOT), AAVSO, CTIO:1.3m,


\software{AstroPy \citep{2013A&A...558A..33A,2018AJ....156..123A}, NumPy \citep{harris2020array}, Montage \citep{2010arXiv1005.4454J}} %  
          %Cloudy \citep{2013RMxAA..49..137F}, 
          %Source Extractor \citep{1996A&AS..117..393B}

%%%%%%%%%%%%%%%%%%%%%%%%%%%%%%%%%%%%%%%%%%
\appendix
%%%%%%%%
\section{Expected H$\alpha$ signals}\label{app:predictions}

This appendix works out the H$\alpha$ emission to be expected in the CGM and IGM around nearby galaxies. The estimates are very uncertain, but they all converge to provide a level of emission that includes the surface brightness of the signals detected in this work. Table~\ref{tab:summary} summarizes the expected signals under the various circumstances and approximations explained in the next paragraphs.
%%%%%
% Table with the signals to be expected.  
\begin{deluxetable}{lccc}[h]
%\tablenum{1}
\tablecaption{H$\alpha$ signal to be expected in the local Universe. \label{tab:summary}}
\tablewidth{0pt}
\tablehead{
\colhead{Based on} & 
\colhead{Surface brightness} & 
\colhead{Region} &
\colhead{Reference}\\
~~~~~~~~~~~~~~~(1)&(2)&(3)&(4)
}
%\decimalcolnumbers
\startdata
Scaled Ly$\alpha$ observations &  1 -- 30 & IGM & \citet{2014Natur.506...63C}\\
Scaled H$\alpha$ observations &  10 & IGM & \citet{2018MNRAS.480.2094L}\\
Fluorescence&0.01 -- 10 & IGM &\citet{2010ApJ...708.1048K}\\
Fluorescence&0.03 -- 0.1&IGM&\citet{2005ApJ...628...61C}\\
CIB\tablenotemark{$a$} Fluorescence& 0.001 -- 0.1 & CGM&\citet{2017ApJ...849...51B}\\
Mechanical Feedback&$\sim 1$&CGM&\citet{2004ApJ...613L..97M}\\
Gravity driven &0.09 -- 5& CGM& \citet{2010MNRAS.407..613G}\\ 
Gravity driven &0.001 -- 2 & CGM & \citet{2010ApJ...725..633F}\\
Gravity driven $z=0$ &0.002 -- 20 & IGM &\citet{2003ApJ...599L...1F}\\
Observed $z=0$&$\geq 0.5$ &CGM&\citet{2016ApJ...832..182C,Herrenz17}\\
Ly$\alpha$ blobs at $z=3$ &1 -- 30 &CGM &\citet{2004AJ....128..569M}\\
\enddata
\tablecomments{
(1) Origin of the empirical data or physical mechanism used to estimate the H$\alpha$ signal (details in App.~\ref{app:predictions}).
(2) In units of $10^{-18}~\cgsunits$.
(3) Whether the signals correspond to the CGM or the IGM.
(4) Main reference used for the estimate.
}
\tablenotetext{a}{Cosmic Ionizing Background.}
\end{deluxetable}

%\smallskip\noindent$\bullet$ {\em Re-scaling of Ly$\alpha$ signals observed at high redshift.}\label{sec:re-scaling}
As we put forward in Sect.~\ref{sec:intro}, diffuse Ly$\alpha$ emission around galaxies is routinely detected at high $z$. Most Ly$\alpha$ emission mechanisms ultimately rely on the recombination of electrons and protons, which grants the production of H$\alpha$ photons together with the Ly$\alpha$ photons. Detecting these H$\alpha$ photons rather than the Ly$\alpha$ photons has pros and cons. Among the cons, the flux ratio between Ly$\alpha$ and H$\alpha$ ($\epsilon$) is expected to be around 8.7 if the lines are produced by recombination \citep[e.g.,][]{1974ARA&A..12..331M}. It is observed to be between  0.5 and 15 \citep{1977MNRAS.178P..67B,1992ApJ...391..608H,2009A&A...506L...1A,2010Natur.464..562H,2018MNRAS.480.2094L}. The emissivity of the lines is expected to increase with increasing density \citep[e.g.,][]{1974agn..book.....O}, therefore, another con has to do with the evolution of the cosmic web gas that decreases in density ($\rho_g$) with decreasing redshift ($z$) following the global expansion of the universe \citep[e.g.,][]{2012MNRAS.423.2991V}, i.e,
\begin{equation}
\rho_g  \propto (1+z)^3.
\end{equation}
Among the pros, the faint Ly$\alpha$ signals have been detected around high redshift galaxies whereas H$\alpha$ can be observed in local galaxies that are closer and thus less affected by cosmological dimming. From redshift 0 to $z$, there is a gain in surface brightness that scales as $(1+z)^4$  \citep[e.g.,][]{2010MNRAS.407..613G,1999astro.ph..5116H}. A final advantage of H$\alpha$ is the fact that Ly$\alpha$ photons are more easily destroyed than H$\alpha$ photons 
(e.g., by dust absorption or collisional de-excitation) and so $\epsilon$ tends to be reduced with respect to the theoretical recombination value. For the sake of having a rough estimate of the expected signal, we will assume that the H emission roughly drops with the mean gas density of the Universe. Then the ratio between the Ly$\alpha$ surface brightness at redshift $z$ ($S\!B_{\rm Ly\alpha}[z]$) and the H$\alpha$ surface brightness if the same structure is observed in the local Universe ($S\!B_{\rm H\alpha}[0]$) would be,
\begin{equation}
S\!B_{\rm H\alpha}(0)= \frac{(1+z)^4}{\epsilon(1+z)^3}S\!B_{\rm Ly\alpha}(z)=\frac{1+z}{\epsilon}S\!B_{\rm Ly\alpha}(z),
\label{thirdeq}
\end{equation}
which yields
\begin{equation}
S\!B_{\rm H\alpha}(0) \sim S\!B_{\rm Ly\alpha}(z),
\label{eq:equal}
\end{equation}
provided that $z$ is in the range between 2 and 4 and $\epsilon$ remains around 5, which is reasonable considering the range of redshifts where extended Ly$\alpha$ emission has been detected (Sect.~\ref{sec:intro} and the references below) and the observed values for $\epsilon$. 
To sum up, the two fluxes are expected to be similar (Eq.~[\ref{eq:equal}]), a result that depends on two key assumptions, namely, (1) $\epsilon$ is not far from the value provided by recombination and (2) the line emission scales with the mean density of the universe.

%%%%%%
%
\smallskip\noindent$\bullet$ {\em Emission expected scaling the signal found by} \citet[][]{2014Natur.506...63C}.
%
% Taken form the notes I wrote down for Amanda's thesis, Pag. 67ii
The work by \citeauthor[][]{2014Natur.506...63C} is used as a reference to quantify the Ly$\alpha$ signals observed at $z$ between 2 and 3, but similar results are obtained using any of the other works mentioned in Sect.~\ref{sec:intro} \citep[e.g.,][]{2014ApJ...786..107M,2016A&A...587A..98W,2018Natur.562..229W}. The peak surface brightness of the Ly$\alpha$ structure detected by \citet[][]{2014Natur.506...63C}  is 
$S\!B_{\rm Ly\alpha}(3)\simeq {3\times 10^{-17}}~\cgsunits$. % {\rm erg\,sec^{-1}\,cm^{-2}\,arcsec^{-2}}
The Ly$\alpha$ signal presents a range of values of around a factor of 30, therefore, considering Eq.~(\ref{eq:equal}) $F_{\rm H\alpha}(0)$ is expected to be in between $1$ and $30$ in units of $10^{-18}~\cgsunits$. 

\citet{2018MNRAS.480.2094L} measured the H$\alpha$ emission corresponding to the Ly$\alpha$ nebula detected by \citet{2014Natur.506...63C}. H$\alpha$ is in the near infrared at the redshift of the source, so, they use the spectrograph MOSFIRE\footnote{\tt https://irlab.astro.ucla.edu/instruments/mosfire/} on the Keck telescope. The H$\alpha$ surface brightness is around $3\times 10^{-18}\,\cgsunits$, with a Ly$\alpha$ to H$\alpha$ flux ratio of 5.5. This observational result is consistent with the above estimates.  If we correct for the variation with redshift, the H$\alpha$ signal would be around  $10^{-17}\,\cgsunits$, as listed in Table~\ref{tab:summary}. 

%%%%%
%
\smallskip\noindent$\bullet$ {\em Fluorescence of UV photons.}
UV photons from the cosmological background or from nearby sources  are expected to photo-ionize the gas clouds existing in the CGM and IGM. The recombination of the photo-ionized H produces Ly$\alpha$ photons that, after internal scattering, can escape producing a diffuse Ly$\alpha$ signal that traces the cloud.  In optically thick clouds, the mechanism very effectively transforms the UV photons into Ly$\alpha$ photons, with an efficiency approaching one \citep[e.g.,][]{1996ApJ...468..462G}. This whole process is often refereed to as {\em fluorescence} \citep[e.g.,][]{2005ApJ...628...61C,1996ApJ...468..462G}. Since photons are finally produced through recombination, fluorescence  produces all the other recombination lines of H, including H$\alpha$. 

There are works estimating the Ly$\alpha$ signals expected due to the illumination by the cosmic UV background and by nearby quasars.   Using cosmological simulations of gas clouds at $z\simeq 3$, \citet[][]{2005ApJ...628...61C} work out the Ly$\alpha$ signals, that turn out to be in the range between   $3\times 10^{-20}~\cgsunits$ and a few times $10^{-19}~\cgsunits$, the latter corresponding to the presence of a nearby quasar. \citet{2010ApJ...708.1048K} carry out a similar exercise rendering larger signals that can reach a value of $10^{-17}\cgsunits$ if the quasar illumination is strong enough.  The signal excited by the UV background is in the range of $10^{-20}~\cgsunits$. For more references on the estimates see, e.g., \citet{2012MNRAS.425.1992C}. 
The scaling between the fluxes in H$\alpha$ and Ly$\alpha$ in Eq.~(\ref{eq:equal}) can also be used in this case to render the range of expected H$\alpha$ fluxes included in Table~\ref{tab:summary}.

\citet{2017ApJ...849...51B} model the H$\alpha$ emission in the outer gas disk of  MW-like galaxies produced by fluorescence of photons coming from the cosmic ionizing background (CIB). The estimated signals are in the range between $10^{-19}$--$10^{-21}~\cgsunits$. 

%%%
\smallskip\noindent$\bullet$ {\em Emission driven by mechanical feedback from galaxies.}
The mechanical energy injected by galaxy winds may also lead to shocks that are powerful sources of ionizing photons produced in the hot post shock plasma as it cools down \citep[e.g.,][]{1995ApJ...455..468D}. They may be responsible for some of the extended Ly$\alpha$ emission found at $z\sim 2$\,--\,3 \citep[e.g.,][]{2000ApJ...532L..13T,2003ApJ...591L...9O}. \citet[e.g.,][]{2004ApJ...613L..97M} model the winds produced by SN explosions in primordial galaxies at $z = 3$,  and work out the emission properties assuming an optically thin gas in collisional ionization equilibrium. The total flux in Ly$\alpha$ is around $10^{-16}~{\rm erg\,sec^{-1}\,cm^{-2}}$, and it is emitted in region of around 10\,arcsec, which renders a surface flux density of some $10^{-18}~{\rm erg\,sec^{-1}\,cm^{-2}\,arcsec^{-2}}$. According to Eq.~(\ref{eq:equal}), this is also the level of signal to be expected if observing the shocks in H$\alpha$ in the local Universe.

%%%
%
\smallskip\noindent$\bullet$ {\em Gravity driven emission.}
The potential energy of the gas falling into a gravitational well is transformed into kinetic and thermal energy, and has to be released for the gas to become gravitationally bound. This energy is partly radiated away in the form of H recombination lines \citep{2009MNRAS.400.1109D}.

There are several studies in the literature that model this process and the resulting line emission. We take the work by 
\citet{2010MNRAS.407..613G} as reference. They use cosmological numerical simulations of galaxy formation to work out the Ly$\alpha$ flux to be expected from the gas streams that fed galaxies at high redshift.  UV background excitation, collisions with free electrons, and dust attenuation are included when computing the Ly$\alpha$ emission.  They emission is mainly driven by the excitation produced by collisions with electrons since the filaments are dense enough to be shielded from the UV background. Most of the Ly$\alpha$ emission comes from extended (50\,--\,100\,kpc) narrow, partly clumpy, inflowing, cold streams of $\sim\,10^4$\,K that feed the growing galaxies. The predicted morphology is irregular, with dense clumps and elongated extensions.  The typical surface brightness increases with decreasing distance from the halo centre.
The peak surface brightness is $2\times 10^{-17}$ $\cgsunits$ for MW-like halos at $z = 2.5$, with typical signals in the range from $3\times 10^{-17}$ to  $5\times 10^{-19}\, \cgsunits$. 
No estimate for the H$\alpha$ flux is given, however, they mention that {\em other H emission 
lines are expected to be two orders of magnitude less luminous in our model}. H$\alpha$ is expected to be the strongest one so we assume $\epsilon$ (i.e. the ratio Ly$\alpha$ and H$\alpha$) to be 20. Using this value for $\epsilon$ in Eq.~(\ref{thirdeq}), we end up with the expected H$\alpha$ surface brightness range reported in Table~\ref{tab:summary}.

\citet{2010ApJ...725..633F} carry out a similar simulation but with more sophisticated radiative transfer. As in the work by \citet{2010MNRAS.407..613G}, the signals are concentrated towards the center of the dark matter halos, but some also emerge from the IGM.  Depending on the treatment of the radiative transfer, the predicted Ly$\alpha$ signals can vary a lot, going from $10^{-17}$ to $5\times 10^{-21}~\cgsunits$. Re-scaling to H$\alpha$ and to the local Universe, as we did for  \citeauthor{2010MNRAS.407..613G} signals, the expected H$\alpha$ surface brightness is in the range from $2\times 10^{-18}$ to $10^{-21}\, \cgsunits$ (Table~\ref{tab:summary}).

\citet{2003ApJ...599L...1F} worked out predictions directly for the local Universe ($z\sim 0.15$), therefore, there is no need to apply high-$z$ corrections in this case. Their Fig.~2 shows structures producing between $10^2$ to $10^6$~photons~cm$^{-2}$~s$^{-1}$~sr$^{-1}$, which properly transformed into the usual units render $S\!B_{\rm Ly\alpha}(0)$ in the range between  $4\times 10^{-20}$ and  $4\times10^{-16}~\cgsunits$. Correcting only for the ratio of emissivities between H$\alpha$ and Ly$\alpha$, $S\!B_{\rm H\alpha}(0)$ turns out to be between  $2\times 10^{-21}$ and  $2\times10^{-17}~\cgsunits$. 

Likewise, \citet{2012MNRAS.423..344R} model the Ly$\alpha$ emission in numerical simulations. The emitting gas is at the center of the dark matter halos radiating away the excess of  gravitational energy explicitly modeled in the previous works. The expected signals are similar to the ones resulting from the previous simulations. Updated numerical simulations by \citet{2021A&A...650A..98W} at $z > 2$ give results consistent with the previous ones.  

%
%%%%%%%%%%%%%%%
%
\smallskip\noindent$\bullet$ {\em Observed diffuse H$\alpha$ emission in the local Universe.}
The existing H$\alpha$ surveys in the local Universe have a surface brightness limit insufficient to detect the faint signals to be expected (Table~\ref{tab:summary}): equal or fainter than $10^{-17}~\cgsunits$. This level is not within reach of the existing surveys \citep[e.g.,][]{1988MNRAS.232..381M,2004A&A...414...23J,2008MNRAS.388..500E,2008AJ....135.1412D,2016MNRAS.462...92J,2021arXiv210712897G}. Moreover, some of them are focused on the emission of the MW, either diffuse \citep[e.g.,][]{2001PASP..113.1326G} or its point-like sources \citep[e.g.][]{2014MNRAS.444.3230B}. 

However, diffuse H$\alpha$ emission have been detected around individual galaxies of the local Universe when the observation is deep enough.  As part of the LARS\footnote{Ly$\alpha$ Reference Sample \citep{2014ApJ...797...11O}.} project to study the visibility, strength, and escape fraction  of the Ly$\alpha$ photons in 14 galaxies the local universe, \citet{2014ApJ...797...11O} and \citet{2014ApJ...782....6H} obtained deep H$\alpha$ maps hinting at the existence of extended emission. Although the sensitivity is good  \citep[{around $10^{-18}~\cgsunits$;}][]{2013ApJ...765L..27H}, the FOV includes only a few effective radii around the galaxies, therefore, missing most of the CGM and the full IGM. Flux-wise, the integrated emission observed in H$\alpha$ and Ly$\alpha$ are comparable.

\citet{Herrenz17} detect filaments of ionised gas in the CGM of the extremely metal-poor galaxy  SBS 0335-052E. The  features are detected in H$\alpha$ and [O{\sc iii}]$\lambda$5007 down to a limiting surface brightness of $5\times 10^{-19}~\cgsunits$. The galaxy diameter is around 3~kpc, with  the filaments extending more than 9 kpc and connecting seamlessly in velocity space to the galaxy. 
Extended H$\alpha$ emission at a level of  $5\times 10^{-18}~\cgsunits$ have been found around 3 dwarf galaxies by \citet{2016ApJ...817..177L}. 
\citet{2016ApJ...832..182C} find a lone H$\alpha$ emitting blob next to an evolved galaxy with a signal around $3\times 10^{-18}~\cgsunits$.
The flux representative of these observations is included in Table~\ref{tab:summary}.

%
%%%%%%%%%%%%%%%%%%%%%%%%%%%%%%
% This comes from the notes I developed for Amanda's thesis. Pag. 67ii
\section{Contamination by background sources}\label{app:a}

Assume that emission line sources with luminosity function $\Phi(L,z)$ are leaking into the band-pass of the filter $f(\lambda)$ used to detect H$\alpha$,
\begin{equation}
f(\lambda)=\Pi\Big(\frac{\lambda-\lambda_{obs}}{\Delta\lambda}\Big),
\end{equation}
with $\lambda_{obs}$ and $\Delta\lambda$ representing the central wavelength and the width of the filter, respectively.  In our case, the filter has a top-hat shape,
\begin{equation}
\Pi(x)=\cases{1&$|x| \le 1/2$,\cr
0&elsewhere,}
\end{equation}
with $\Delta\lambda= 20$\,\AA\  and $\lambda_{obs}$ the wavelength of H$\alpha$ ($\lambda_{\rm H\alpha}$) at the redshift of the central galaxy $z_g$, i.e., $(1+z_g)\,\lambda_{\rm H\alpha}$. If the contaminants are Ly$\alpha$ emitters, the window function defining the range of redshifts $z$ where they contribute is
\begin{equation}
W(z) = f\Big[(1+z)\lambda_{Ly\alpha}\Big].
\label{eq:redshift}
\end{equation}
Given any window function, the total number of targets leaking in is  (e.g.,  \citeauthor{1993ppc..book.....P}~\citeyear{1993ppc..book.....P}, Eqs. [13.60] and [13.61]; \citeauthor{2001A&A...367..788F}~\citeyear{2001A&A...367..788F}, Eq.~[3]),
\begin{equation}
N=\Omega \int_0^\infty\, W(z) \Big(\int_{Lmin(z)}^{Lmax(z)}\, \Phi(L,z)\,dL\Big)\,{{dV(z)}\over{dz}}\,dz,
\label{eq:main}
\end{equation}
where  $dV(z)/dz$ represents the variation with redshift of the comoving volume of the universe per unit solid angle, and  $\Omega$ is the solid angle of the field covered by the measurement. 
If $F_{min}$ is the minimum flux density that we can observe (with $[F_{min}]=$\,erg\,s$^{-1}$\,cm$^{-2}$), then
\begin{equation}
L_{min}(z)=4\,\pi\,D_L^2(z)\,F_{min},
\end{equation}  
with $D_L$ the luminosity distance  \citep[e.g.,][]{1993ppc..book.....P,1999astro.ph..5116H}. On the other hand, we do not include in our sampling sources brighter than $F_{max}$, otherwise they would be detected as individual galaxies and excluded from the search. The value of $F_{max}$ sets $L_{max}$, explicitly, 
\begin{equation}
L_{max}(z)=4\,\pi\,D_L^2(z)\,F_{max}.
\end{equation}  
Since the redshift window $W(z)$ is very narrow, all other functions of $z$ in Eq.~(\ref{eq:main}) can be regarded as constant and so can be pulled out of the integral, rendering  
\begin{equation}
N\simeq C\, \Omega\, \frac{\Delta \lambda}{\lambda_{\rm Ly\alpha}} {{dV(z_{\rm Ly\alpha})}\over{dz}}\,\int_{Lmin(z_{\rm Ly\alpha})}^{Lmax(z_{\rm Ly\alpha})}\, \Phi(L,z_{\rm Ly\alpha})\,dL,
\label{eq:approx}
\end{equation}
with 
\begin{equation}
z_{\rm Ly\alpha} = \frac{\lambda_{\rm H\alpha}}{\lambda_{\rm Ly\alpha}} (1+z_g)-1.
\end{equation}
A global fudge factor $C$ has to be included in Eq.~(\ref{eq:approx}) to account for the completeness of detections. This factor can be far from unity when the expected signals are close to the noise level, which is our case \citep[e.g.,][]{2005ApJ...631..208B,2019A&A...621A.107H}. 

All in all, the number of Ly$\alpha$ emitters leaking in ($N$ in Eq.~[\ref{eq:approx}]) depends on the luminosity function $\Phi$ at redshift $z_{\rm Ly\alpha}$, the flux limits ($F_{min}$ and $F_{max}$), the angular extend of the field-of-view around each central galaxy ($\Omega$), and the cosmological parameters that define the Universe (through $dV/dz$ and $D_L$).

\begin{figure*}
\centering
\includegraphics[width=0.45\linewidth]{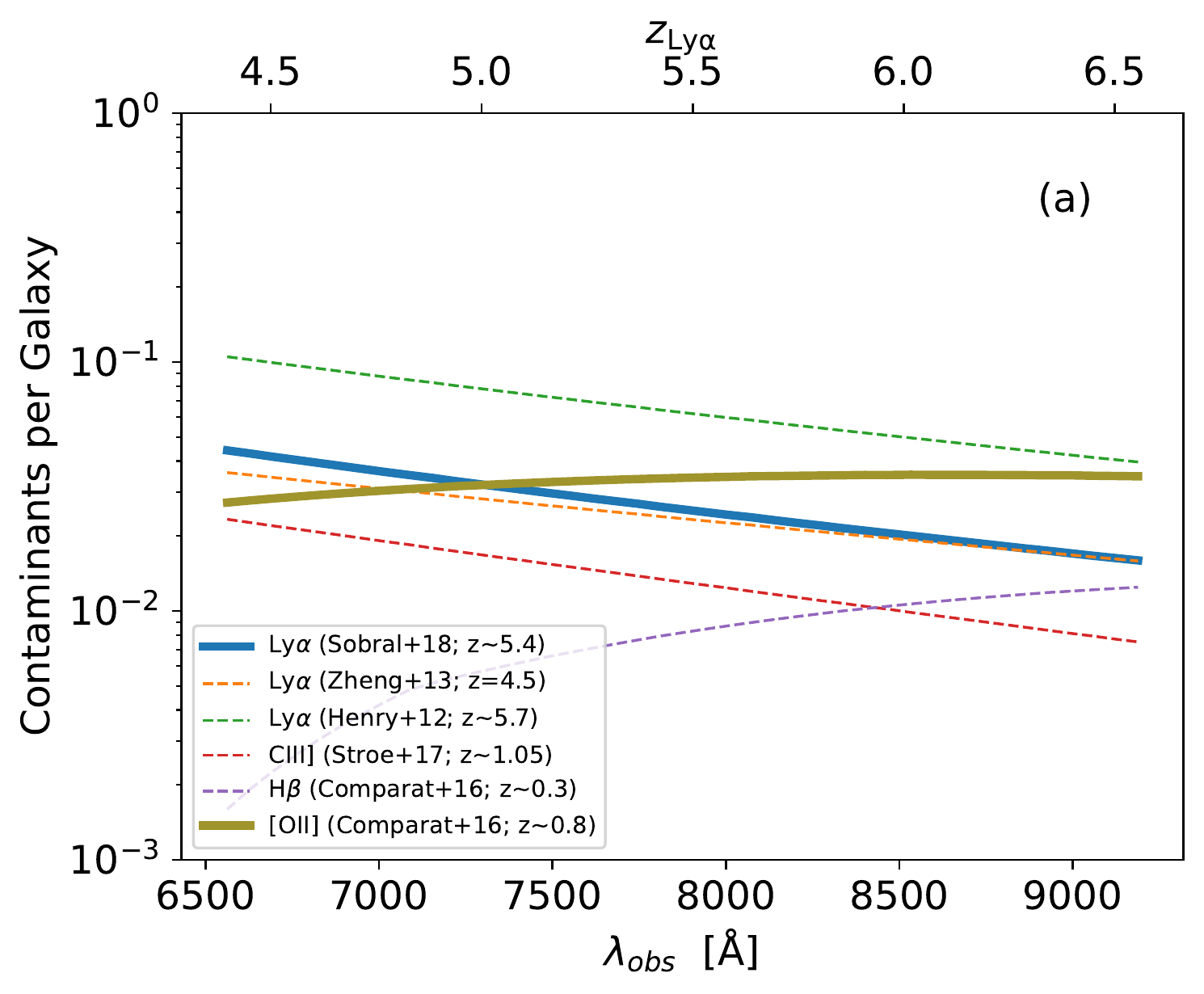}
\includegraphics[width=0.45\linewidth]{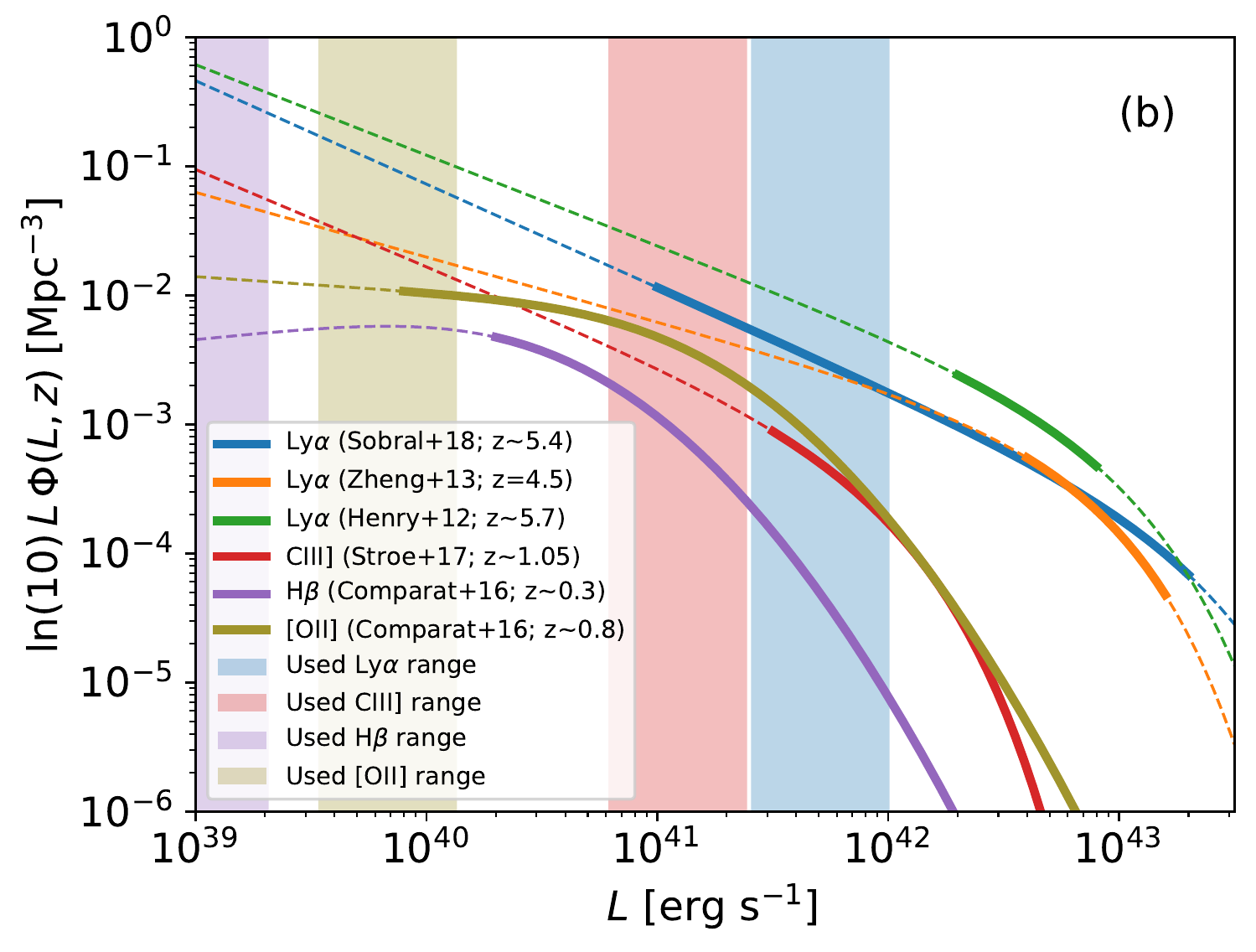}
\caption{(a) Expected number of background sources contaminating our signals. We give the number of contaminants per observed galaxy as function of the observed wavelength ($\lambda_{obs}=[1+z_g]\lambda_{\rm H\alpha}$). The axis on top shows the redshift if the contaminants were Ly$\alpha$ emitters. The dashed lines correspond to estimates based on luminosity functions extrapolated outside the observed range of luminosities, whereas the solid lines use the range of observed luminosities.  (b) Luminosity functions used to compute the number of contaminants in (a). They are either Schechter functions or \citeauthor{1990MNRAS.242..318S} functions based on luminosities observed within the range shown as solid lines, with the luminosity function outside this range shown as dashed lines. The color code is the same as in panel (a). The colored regions indicate the range of luminosities used to estimate the number of contaminants. As indicated in the inset, we include several high redshift Ly$\alpha$ luminosity functions \citep[][]{2012ApJ...744..149H,2013MNRAS.431.3589Z,2018MNRAS.476.4725S}, a C{\sc iii}]$\lambda$1909 luminosity function \citep[][at redshift 1]{2017MNRAS.471.2575S}, 
a [OII]$\lambda\lambda3726,3729$ luminosity function \citep[][at redshift 0.8]{2016MNRAS.461.1076C}, and a H$\beta$ luminosity function \citep[][at redshift 0.3]{2016MNRAS.461.1076C}. As usual, rather than the actual luminosity function $\Phi(L,z)$, the plot shows the luminosity function per $\log L$ interval ($=\ln[10]\times L\times \Phi[L,z]$).}
\label{fig:leaking_in}
\end{figure*}

Figure~\ref{fig:leaking_in}a uses  Eq.~(\ref{eq:approx}) to work out the expected number of contaminants per galaxy assuming typical values for the parameters describing our observation: $F_{min}=10^{-17.9}$~erg\,sec$^{-1}$\,cm$^{-2}$,  $F_{max}= 4\times F_{min}$ (Fig.~\ref{fig:Gaussian_Fits_Histogram}), $C=0.2$ \citep[][]{2019A&A...621A.107H}, and $\Omega = \pi r_{obs}^2$ with $r_{obs}= 100$\,arcsec (Sect.~\ref{sec:data_analysis}).  The derivative of the comoving volume ($dV/dz$) and $D_L$ have been computed using the {\tt astropy cosmology} package \citep{2013A&A...558A..33A,2018AJ....156..123A} and the cosmological parameters adopted in the text (Sect.~\ref{sec:intro}). The Ly$\alpha$ luminosity functions used in the estimate are represented in  Fig.~\ref{fig:leaking_in}b \citep[][]{2012ApJ...744..149H,2013MNRAS.431.3589Z,2018MNRAS.476.4725S}.  Figure~\ref{fig:leaking_in}b also shows the range of luminosities used to estimate the number of contaminants (the blue band). Only the luminosity function worked out by \citet{2018MNRAS.476.4725S} reaches  faint enough luminosities for the estimate to be based on actual Ly$\alpha$ emitter counts (i.e., the blue {\em solid} line and the blue band in Fig.~\ref{fig:leaking_in}b overlap). This luminosity function predicts less than 0.03 contaminants per central galaxy at $\lambda_{obs}\simeq 8500$\,\AA (Fig.~\ref{fig:leaking_in}a), which corresponds to $z_g\simeq 0.3$ where most signals appear (Fig.~\ref{fig:histo_redshifts}). This level of contamination is insufficient to account for the number of observed signals, of the order of one per galaxy (Sect.~\ref{sec:statistics}).   

Note that above formalism holds for any other emission line replacing $\Phi$ and $\lambda_{\rm Ly\alpha}$ with the corresponding luminosity function and rest-frame wavelength of emission. We expected that these other contaminants are less important than the Ly$\alpha$ emitters because they are less numerous for the same $L$ and because they sample a smaller redshift range (i.e., they sample a smaller volume of Universe). To check it out, we also include in Fig.~\ref{fig:leaking_in} the luminosity function and the expected number of counts for C{\sc iii}]$\lambda$1909, [O{\sc ii}]$\lambda\lambda3726,3729$, and H$\beta$ emitters \citep[][]{2017MNRAS.471.2575S,2016MNRAS.461.1076C}. The largest number of expected counts correspond to [O{\sc ii}] emitters, and it is similar to the (low number of) Ly$\alpha$ contaminants and so insufficient to account for observation presented in the paper.   

%%%%%%%%%%%%
\section{Line shapes expected from partly corrected telluric lines} \label{app:b}
\begin{figure*}
\centering
\includegraphics[width=0.8\linewidth]{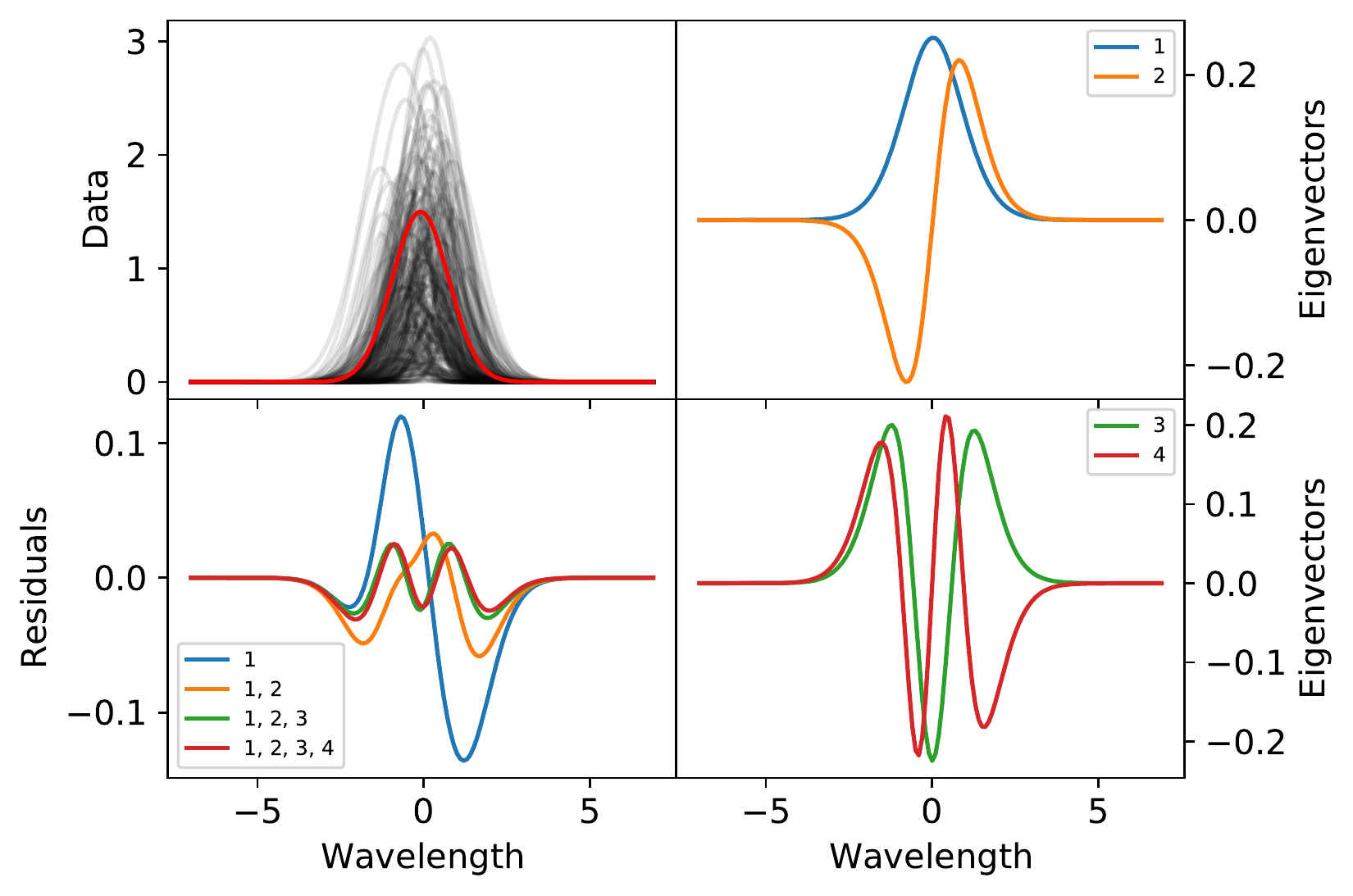}
\caption{Monte Carlo simulation showing the residuals to be expected under insufficient telluric line contamination correction. Top left panel: mock emission lines used to construct the PCA database that reproduce them all (black solid lines). Right panels: first 4 eigenvectors of the resulting PCA database. Bottom left panel: residual left when the profile shown as the red solid line in the top left panel is represented in the PCA database using  different sets of eigenvectors (as indicated in the insets). Intensities and wavelengths are given in arbitrary units.}
\label{fig:pca}
\end{figure*}
Telluric line contamination is one of the main concerns to interpret the detected signals (Sect.~\ref{skylines}). This appendix describes a schematic Monte Carlo simulation carried out to show the shapes to be expected if telluric line residuals were present in the observed spectra. The telluric line removal is based on a principal component analysis (PCA) decomposition of the spectra in the part of the FOV assumed to have no astronomical signal and thus assumed to represent {\em sky emission} \citep{2016MNRAS.458.3210S}. Simply put, such sky emission signal is used to compute the eigenvectors that characterize the telluric contamination. Then the spectrum of each spaxel of the full FOV is decomposed in this PCA database, and the sky emission spectrum resulting for every spaxel is subtracted out from the observed spectrum. In a very schematic representation to illustrate the procedure, we carried out the PCA decomposition of $10^4$ mock Gaussian emission line spectra with various strengths, widths, and shifts: the grey lines in the top left panel of Fig.~\ref{fig:pca}. They mock any of the telluric emission lines present in the \muse\ FOV. The first four eigensvectors of this decomposition are shown in Fig.~\ref{fig:pca}, right panels. Changes in line width are described by odd eigensvectors whereas changes in line shift are represented by even eigenvectors. This PCA database is used to represent the {\em problem} profile shown as the red solid line in Fig.~\ref{fig:pca}, top left panel. The residuals left when 1, 2, 3, and 4 eigenvectors are included in the representation of the {\em problem} profile are shown in the bottom left panel of Fig.~\ref{fig:pca}. As the number of eigenvectors increases, the residuals have more wiggles. The lobes of the residuals are positive and negative. Importantly, the observed H$\alpha$ shapes (Sect.~\ref{sec:results}) do not have any of the properties of these mock residuals. % On the contrary, the residuals left in H$\alpha$ when integrating the sky without emission do have these properties (see Fig.~\ref{fig:super_residuals}). 
Although this Monte Carlo simulation is very schematic, and in reality the PCA decomposition of the sky is based on many emission lines, the shape of the true residuals do not vary much from our simple modeling  \citep[see Figs.~2 and 3 in][]{2016MNRAS.458.3210S}. Thus, we use the difference in shape between the observed H$\alpha$ profile and the expected residuals as an additional argument that disfavor the observed signals to be caused by telluric line contamination.    
Similar results and identical conclusions are reached when using other Monte Carlo realizations and problem profiles.  

%
%%%%%%%%%%%%%%%%%%%%%%%%%
\section{X-ray stacking}\label{sec:xray_stacking}

The cross-matching of our sources with the available Chandra X-ray catalogue for the CDFS field did not yield positive results (Sect.~\ref{sec:Xrays_radio_match}). In order to improve the X-ray sensitivity, we resort to stacking techniques. We make use of the images publicly available from \cite{2017ApJS..228....2L} and create stacks of our candidates using mean and median statistics.

\begin{figure} %%%%%
\centering
\includegraphics[width=0.9\linewidth]{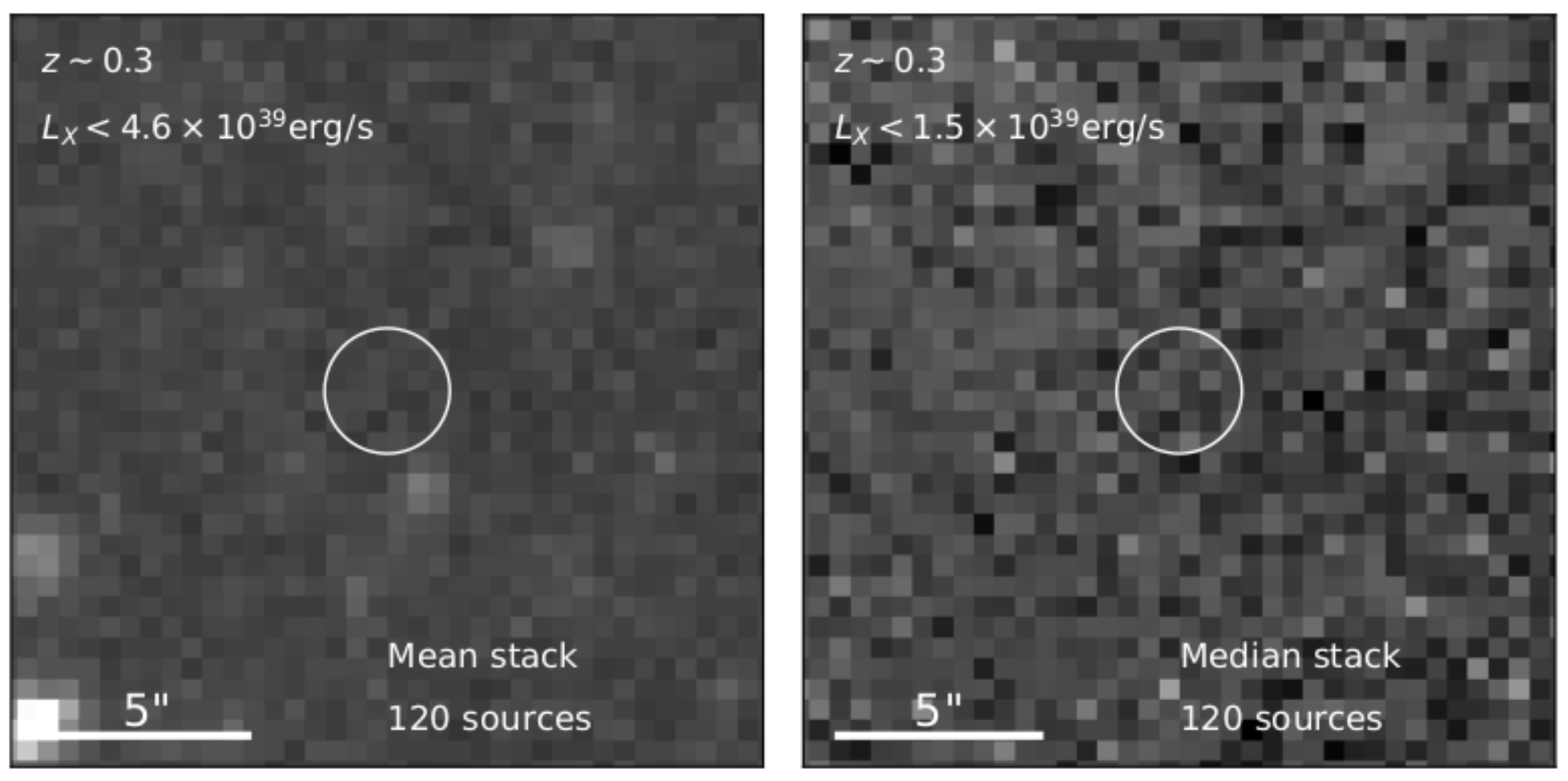}
\caption{
The result of stacking our candidates in full band (0.5-7 keV) 7Ms X-ray Chandra images. The left panel shows mean stacking whereas the right panel shows median stacking. The mean stacking reveals the presence of some clear sources, likely in the vicinity of several of our clumps, but no actual detection in the position of our candidates (marked with a circle), a lack of signal also supported by the median stack. Our mean stacking reveals an X-ray non-detection with an upper limit luminosity of $\rm L_X < 4.6 \times 10^{39} \, erg \, s^{-1}$. The  images have been scaled to their minimum  (black) and $\rm 0.85 \times maximum~(white)$ values.
%\comment{Scale of grays used in the images? are they scaled between min(black) and max (white)? Are the scales the same for the two panels?}
}
\label{fig:x-ray_stack}
\end{figure}

We start by creating cut-outs of $\rm 100 \times 100\,px^2$ centered around each gas candidate, then we stack the associated exposure maps to create images with the mean and the median counts\,s$^{-1}$ in every pixel (Fig.~\ref{fig:x-ray_stack}). The signal is measured as the central count rate in the stack using an aperture of $\sim 1.5 \arcsec$ ($\sim 3$\,px) radius, which is large enough to account for the sizes of the majority of our clumps (Sect.~\ref{sec:physical_properties}). We estimate the noise by randomly sampling the signal in the stacked image with the same aperture and avoiding sources. The distribution of count rates  of 2000 such random apertures is used to compute the noise. Specifically, we use as noise three times the difference between the mean and the 16th percentile of the distribution, a value equivalent to the 3-$\sigma$ error bar for a Gaussian distribution.
%The count/s in each aperture are summed and the median value of $\rm \sim 2000$ apertures is taken as the background value for the stack. The uncertainty is measured through asymmetric errors, with the lower and upper errors taken from the 16th and 84th percentile of the backgrounds. The signal-to-noise ratio (S/N) is defined as the ratio between the net counts and the lower error for each stack. A stack is considered to have a detection if $\rm S/N \geq 3$. 

We stack the X-ray cutouts of our candidates using mean and median statistics for two different groups. The first one uses all level~2 and 3 candidates whereas the second one includes only the candidates who present evidence of multiple peaks. We find no evidence of X-ray emission in any of the stacks (Fig.~\ref{fig:x-ray_stack} shows the result of stacking all sources with mean and median statistics). 
%, and although the stack using only the multiple peak candidates does appear to show a stronger signal in the median stack, it is not strong enough to classify as a detection ($\rm \sigma \sim 1.01$). 
However, we are still in a position to provide upper limits for the X-ray flux and luminosity of our stacked candidate.

%We estimate the x-ray flux associated with the count rate b( using a conversion factor to transform the count rate into flux quantities. 
We determine the conversion factor between count rate and X-ray flux (in the Chandra 0.5-7 keV band) using all the sources in \cite{2017ApJS..228....2L}. The mean of the ratio between the flux of each source and its count rate is used as conversion factor\footnote{ $\rm (1.5\pm 0.5) \times 10^{-11}\,\, erg \, s^{-1} \, cm^{-2}º\,(counts\,s^{-1})^{-1}$.}. 
Using this calibration, the flux corresponding to the 3-$\sigma$ noise level of the mean stack considering all candidates and those with multiple peaks are $\rm 1.18 \times 10^{-17} \, erg \, s^{-1}\,cm^{-2}$ and $\rm 6.5 \times 10^{-18} \, erg \, s^{-1}\,cm^{-2}$, respectively   %\comment{complete}. 
The values using median stacking are similar. The corresponding X-ray luminosity is computed using Eq.~(\ref{eq:lumha}), with the luminosity distance $D_L$ set by the median redshift of the clumps in the stack.
The luminosity equivalent to the noise level ($L_X$) is of the order of $\rm 4.6 \times 10^{39} \, erg \, s^{-1}$, for the average stacking of the full sample, and $\rm 2.1 \times 10^{39} \, erg \, s^{-1}$, for the stacking using only multiple peaks. 

%\joao{17/11/2021 - Downloaded an actual backup copy (Images, tex file, biblio) because Overleaf gave me a heart attack yesterday by going 404 sometime in the afternoon...}

%\jorge{
%\comment{From Mar Mezcua}
%From the X-ray luminosity upper limit of $L_\mathrm{X} < 10^{40}$ erg s$^{-1}$ we can estimate an upper limit on the black hole of the putative IMBHs as $M_\mathrm{BH}$ = $L_\mathrm{bol}$/($\lambda_\mathrm{Edd} \times 1.3 \times 10^{38}$), where $\lambda_\mathrm{Edd}$ is the Eddington ratio, $L_\mathrm{bol}$ is the bolometric luminosity $L_\mathrm{bol} = k L_\mathrm{X}$ and $k$ is the bolometric correction factor. Low-mass AGN have been commonly found to have near-Eddington accretion rates (e.g. Greene \& Ho 2007; Baldassare et al. 2016; Mezcua et al. 2018). Assuming $\lambda_\mathrm{Edd}$ = 1 and $k = 10$ typical of low-mass AGN (Marconi et al. 2004; Mezcua et al. 2018) we find an upper limit on the black hole mass of $M_\mathrm{BH}$ = 769 M$_{\odot}$. Taking the $L_\mathrm{X}$ upper limit of 2.1 $\times$ 10$^{39}$ erg s$^{-1}$ derived from the CDFS stacking we would obtain $M_\mathrm{BH}$ = 154 M$_{\odot}$. 
%}

%

%%%%%%%%%%%%%%%%%%%%%%%%%%%%%%%%%%%%%%%%%%%%%%%%%%%%%%%
%\bibliography{biblio}{}
%\bibliographystyle{aasjournal}



\end{document}